\documentclass[longauth]{aa} 
\usepackage{longtable,lscape}
\usepackage{natbib}
\usepackage{graphicx}
\usepackage{amsmath}

\usepackage{tabularx}
\usepackage[labelformat=simple]{caption,subcaption}
\usepackage{multirow}
\usepackage{ltablex}

\usepackage[colorlinks=true,citecolor=blue]{hyperref}
\def\m2s2{\,m$^{2}$\,s$^{-2}$} 

\begin{document}
\title{The SPHERE infrared survey for exoplanets (SHINE)\thanks{Tables 5-11 are only available in electronic form at the CDS via anonymous ftp to cdsarc.u-strasbg.fr (130.79.128.5) or via http://cdsweb.u-strasbg.fr/cgi-bin/qcat?J/A+A/}}
\subtitle{I Sample definition and target characterization}

\author{
S. Desidera \inst{1}, G. Chauvin \inst{2}, M. Bonavita \inst{3,4,1},
S. Messina \inst{5}, H. LeCoroller \inst{6}, T. Schmidt \inst{7,8}, 
R. Gratton \inst{1}, C. Lazzoni \inst{1,9}, M. Meyer \inst{10,11}, 
J. Schlieder \inst{12,13}
% ...
}

\institute{
INAF-Osservatorio Astronomico di Padova, Vicolo dell'Osservatorio 5,  35122 Padova, Italy 
\and
Univ. Grenoble Alpes, CNRS, IPAG, F-38000 Grenoble, France 
\and SUPA, Institute for Astronomy, University of Edinburgh, Blackford Hill, Edinburgh EH9 3HJ, UK 
\and  Centre for Exoplanet Science, University of Edinburgh, Edinburgh EH9 3HJ, UK 
\and
INAF - Osservatorio Astrofisico di Catania, Via S. Sofia 78, I-95123, Catania, Italy 
\and Aix Marseille Univ, CNRS, CNES, LAM, Marseille, France 
\and Hamburger  Sternwarte,  Gojenbergsweg  112,  D-21029  Hamburg, Germany 
\and LESIA, Observatoire de Paris, Universit\'e PSL, CNRS, Sorbonne Universit\'e, Universit\'e de Paris, 5 place Jules Janssen, 92195 Meudon, France 
\and Dipartimento di Fisica e Astronomia Galileo Galilei, Universit{\'a} di Padova, Vicolo dell'Osservatorio 3, I-35122, Padova, Italy 
\and Department of Astronomy, University of Michigan, Ann Arbor, MI 48109, USA
\and Institute for Particle Physics and Astrophysics, ETH Zurich, Wolfgang-Pauli-Strasse 27, 8093 Zurich, Switzerland  
\and Exoplanets and Stellar Astrophysics Laboratory, Code 667, NASA Goddard Space Flight Center, 8800 Greenbelt Rd., Greenbelt, MD 20771, USA 
\and Max Planck Institute for Astronomy, K\"onigstuhl 17, D-69117 Heidelberg, Germany 
% ...
}

\date{}
 
\abstract
{Large surveys with new-generation high-contrast imaging instruments are needed to derive the frequency and properties of exoplanet populations with separations from $\sim$5 to 300\,au. A careful assessment of the stellar properties is crucial for a proper understanding of when, where, and how frequently planets form, and how they evolve. The sensitivity of detection limits to stellar age makes this a key parameter for direct imaging surveys.} 
{We describe the SpHere INfrared survey for Exoplanets (SHINE), the largest direct imaging planet-search campaign initiated at the VLT in 2015 in the context of the SPHERE Guaranteed Time Observations of the SPHERE consortium. In this first paper we present the selection and the properties of the complete sample of stars surveyed with SHINE, focusing on the targets observed during the first phase of the survey (from February 2015 to February 2017). This early sample composed of 150 stars is used to perform a preliminary statistical analysis of the SHINE data, deferred to two companion papers presenting the survey performance, main discoveries, and the preliminary statistical constraints set by SHINE.} 
{Based on a large database collecting the stellar properties of all young nearby stars in the solar vicinity (including kinematics, membership to moving groups, isochrones, lithium abundance, rotation, and activity), we selected the original sample  of 800 stars that were ranked in order  of  priority according to their sensitivity for planet detection in direct imaging with SPHERE. The properties of the stars that are part of the early statistical sample were revisited, including for instance measurements from the GAIA Data Release 2. Rotation periods were derived for the vast majority of the late-type objects exploiting TESS light curves and dedicated photometric observations.} 
{The properties of individual targets and of the sample as a whole are presented.} 
{} 

\titlerunning{The SPHERE infrared survey for exoplanets (SHINE). I.}
\authorrunning{S. Desidera, G. Chauvin, M. Bonavita et al.}

\maketitle

\section{Introduction}
\label{s:intro}

Today's success in direct imaging of exoplanets is intimately connected to the pioneering work in the previous decades to develop adaptive optics (AO) systems, infrared detectors, coronagraphs, and differential imaging techniques for ground-based telescopes \citep{mawet2012,chauvin2018}. With the advent of dedicated instruments on 5-10 m telescopes (e.g., LBT, Palomar, Subaru, Keck, VLT, Gemini, and Magellan), high-contrast imaging (HCI) demonstrated the ability to detect and characterize exoplanets and planetary systems, confirming that ground-based instrumentation may reach performance levels that could compete with those from space. Early large systematic surveys of young nearby stars led to the discovery of the first planetary-mass companions at large separations ($>100$\,au) or with a low mass ratio relative to their stellar host \citep{chauvin2004,chauvin2005, neuhauser2005,luhman2006,lafreniere2008}, followed by the  breakthrough discoveries of closer  planetary-mass companions such as HR\,8799\,bcde \citep{marois2008,marois2010}, $\beta$\,Pictoris\,b \cite{lagrange2009}, $\kappa$~And\,b \citep{carson2013}, HD\,95086\,b \citep{rameau2013b}, and GJ\,504\,b \citep{kuzuhara2013}. This allowed a first systematic characterization of the giant planet population with separations typically $\ge$ 20-30 au \citep[e.g.,][]{biller2007,lafreniere2007,heinze2010,chauvin2010,vigan2012,rameau2013,nielsen2013,wahhaj2013, brandt2014,nacolp,galicher2016,bowler2016,vigan2017}. These early results confirmed that direct imaging is an important complementary technique in terms of discovery space with respect to other planet hunting techniques like radial velocity, transit, $\mu$-lensing and  astrometry  \citep[e.g.,][]{johnson2010,howard2010, mayor2011, sumi2011,cassan2012,bonfils2013,meyer2018,fernandes2019}. Nowadays, direct imaging brings a unique opportunity to explore the outer part of exoplanetary systems 
with separations beyond $\sim$5\,au to complete our view of planetary architectures, and to explore the properties of relatively cool giant planets. The advent of the new generation of extreme-AO planet imagers like SPHERE \citep{sphere} and GPI \citep{macintosh2014} connected to systematic surveys of hundreds young nearby stars led to new discoveries like 51\,Eri\,b \citep{macintosh2015}, HIP\,65426\,b \citep{chauvin2017}, and PDS\,70\,b and c \citep{keppler2018,haffert2019,mesa2019}. However, the low rate of discoveries despite the unprecedented gain in detection performance achieved with SPHERE and GPI showed that massive giant planets with an orbital semi-major axis beyond 10\,au are rare \citep{nielsen2019}.

On the other hand, the gain in performance allows a much better characterization of exoplanetary systems and exoplanets themselves. In direct imaging the exoplanet’s photons can indeed be spatially resolved and dispersed to directly probe the atmospheric properties of exoplanets and brown dwarf companions.  In comparison to generally older free-floating substellar objects, exoplanets discovered by high-contrast imaging are younger, hotter, and brighter. Their atmospheres show low-gravity features, and the presence of clouds and non-equilibrium chemistry processes. These physical conditions are very different and complementary to those observed in the atmospheres of hot Jupiters (observed in transmission or via secondary-eclipse). A large number of young brown dwarf and exoplanet atmospheres have been systematically characterized to test our current understanding of the processes at play in the atmospheres of substellar objects and to test evolutionary models including HR\,8799\,bcde \citep{ingraham2014,bonnefoy2016,greenbaum2018}, 51\,Eri\,b \citep{rajan2017,samland2017}, $\beta$ Pictoris\,b \citep{chilcote2017}, HD\,95086\,b \citep{derosa2016,chauvin2018b}, HIP\,65426\,b \citep{chauvin2017,cheetham2019}, HIP\,64892\,B \citep{cheetham2018}, and PDS\,70\,b and c \citep{muller2018,mesa2019}. 

Regarding planetary architectures, relative astrometry at 1-2\,mas precision with SPHERE and GPI opens up  a new parameter space to carry out a precise monitoring of the orbital motion of a handful of exoplanets and brown dwarfs. This constrains their orbital properties and allows  the exploration of the dynamical stability of the whole architecture. Examples of systems for which this analysis was done include $\beta$ Pictoris \citep{wang2016, lagrange2019}, HR\,8799 \citep{wang2018}, HD\,95086 \citep{rameau2016,chauvin2018b}, HR\,2562 \citep{maire2018}, 51 Eri \citep{maire2019,derosa2020}, and GJ\,504 \citep{bonnefoy2018}. 

The current instrumentation does not yet allow us to detect mature planets reflecting star light, except perhaps for the closest and brightest stars. The focus of direct imaging programs is therefore on thermal emission from young planets because they are expected to be brighter than their older counterparts. For this same reason they are very useful to directly probe the presence of planets within the environment where they form, the circumstellar disks. This allows us to connect the spatially resolved structures of circumstellar disks (e.g., warp, cavity, rings, vortices) with imaged or unseen exoplanets. This is a fundamental and inevitable path to understand the formation of giant planets, and more generally planetary architectures favorable to the formation of smaller rocky planets with suitable conditions to host life \citep{barbato2018,bryan2019}. Studying the demographics of exoplanets is particularly important in order to understand the architecture and the formation and evolution of exoplanets. Giant planets dominate the architecture of planetary systems from a dynamical point of view, with impact on the subsequent
formation and evolution of smaller planets, the distribution of water in the system, and thus the chances for habitability. 

Within this framework, we planned the SpHere INfrared survey for Exoplanets (SHINE; \citealt{chauvin2017b}). This survey uses a total of 200 nights that were allocated in visitor mode (typically affected by 20\% of poor conditions for AO) and makes up a large fraction of the SPHERE consortium Guaranteed Time Observations allocated by ESO for the design and construction of the SPHERE instrument \citep{sphere}. SHINE has been designed by the SPHERE consortium to: (i) identify new planetary and brown dwarf companions and provide a first-order characterization; (ii) study the architecture of planetary systems (multiplicity and dynamical interactions); (iii) investigate the link between the presence of planets and disks (in synergy with the GTO program aimed at disk characterization); (iv) determine the frequency of giant planets with semi-major axes beyond 5\,au; and (v) investigate the impact of stellar mass (and even age if possible) on the frequency and  characteristics of planetary companions over the range $\sim$0.5 to 3.0 $M_{\odot}$.
 
SHINE started in February 2015 and is planned to be completed by July 2021, with observation of a total of about 500 young nearby stars. This is the first in a series of three papers describing early results obtained from the analysis of about one-third of this very large sample, in which we consider only those targets whose first observation was done before February 2017. 
We chose this cut-off date as second-epoch observations with time separations of 1-2 years were required for a large number of candidates to vet physical companions from field stars (mainly background).
This sample, hereafter referred to as F150 (as it consists of 150 stars), is already large enough for a first statistical discussion of the incidence of massive planets at a separation $\ge 5$~au, and to have a first indication of the formation scenarios for giant planets. This paper describes the general characteristics of the survey and the observed sample. A second paper (Langlois et al. 2020, A\&A, in press; Paper II) describes the observations and analysis methods, and presents the results in terms of detection and upper limits, while a third paper
 \citep[][Paper III]{vigan2020} presents the statistical analysis and a discussion of the implications, as derived from the F150 sample.

The paper is organized as follows: Section \ref{s:design} describes the SHINE survey, its science goals, and the target selection criteria. Section \ref{s:sample} presents the selection of the complete SHINE sample and the priority ranking. Section \ref{s:f100} describes the F150 subsample used for the early statistical analysis. In Sect. \ref{s:param} we derive the most relevant parameters of individual targets and in Sect. \ref{s:properties} we present the ensemble properties of the sample. Section \ref{s:conclusion} summarizes the results. Appendix \ref{a:notes} includes notes on individual targets.

\section{Design of the SHINE survey}
\label{s:design}

In order to achieve its scientific goals, the design and selection of the SHINE sample was of prime importance to optimally exploit a total of 200 nights of Guaranteed Time Observations with SPHERE at VLT dedicated to this campaign. Since massive planets at large separations are rare with a typical frequency of a few percent \citep[see, e.g.][]{vigan2017, nielsen2019}, several hundred targets must be surveyed in order to lead to new discoveries and set precise constraints on the occurrence of giant planets beyond 5\,au. Having a sample that is complete (in terms of distance, age, or limiting magnitude) likely implies a low efficiency. On the other hand, a proper statistical discussion requires well-defined selection criteria. The approach we considered to combine these apparently conflicting issues was to start from a very large sample of potential targets for which a wide set of properties, including magnitude, distance, mass, and age, was known (determined by us). We then divided them into priority groups according to a figure of merit (FoM) determined from these properties. A higher value for this FoM implies a higher probability that a star has a planet possibly detectable by SPHERE according to a specified model describing the planet distribution. To reduce the possibility that results will be poor because the selected model is not appropriate, the final priority list was actually obtained combining rankings given by two completely different models (see Sect.~\ref{sec:simul}). This approach allows a reasonably high efficiency in detecting planets combined with the requirement of well-defined selection criteria, that can be finally considered in the statistical analysis.

The survey design included the optimization of the number of visits versus observing time per visit. We adopted as a compromise a visit of about 1.5 hours including pointing and AO setup overheads. This ensures a field rotation of $\ge30$ degrees for most declinations in the case of observations including the meridian passage, allowing good removal of the speckle patterns using angular differential imaging (ADI; \citealt{2006ApJ...641..556M}). Longer exposures provide little improvement because of the limited additional field rotation. Shorter exposures would allow us to observe more targets, but would imply significant degradation of the achievable contrast for individual observations. Considering that the available number of high-merit targets (nearby very young stars) is not particularly large (see below), this would imply a smaller number of expected detections.

In the original survey design we planned to devote 70\% of the time to first-epoch observations, 20\%  to second epochs for common proper motion confirmation, and 10\% to additional characterization observations, exploiting the variety of observing modes available for SPHERE. The estimate of the amount of time needed for second-epoch observations was based on predictions of the background star contamination rate in the SPHERE field of view using Galactic population models \citep[e.g., ][]{besancon} and our target coordinates.

The selected setup for the survey used the IRDIFS mode, allowing simultaneous observations in YJ range ($0.95-1.35\,\mu m$ using IFS \citep{ifs} over a small field of view ($1.77\,\!'' \times 1.77\,\!''$) and observations in two narrowband filters in H-band using IRDIS \citep{irdis} over a $11\,\!'' \times 11\,\!''$ field of view. The two narrowband filters \citep[H2 and H3, ][]{vigan2010} were selected for their sensitivity to methane-dominated objects (H2-H3 $\le$ 0.0 mag) and to very red, late L objects (H2-H3 $\sim$ 0.0-0.5 mag). Field stars have an H2-H3 color close to zero, allowing the implementation of a robust priority scheme for the confirmation of the candidates (see Paper II for details).

\subsection{The SHINE database }
\label{s:database}

Over the past ten years we assembled a large sample of young stars with the main goal of preparing the SHINE survey. 
This work included the determination of several stellar properties, either from the analysis of new observations or from the literature. The database was also used to select samples for other programs, such as the NaCo Large Program for Exoplanet Imaging (NaCo-LP; \citealt{nacolp}), the SPOTS survey for circumbinary planets \citep{spots3}, the HARPS Large Program for planets around young stars \citep{grandjean2020}, the search for planets around young stars in the framework of the GAPS program at TNG \citep{carleo2020}, and several other programs. 

The determination of target parameters is mostly based on the methods described in the NaCo-LP target characterization paper \citep{desidera2015}. Briefly, stellar ages
were obtained from a combination of age methods (membership to groups, lithium, rotation,
activity, kinematics, isochrone fitting).
Moving groups (MGs) membership was taken from \citet{torres2008}, with updates from the literature 
in the following years. Ages of moving groups were those adopted by \citet{desidera2015} (their Table 8).
Stellar distances were taken from Hipparcos trigonometric parallax when available \citep{vl07}, otherwise the (age-dependent) photometric distances derived as done in \citet{desidera2015} were adopted. 
Stellar masses were derived following the \citet{reid2002} calibrations.
The original values of stellar mass, distance, and age are listed in Table
\ref{t:original}.
Comparison with the updated values derived in this paper (Sect. \ref{s:param})
shows a nice agreement for the distance (mean difference 3.7 pc, rms 9.4 pc), a small offset in stellar age (0.13 dex with rms 0.21 dex, with the original ages being younger), and a systematic difference in stellar masses above
1.8-2.0 $M_{\odot}$ (the original mass being smaller) and a fairly good agreement below this value. This last difference has some impact on the actual upper limit in mass for the sample (see below), but otherwise the use of the original parameters with respect to the updated ones derived in this paper should have a minor impact on the target selection and the priority scheme  defined below.
Originally the sample was limited to distances closer than 100 pc; 
it was complemented with stars in the Sco-Cen OB association \citep[e.g., ][]{dezeeuw1999} to reach a suitable number of young, early-type stars (from early F to late B) at slightly larger distances.

For the final selection of the SHINE sample, we first identified some general selection criteria,  driven by the characteristics of the SPHERE instrument (coordinates of the site; magnitude limit for good performance of the AO system) or by the science goals described above.
The following selection criteria were set:
\begin{enumerate}
\item  Declination limits between -84 and +21 degrees to ensure observations at airmass values of less than  2;
\item Wavefront sensor (WFS) flux\footnote{Flux as seen by the SPHERE WFS sensor; see \citet{sphere} for details. }
 $>$ 5 e-/subpup/frame or $R \leq 11.5$ (to ensure good quality of AO correction, being on the conservative side of estimates from the SPHERE performance simulations available at the time of target selection);
\item Exclusion of known spectroscopic and close visual binaries (projected separation $<$ 6 arcsec, i.e., within the IRDIS field of view)\footnote{Preparatory observations of part of the sample were performed with FEROS \citep{mouillet2010,desidera2015} and with AstraLux \citep{astralux}.}. This is motivated by the technical limitations of AO working and ADI processing for spatially resolved binaries and by our scientific choice of having a homogeneous sample of single stars or components of wide binaries without the complications of the severe dynamical influence of stellar companions or of uncertain process of planet formation and evolution in circumbinary disks;
\item Age  $<$ 800 Myr because planets are too faint at older ages for wavelengths $<2.3 \mu m$;
\item Distance $<$ 100 pc to probe the smallest physical separations, except for Sco-Cen members, as this region is rich in young early-type stars;
\item $M_*< 3~M_{\odot}$. This limit was set because the frequency of planets detected using radial velocities (RV) appears to drop above this mass \citep{reffert2013}. We did not set any explicit lower-mass cutoff, though an implicit limit was set by the requirement on AO flux. Our survey then covers the full range 0.5 to 3.0 $M_{\odot}$, which allows us to explore the influence of stellar mass.
\end{enumerate}

The sample has no explicit biases related to the presence of disks nor to metallicity.  At young ages metallicity determinations are quite sparse \citep{biazzo2012,vianaalmeida2009}. The available results point toward a metallicity close to solar for stars younger than $\sim$200 Myr in the solar vicinity. The metallicity dispersion becomes significant for older stars, which constitute a very small fraction of our sample.

The resulting list satisfying the above criteria at the time of  freezing the SHINE sample (May-August 2014) included 1224 targets; this is larger by about 50\% with respect to the sample to be selected (800 stars), which in turn is almost twice the number of stars that could be effectively observed. 

\subsection{Target priorities}

A well-defined priority scheme is necessary in order to run an efficient survey on a sample of several hundred targets because the potential for exoplanet discovery is very different from star to star, depending on age, distance, mass, and magnitude. Ideal targets are very young stars close to the Sun. Simulations described in Sect. \ref{sec:simul} were performed on this extended list in order to estimate the potential for planet detectability of each target and drive the final target selection. 

The priority assignment applied to the SHINE database is the following:
\begin{enumerate}
\item  Build a FoM that allows the stars to be ranked according to the probability that they host planets potentially detectable by SPHERE. To estimate this probability, we used  \it i) \rm planet population models based on power laws in planet mass and separation, with cutoffs at large separations; combined with stellar ages this enables the planet luminosity to be estimated via suitable evolution models; \it ii) \rm the MESS code \citep[Multi-purpose Exoplanet Simulation System, see][for details]{mess, qmess}\footnote{An updated version of the MESS code, now called Exoplanet Detection Map Calculator \citep[Exo-DMC][]{Bonavita:2020ascl}, is available for download at \url{https://github.com/mbonav/Exo_DMC}} that allows the planet position to be projected on the sky at the epoch of observation;  and  \it iii) \rm expected contrast limits appropriate for each star given the stellar properties (distance, age, magnitude). These three different aspects of the construction of the FoM are described in the next subsection. 
\item  Sort the parent sample according to this FoM and construct an optimal 400 star sample as high priority (List 400), and complete it with additional targets to fill in according to rank order, lower priority, up to 100\% over-subscription (List 400+). The final sample reaches 800 stars\footnote{This is significantly larger than the sample size allowed by the available observing time, but we decided to oversize the sample in order to have enough flexibility on the scheduling, considering the requirement of observing the targets including the meridian passage.}. 
\item  Check that this  optimized-for-detections sample covers a reasonably wide distribution in stellar mass.
\item  Consider some adaptions of this sample, such as \it i) \rm inclusion of homogeneous subsamples, as
volume-limited members of nearby young moving groups
and \it ii) \rm  limiting the number of Sco-Cen members to 20\% of the sample because these stars have a limited span in right ascension, and the background contamination is often large  due to their low galactic latitude, which has a severe impact on the expected need for second-epoch observations.
\end{enumerate}

\subsection{Simulations of planet detectability}
\label{sec:simul}

In order to define the FoM to be used to select the final SHINE sample, we estimated the expected survey yield as a function of the target list as well as the assumptions on the underlying characteristics of the planet population. For this purpose we used the MESS \citep[see][for details]{mess} code to evaluate the probability of detecting a companion given the expected instrument performance. 
The synthetic companions were generated according to two different models, both based on the one described by \cite{cumming2008}, that is, adopting power-law distributions for the companion mass and semi-major axis.

Both models are defined so that the resulting frequency of companions $F$ is consistent with the value of the fraction of planets per star retrieved by \cite{cumming2008} ($F_0 = 0.0394$) if calculated over the same parameter space (mass: $1-13~M_{Jup}$, semi-major axis: $0.3-2.5$\,au, stellar mass: $0.7-1.6~M_{\odot}$, [Fe/H]: $-0.5-+0.5$.). This implies the use of a normalization factor $C_0$ defined as follows:

\begin{equation}\label{eq:norm_mod1}
    \begin{aligned}
F_0 =  C_0 \int_{0.7M_{\odot}}^{1.6M_{\odot}}~dM*\int_{1M_{Jup}}^{13M_{Jup}}m^{-1.3}~dm\int_{0.3au}^{2.5au}a^{-0.61}~da 
\end{aligned}
\end{equation}

The companion mass range was fixed at $m_{max}=75~M_{jup}$ for all the targets in the first model (hereafter Mod01). For the second model (hereafter Mod02) we instead adopted $m_{max}=0.03 \times M_*\frac{M_{Jup}}{M_{\odot}}$. Mod02 also includes a dependency on the stellar mass for the expected value of the frequency. For this model the normalization factor is in fact expressed as $C_0f(M_*)$, where $f(M_*)$ is a mass function that ensures that the resulting number of companions increases linearly with stellar masses up to 2~$M_{\odot}$, and is zero for stellar masses higher than 3~$M_{\odot}$, following the findings of \citet{reffert2013} for close-in planets. 

The main difference between the two models is therefore that for Mod02 the properties of the generated planet population are not fixed, but change for each target. For this reason, a ranking based only on Mod02 would introduce a bias towards A-F stars. The use of both models combined instead supports the selection of a more balanced sample, which is necessary to assess the dependence of planet frequency on stellar mass, which is one of the main scientific drivers of the survey.

\begin{figure*}
\centering
\includegraphics[width=8.5cm]{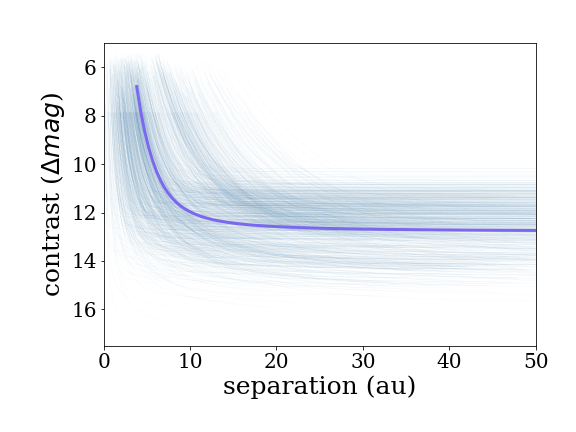}
\includegraphics[width=8.5cm]{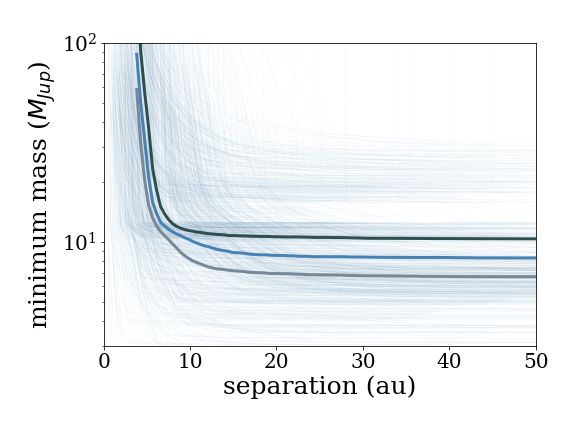}
\caption{Expected IFS detection limits (light blue curves), evaluated using the method described in \cite{mesa2015}, expressed in terms of contrast (left panel) or minimum companion mass (right panel, calculated assuming the best age for each target). The solid blue line shows the average over the full input sample. The light and dark gray curves in the right panel show the average mass limit obtained adopting the minimum and maximum values of the age, respectively.}
\label{fig:detlims}
\end{figure*}

For all the stars in the initial list we evaluated the expected SPHERE detection limits, expressed in terms of luminosity contrast versus projected separation (shown in the left panel of Fig.~\ref{fig:detlims}), using the method described in \cite{mesa2015}. The models of \cite{baraffe2003} were then used to estimate the corresponding minimum detectable companion mass $M_{lim}$ \footnote{ Alternative sets of models available at the time of sample building were considered but not used because of the lack of grid covering our space of parameters of our interest and/or because of counter-examples already available at that time against some of these models, such as the extreme cold-start scenario by \citet{marley2007}.   } (see Fig.~\ref{fig:detlims}, right panel), under the assumption that any companion would be coeval with its host star, hence using the age of the star to select the appropriate evolutionary track. 
To take into account the fact that the detection limits are expressed in projected separation, while the models produces semi-major axis values, the code evaluates the probability that a companion with a given semi-major axis can have a projected separation which would put it inside the field of view (FoV). This is done assuming a random orbital phase and eccentricity drawn from a Gaussian distribution \citep[see][for details]{mess}.

For each model we performed six MESS runs, using the list of 1244 targets from the SHINE database, but changing
the following parameters: 
\begin{itemize}
\item the value of the age used for the magnitude to mass conversion of the detection limits (the adopted age of each star, as well as the minimum and maximum values) in order to properly estimate the impact of the uncertainty on the age on the survey results.
\item the cutoff of the semi-major axis distribution, which was set at 15 or 30\,au (a value of 50 au was used for the Sco-Cen members to take into account the larger distance of this region with respect to the other targets in the list).\footnote{This allowed us a proper ranking of the Sco-Cen targets, which would not be reliable for closer cutoff values. A cutoff as large as 50 au was already ruled out at the time of the sample selection for solar-type stars. The situation was less clear for more massive stars, with possible scaling of planet distribution in mass (following the mass ratio) and in separation (following the loction of the snow line). 
Considering the roughly flat sensitivity curve of SPHERE at separation larger than 0.3", this simulation was roughly equivalent from the point of view of the ranking the Sco-Cen targets to a more realistic one with a low-frequency but broad distribution at wide separation. This inhomogeneity has no consequences in the final
sample selection, considering that we included a fixed number of Sco-Cen targets for
each priority bin (see Sect. \ref{s:sample}).  }
\end{itemize}

\noindent Table~\ref{tab:ndet} summarizes the outcome of the simulations, in terms of expected detection yield, assuming a sample made of the best 400 targets ranked according to each model. The values obtained using the 15\,au and the 30\,au cutoff values are both shown, and those obtained using the minimum and maximum values of the stellar age for the magnitude to mass conversion of all the detection limits. 

\begin{table}[htbp]
\caption{Summary of the expected detection yield for the best 400 targets ranked according to the different models. The same values obtained considering the minimum and maximum estimated age of each target are also shown. The number of Sco-Cen members (for which the cutoff is always 50 au) is reported in the last column. }
\centering
\begin{tabular}{l|ccc|c }
\hline
Model, cutoff   & \multicolumn{3}{|c|}{Number of expected detections} & Sco-Cen  \\ 
                 & adopt. age & min. age & max age  & Members \\
\hline
 Mod01, 15 au  & 24.73  & 28.29  & 22.41 & 15 \\
 Mod01, 30 au  & 48.27  & 53.65  & 44.82 & 35\\
\hline
 Mod02, 15 au  & 16.12  & 18.89  & 14.25 & 2\\
 Mod02, 30 au  & 30.22  & 34.25  & 27.12 & 12\\ 
 \hline
\end{tabular}
\label{tab:ndet}
\end{table}

The expected number of detections in Table \ref{tab:ndet} is larger than the actual number (although the final number of detection from SHINE is not yet available, pending the completion of the second-epoch observations).
This indicates that some of the assumptions we made in the simulations are not adequate.
The dependency of planet frequency on semi-major axis is likely the most prominent case.
There were already hints at the time of the sample selection that the extrapolations of RV-based power laws with outer cutoff are not ideal, considering both the few individual detections at very wide separations, and low frequency of substellar companions resulting from various surveys. However, alternative distributions such the log-normal distribution by \citet{meyer2018} were published a few years later.
Therefore, we were expecting the outcome of our simulations to be overestimating
the number of new discoveries
since the beginning of the survey, and in this perspective we planned only a limited amount of time for characterization observations in the original survey design (see Sect. \ref{s:design}). 
In spite of these limitations, we consider our simulations appropriate for the main aim of the priority ranking, which is to build a good sample to answer our scientific questions (e.g., including a broad range of stellar masses).

\section{The final SHINE target list}
\label{s:sample}

After the simulations described above, the SHINE list is composed of 800 stars rank ordered following the proposed scheme for two bins of masses Bin-1 (early-type; $\ge1.5$~M$_\odot$) and Bin-2 (low-mass $\le1.5$~M$_\odot$) as summarized in Table~\ref{tab:priority}.  For operational reasons linked to the scheduling of the observations, we choose to have four priority classes, which we labeled P1 (highest priority), P2, P3, and P4 (lowest priority), rather than individual priorities that are different for each target. We considered the relevance of  members
of nearby young MGs, aiming at a complete sample
of members, and considering the lower errors in stellar age compared to isolated objects and the better availability of information in the literature. For this reason, in a few cases we overruled the priority ranking resulting from simulations to have well-defined distance limit for MG members in the first two
priority bins. We also limited the number of Sco-Cen targets to fulfill observational constraints.
Overall, 200 targets were included in each bin, as detailed below.
~\\

{\bf P1 sample}:
\begin{itemize}
\item all the known members of nearby  young MGs ($\beta$ Pic, Tucana, Columba, Carina, TW Hya, Argus, AB Dor) within 60 pc;
\item the known members of the youngest groups (those listed above except AB Dor, which is significantly older) between 60 to 80\,pc and the two early-type members of $\eta$ Cha cluster;
\item the first 40 Sco-Cen stars from ranking of the simulation performed using the 50\,au outer cutoff;
\item the remaining stars to achieve the planned size of the sample (about 20 stars)  taken from the ranking of the 15 and 30\,au simulations, taking stars alternatively from the two lists sorted according to planet detection probabilities.
\end{itemize}

{\bf P2 sample}:
\begin{itemize}
\item all the known members of nearby  young MGs ($\beta$ Pic, Tucana, Columba, Carina, TW Hya, Argus,  AB Dor) in the distance range between 80 to 100\,pc;
\item the known members of AB Dor MG from 60 to 80\,pc;
\item a handful of stars whose membership status in the above MGs is controversial in the literature;
\item the next 40 Sco-Cen stars from ranking of the simulation performed using the 50\,au outer cutoff;
\item the remaining stars to achieve the planned size of the sample taken from the ranking of the 15 and 30\,au simulations, taking stars alternatively from the two lists sorted according to planet detection probabilities.
\end{itemize}

{\bf P3 sample}:
\begin{itemize}
\item the next 40 Sco-Cen stars from ranking of the simulation performed using the 50\,au outer cutoff;
\item the remaining stars to achieve the planned size of the sample (160 stars) taken from the ranking of the 15 and 30\,au simulations, taking stars alternatively from the two lists sorted according to planet detection probabilities.
\end{itemize}

{\bf P4 sample}:
\begin{itemize}
\item the next 40 Sco-Cen stars from ranking of the simulation performed using the 50\,au outer cutoff;
\item the remaining stars to achieve the planned size of the sample (160 stars) taken from the ranking of the 15 and 30\,au simulations, taking stars alternatively from the two lists sorted according to planet detection probabilities.
\end{itemize}

\begin{table}[htbp]
\caption{Priority distribution of the SHINE sample}
\centering
\begin{tabular}{l|cc}
\hline \hline
Priority                & Early-type  & Solar and Low-mass \\
\hline
P0                      &  \multicolumn{2}{c}{Special targets}  \\
P1                      & 20 MGs + 40 ScoCen & 120 MGs + 20 Field \\
P2                      & 20 Field + 40 ScoCen & 50 MGs + 90 Field \\
P3                      & 20 Field + 40 ScoCen & 140 Field \\
P4                      & 20 Field + 40 ScoCen & 140 Field \\
P5                      &  \multicolumn{2}{c}{Bad weather backup or filler} \\
\hline
\end{tabular}
\label{tab:priority}
\end{table}

We also selected bright targets to be observed with short observations in case of bad weather conditions or short gaps in the schedule (labeled P5 targets). They include known binary systems of specific interest for orbit determinations \citep{rodet2018} and stars with known RV planets to look for stellar companions (Hagelberg et al., in preparation.). They are not intended to be deep enough to allow us the detection of planetary companions and will not be considered in this paper.

\subsection{New targets added after the original definition}

After evaluating of the actual on-sky performances of SPHERE, the results of which showed a reasonably good performance on stars as faint as R $\sim$ 12 \citep[][see their Fig. 7]{sphere}, the original sample was complemented starting from ESO period P98 (April-September 2016) with about 50 M-type objects proposed as members of young moving groups. The targets were assigned to priority bins P1 and P2 following the distance limits for individual groups, as defined above.

\subsection{Special targets}
\label{sec:P0}

A number of special targets were identified and promoted to higher priority for observation, defined as P0 priority. The science motivation for special priority is related to the known presence of planets or brown dwarfs, amenable to detailed characterization studies \citep[e.g., ][]{bonavita2017,lagrange2019}, stars with spatially resolved disks, especially when the disk properties suggest the presence of planets \citep[e.g., PDS 70][]{keppler2018,mesa2019}, or stars with long-period RV planets potentially detectable with SPHERE \citep{zurlo2018}.

Other targets were promoted as P0 during the survey, thanks to results of observations of other components of the SPHERE-GTO (DISK program) or discoveries by other groups \citep[e.g 51 Eri after planet detection,][]{macintosh2015}. Targets that were not included in the original statistical sample, as defined above, are not considered in this paper as their inclusion would considerably bias the overall frequency of substellar objects. This is, for instance, the case of PDS\,70 not originally selected given its distance, but observed in the context of a SHINE follow-up of the DISK program. Only targets included in the original statistical sample are kept for the present analysis. Nevertheless, some bias is still present, as the increase in priority implied a greater probability of being actually observed. This is further discussed in Paper III \citep{vigan2020}.

Table \ref{t:P0targets} lists the targets in the F150 sample promoted to P0, the original priority class, the motivation for the priority upgrade, and the reference to individual papers based on SHINE data, if any. $\eta$ Tel and CD-35 2722 were not flagged as special objects in spite of the previously known BD companions.

\begin{table*}[htbp]
\caption{Stars in the SHINE statistical sample observed as special targets (highest priority). The original priority in the selection of the statistical sample, the motivation for priority upgrade, and the references to discovery papers and individual SPHERE papers are listed. SAM stands for sparse aperture masking
\citep[e.g., ][]{tuthill2006}. }
\centering
\begin{tabular}{lclll}
\hline \hline
Target                & Priority  &   Remarks  & Discovery paper & SPHERE paper \\
\hline
$\beta$ Pic             & P1 & known planet and disk & \cite{lagrange2009} & \cite{lagrange2019}  \\
HR 8799                 & P1 & known planet & \cite{marois2008} & \cite{zurlo2016} \\
HD 95086                & P1 & known planet & \cite{rameau2013} & \cite{chauvin2018b} \\
Fomalhaut               & P2 & known planet and disk & \cite{kalas2008}& -- \\
FomalhautB              & P3 & companion to P0 star  & -- & -- \\
PZ Tel                  & P1 &  known brown dwarf & \cite{biller2010} & \cite{maire2016} \\
HIP 107412              & P4 &  known brown dwarf & \cite{milli2017}&  \cite{delorme2017,grandjean2019} \\
51 Eri                  & P1 & known planet & \cite{macintosh2015} & \cite{samland2017,maire2019} \\
AB Pic                  & P1 & known brown dwarf & \cite{chauvin2005} & -- \\
TYC 8047-0232-1         & P1 & known brown dwarf & \cite{chauvin2005gsc} & -- \\
HIP 78530               & P1 & known brown dwarf & \cite{lafreniere2011} & -- \\
HD 61005                & P1 & known disk & \cite{hines2007} & \cite{olofsson2016} \\
HR 4796                 & P1 & known disk & \cite{schneider1999} & \cite{milli2017, Milli2019}\\
AU Mic                  & P1 & known disk & \cite{liu2004} & \cite{boccaletti2015,boccaletti2018} \\
HD 30477                & P1 & known disk & \cite{soummer2014} & \\
TWA 7                   & P1 & known disk & \cite{choquet2016} & \cite{olofsson2018} \\
HD 141943               & P2 & known disk & \cite{soummer2014} & \cite{boccaletti2019}\\
$\zeta$ Lep             & P2 & known disk & \cite{moerchen2010} & \\
$\rho$ Vir              & P1 & known disk & \cite{booth2013}    & -- \\
HIP 71724               & P3 & known low-mass comp. & \cite{Hinkley2015} & -- \\
HIP 73990               & P3 & known low-mass comp. & \cite{Hinkley2015} & -- \\
HD 115600               & P3 & known disk & \cite{currie2015} & -- \\
HD 377                  & P2 & known disk & \cite{choquet2016} & -- \\
\hline
\end{tabular}
\label{t:P0targets}
\end{table*}

\section{The F150 sample}
\label{s:f100}

The aim of the present series of papers is to present a preliminary statistical analysis from the first half of targets observed in the SHINE survey. The resulting sample is then necessarily incomplete. The optimal observing procedure for ground-based direct imaging surveys requires the target to be observed at meridian to maximize field rotation. A dedicated program to optimize the scheduling of an extended list of targets over individual nights or runs and even full semesters considering the actual time allocation has been built and is routinely used to prepare the SHINE observing nights \citep{spot}. As a result, the actual targets included in the schedule is a compromise between the scientific priorities, the constraints of meridian passage, and the maximization of the number of targets to be observed.

To build the sample considered in the present paper, we included targets observed until February 2017 (the first two years of the survey), considering only the targets that are part of the statistical sample, as defined in Sect.~\ref{s:sample}. Targets observed in poor conditions (not validated following quantitative criteria of achieved contrast with respect to the expected level considering stellar magnitude and declination) were removed. Details on data reduction are provided in Paper II (Langlois et al.).

We also removed targets identified as new close visual binaries from SPHERE observations in order to be homogeneous with the original selection of single stars or members of wide binaries\footnote{In some cases the physical link between the central star and the companion remains to be demonstrated, but the chance of background objects is small because of the bright magnitude of the candidates. Details will be presented in Bonavita et al., in preparation. Cases of very low-mass star companion candidates were analyzed with special care. A common proper motion test was performed, and stars with moderately bright background objects were kept in the F150 sample. }. The new binaries will be presented in a dedicated paper (Bonavita et al. 2020, submitted). 

We removed four targets (HIP 37288, HIP 39826, HIP 64792=GJ504, HIP 82588)
which resulted to be older than 1 Gyr from the revised age analysis described in 
Sect.~\ref{s:age}.
Finally, we removed four targets (HD 100546, TW Hya, MP Mus, and EP Cha) because of the presence of gas-rich disks. Planets may still be forming in these disks, but are probably heavily obscured by disk features. The case of HD\,100546, formally in our sample, is probably the best example of this \citep{sissa2018}.

The sample built as described above is then formed by 150 targets, listed in Table \ref{t:targetlist}, together with available broadband photometry in several filters.

\section{Updated stellar properties}
\label{s:param}

The target parameters were determined using the methods and procedures described in \cite{desidera2015}. A major improvement is the availability of Gaia DR2 astrometric parameters \citep{gaiadr1, gaiadr2}. As a result, all the targets have  trigonometric parallaxes.

\subsection{Distance and proper motion}

Trigonometric parallaxes and proper motions were mostly taken from Gaia DR2. For two very bright stars not included in Gaia DR2 (Fomalhaut and $\beta$ Leo), we used the Hipparcos results  derived by \cite{vl07} (hereafter VL07). For an additional 13 very bright stars (V$<5$), the VL07 errors are smaller than the Gaia DR2 values. We adopted VL07 parallaxes and proper motions for these targets; they are listed in Table \ref{t:kinparam}, together with other kinematic parameters.

\subsection{Radial velocity}

Radial velocity (RV) is a key input for the kinematic assignment (membership to groups) and for checking multiplicity. RV measurements were mostly taken from the literature (sources listed in Table \ref{t:kinparam}). We obtained new RVs for 32 stars from spectra available in public archives. For 31 stars observed with HARPS we exploited the reduced spectra and RVs provided by the instrument pipeline available on the ESO archive\footnote{\url{http://archive.eso.org/wdb/wdb/adp/phase3_spectral/form}}. 
The mean value was adopted when multiple epochs were available\footnote{In a few cases the RVs available on the ESO archive were obtained with different masks for the same star. In these cases we adopted the value obtained from the mask that is closer to our adopted spectral type. }.
For one star (HIP 69989) we took the RV from public observations obtained with SOPHIE, available in the archive of the Haute Provence Observatory  \citep{sophiearchive}\footnote{ \url{http://atlas.obs-hp.fr/sophie/}}.

\subsection{Multiplicity}

As discussed above, objects with known stellar companions (mass $>75~M_{Jup}$) within 6 arcsec (approximate size of IRDIS field of view) were removed from the sample, including the previously unknown binaries discovered by our observations. The multiplicity search was based on the SPHERE images themselves, Gaia DR2, and other recent literature, including the evaluation of RV variability. 

We opted to be conservative in the removal of targets due to binarity, aiming at avoiding spurious rejections in our sample. For this reason, a few stars with low-amplitude RV variability or indication of multiplicity derived only from the $\Delta \mu$ signature\footnote{Proper motion difference between Gaia DR2 and other catalogs such as Gaia Dr1, Hipparcos, and Tycho2.} were kept in the sample. RV variability of up to 1-2 km/s may be linked to stellar activity for our very young targets \citep[e.g., ][]{Carleo18,brems2019} or to pulsations for early-type stars \citep[e.g., ][]{lagrange2009}, 
or, when combining data from different instruments, due to zero-point offsets. The $\Delta \mu$ signature has instead some ambiguity in terms of physical mass and orbital parameters responsible for the dynamical signature. We also checked  non-coronagraphic images taken with SPHERE at the beginning and end of the observing sequence, looking for the presence of stellar companions \citep{engler2020}.
Some ambiguous cases are discussed individually in Appendix \ref{a:notes}.
It should be noted that several targets in our sample lack RV monitoring, making the rejection of spectroscopic binaries incomplete.

In order to have a more complete view of the properties of the targets in the sample, we also looked for companions at projected separations larger than 6 arcsec from our targets (outside the IRDIS field of view).
They are listed in Table \ref{t:wide}. Dedicated checks using Gaia DR2 have been performed to confirm the physical association of previously known companions and to look for new ones through the evaluation of the parallax and proper motion of individual objects. Forty-one targets are found to have companions outside the IRDIS field of view. In two cases (HIP 95270/$\eta$ Tel and Fomalhaut/Fomalhaut B) we observed both components of the system. Two stars, namely HN Peg and $\beta$ Cir, have brown dwarf companions at wide separation. For MG members, especially for the well-populated Sco-Cen groups, there is some ambiguity between very wide binaries and co-moving members of the associations. We opted to be conservative and then inclusive in Table \ref{t:wide}, requiring tight common proper motion and parallax values (within 2 mas/yr and 1 mas, respectively). 
We note some cases of extremely wide binaries in our sample, with separation even larger than the typical limit for bound objects adopted in the literature \citep[$\sim$ 20\,000\,au, ][]{abt1988,allen2000}. However, at young ages the occurrence of multiple systems at wider separations is expected \citep{caballero2009}. Among the extremely wide multiple systems not previously noted in the literature we mention HIP 22226 + 2MASS J04463413-262755 (the latter being itself a close binary) at a projected separation of about 50\,000\,au. 

The presence of wide companions, an important environmental property of our targets, is also a useful resource for age dating systems when complementary diagnostics can be applied to the 
individual components, depending on their spectral types.

\subsection{Kinematics and moving group membership}
\label{s:kin}

We exploited the updated kinematic data, particularly the high-accuracy parallaxes and proper motions from Gaia DR2, to evaluate the membership of our targets in young moving groups. We derived the space velocities U, V, W following prescriptions by \citet{uvw}, and we exploited the BANYAN $\Sigma$ tool\footnote{\url{http://www.exoplanetes.umontreal.ca/banyan/banyansigma.php}  } \citep{banyansigma}. Unambiguous assignment to groups or field can be made for nearly all targets thanks to the improved kinematics data and group definition. There are a few cases of different membership assignment with respect to the literature, mostly a few exchanges between Tuc-Hor and Columba and Carina or between Sco-Cen subgroups. A few targets classified in the literature as members of AB Dor MG have
low membership probability according to  BANYAN $\Sigma$. However, their properties (lithium, activity, rotation; see below) are fully consistent with those of confirmed members. 
Targets worthy of individual discussion are included in Appendix \ref{a:notes}.

\subsection{Chromospheric and coronal activity}

X-ray luminosities for all the targets were derived from the ROSAT All Sky Catalogs \citep{rosatbright,rosatfaint} using the calibration by \cite{hunsch1999}; 89 stars in the sample have detected X-ray emission.
Chromospheric emission was retrieved from the literature or measured by us on archive HARPS spectra as in \citet{desidera2015}. Overall $\log R^{'}_{HK}$ is available for 46 stars. 
Both quantities are reported in Table \ref{t:ageparam}.
These indicators have limited dependency on stellar age below the Pleiades age (for G-K stars). Therefore, low weight was assigned to these indicators to quantify the age of very young stars, while for older objects we relied on the \citet{mamajek2008} calibrations. 
The possibility of old active stars misclassified as young being tidally locked binaries
or because of the occurrence of some kind of accretion of angular momentum  is unlikely, considering the complementary age diagnostics (e.g. lithium) and the search for close
binaries.

\subsection{Lithium}

The equivalent width of the available age-sensitive diagnostic Li 6707\,AA\  was gathered from the literature and included in  Table \ref{t:ageparam}, to be used as an age diagnostic.

\subsection{Photometric variability and rotation period}

In our sample 103 stars have spectral type later than A. They are all young, and exhibit clear evidence of magnetic activity and photometric variability. The last property is exploited to make measurement of the stellar rotation period and of the level of magnetic activity. 

As the first step in our rotation period study, we explored the literature for all 103 stars and found existing measurements of the rotation period for 64 stars (62\%). As the second step we explored the public photometric time series archives of all 103 stars and found (mostly in the TESS archive) suitable data for rotation period search for 58 stars (56\%). Finally, we planned and collected our own photometric time series data for the remaining targets lacking suitable data and measured the rotation period for six of them. Observations were carried out at the Remote Observatory Atacama Desert (ROAD) in Chile, the Perth Exoplanet Survey Telescope  Observatory (PEST) in Australia, and the York Creek Observatory (YCO) in Tasmania.

As the results of our study we obtained the rotation period for 101 of the 103 late-type stars in F150. They are reported
in Table \ref{t:ageparam}, together with amplitude of the photometric rotational modulation.

Ages from rotation were estimated comparing the observed rotation period with that of stars of similar colors in groups or clusters of known age, considering those compiled by \citet{desidera2015} and more recent literature results (e.g., \citet{rebull2016} for the Pleiades and \citet{Messina17} for $\beta$ Pic MG). 
The ambiguity in rotation age due to the non-monotonic evolution of the rotation (minimum period reached close to the zero age main sequence, ZAMS)
can usually be eliminated by considering additional indicators such as the Li equivalent width (EW) and the position on the color-magnitude diagram (CMD) because very young stars that are in the acceleration phase are above the ZAMS. The number of 
targets involved in these ambiguities is small, considering the large
fraction of members of moving groups, especially at young ages.
The additional ambiguity represented by the small fraction of 
very fast rotators \citep["C" sequence in the nomenclature by][]{barnes2007} is more subtle. As the majority of these fast rotators are found in binary systems \citep[e.g.,][]{Messina17}, which are rejected in our study, and considering the low
fraction of such targets in well-studied clusters and the help of the other age diagnostics, we expect a very limited impact on our age classification.

\subsubsection{Period search methods}

We followed the approach outlined in Messina et al. (\citeyear{Messina10}, \citeyear{Messina11}), to search for the stellar rotation period of our targets.

Briefly, the period search was carried out by computing the Lomb$-$Scargle periodogram (LS; \citealt{Press02}; \citealt{Scargle82}; \citealt{Horne86}) and the CLEAN algorithm \citep{Roberts87}. The false alarm probability (FAP) that a peak of given height in the periodogram is caused by statistical variations was computed through Monte Carlo simulations, by generating 1000 artificial light curves obtained from the real light curve, keeping the time sampling but permuting the magnitude values (see, e.g., \citealt{Herbst02}). We considered only rotation periods that were measured with FAP $<$ 0.01\%. Only a very few targets with FAP in the range 0.01--1\% had their rotation periods  confirmed by independent measurements, for example from the literature.

When data from more observation seasons (or sectors in the case of TESS) were available, we computed LS and CLEAN periodograms  for each season(sector); the most accurate results were reported as adopted values, and they are shown in   Figs.\,\ref{fig:HIP490_fig}--\ref{fig:HIP118008}. We followed the  method used by \citet{Lamm04} to compute the errors associated with the period determinations (see, e.g., Messina et al. \citeyear{Messina10} for details). To derive the light curve amplitude, we fit the data with a sinusoidal function whose period is equal to the stellar rotation period. As a result of our photometric analysis we obtained our own rotation period measurements and photometric variability amplitudes for 82 of the 103 target stars: 76 from data of one or more public archives and 6 from our own photometry. We confirmed 44 previously known rotation periods. Finally, we adopted the rotation periods retrieved from the literature for 18 targets (of which 8 periods were retrieved from Messina et al. \citeyear{Messina10}, \citeyear{Messina11}, \citeyear{Messina17}). We produced plots for all the photometric time series (either new or from archives) analyzed in this work. These plots are available as online material in Figs.\,\ref{fig:HIP490_fig}--\ref{fig:HIP118008}.

\subsection{Isochrone fitting}

Isochronal ages were derived using the models by \citet{bressan2012} and the web interface PARAM
\footnote{Version 1.3 available at \url{http://stev.oapd.inaf.it/cgi-bin/param_1.3}}.
For this determination, we used the V-band magnitude listed in Table \ref{t:targetlist},
the effective temperature listed in Table \ref{t:massteff}, the parallax listed in Table \ref{t:kinparam}, and adopted solar metallicity.
The effective temperatures were obtained through \citet{casagrande2010} for late-type stars,
using the combination of colors adopted in \citet{desidera2015}, and through the
Mamajek tables \citet{pecaut2013} for early-type stars.
A few cases of ambiguities between pre-main sequence and post-main sequence evolution
\citep[see, e.g.][]{bonnefoy2018} are discussed individually in the Appendix. The availability of
other indicators allows us to solve the ambiguity between the two alternatives.
In addition to the systematic uncertainties in the stellar models, possible biases of the isochrone method are linked to photometric variability of most of the late-type targets and to the possibility of unrecognized binaries. The first item is included in the error bar of the input parameters. For the second item, the sensitivity of our SPHERE observations to stellar companions, especially those bright enough to bias the photometry, allows us to rule out cases at separations larger than a few tens of mas. For closer companions we inspected the available data for spectroscopic binaries, although a significant fraction of the targets lack suitable RV monitoring. Finally, after Gaia Dr2, parallax is no longer the dominant source of uncertainty for isochrone age determination.  

\subsection{Adopted ages}
\label{s:age}

The primary age method adopted in this work is the group membership because the age of an ensemble is usually better determined than any individual measurement. For bona fide members the group age was adopted. The ages for the MG and their errors are those described in \citet{bonavita2016} and were also adopted by \citet{vigan2017}. They are summarized in Table
\ref{t:mgage}\footnote{In our sample there are no members of Octans, Octans-Near, Carina-Near, or Pisces-Eridanus MG, and is why these groups are not included in Table \ref{t:mgage}.}.
The \citet{bonavita2016} MG ages were mostly taken from \citet{bell2015}.
The main motivation for this choice is to ensure the best homogeneity for the whole list of targets as 
no comparable studies were published in the following years.
For groups not included in \citet{bell2015}, we checked (using indirect methods such as lithium and rotation) that our adopted age ranking is correct. 
Improved ages for individual groups have been published \citep[e.g., ][ for $\beta$ Pic MG]{miretroig2020},
but adopting them without revision for other groups would imply inconsistencies in the relative ages, which we want to avoid,
We have started to work on comprehensive updates of the MG ages in the perspective of the final analysis of the SHINE sample.

We recognize that there are indications for a significant age spread in some of these groups, such as the Sco-Cen association \citep{pecaut2016}. However, a common age is adequate for the statistical purposes of this series of papers. In addition, the errors associated with the age determination for individual stars are usually comparable or even larger than the age dispersion of the groups. Finally, possible members
with ages that are somewhat discrepant from the bulk of an association are often subject to a dedicated analysis as typically their kinematic membership probability is lower (see, e.g., the case HD 95086
discussed in Appendix \ref{a:notes}).
For late-type bona fide members, lithium, rotation, and secondary rotation-activity
indicators are typically consistent with the age assigned to the MGs.

\begin{table}[htbp]
\caption{Adopted MG ages}
\centering
\begin{tabular}{lccc}
\hline \hline
   Group                    &  Age  &  Min  & Max  \\
                            & (Myr) & (Myr) & (Myr) \\
\hline
TW Hya (TWA)                &  10 &   7 &  13 \\
$\eta$ Cha OC (ETAC)        &  11 &   8 &  14 \\
Upper Scorpius (US)         &  11 &   4 &  12 \\
Lower Centaurus-Crux (LCC)  &  16 &  12 &  20 \\
Upper Centaurus-Lupus (UCL) &  17 &  15 &  20 \\
$\beta$ Pic (BPIC)          &  24 &  19 &  29 \\
Columba (COL)               &  42 &  35 &  50 \\
Tuc-Hor (TUC)               &  45 &  35 &  50 \\
Carina (CAR)                &  45 &  35 &  50 \\
Argus (ARG)                 &  50 &  40 &  70 \\
AB Doradus (ABDO)           & 149 & 100 & 180 \\
\hline
\end{tabular}
\label{t:mgage}
\end{table}

For field objects, the age determination is based on isochrone fitting for early-type stars (spectral type earlier than mid-F), while for late-type stars (spectral type later than mid-F) it is based on  indirect methods (lithium, rotation, chromospheric activity, X-ray emission), complemented by isochrone fitting and kinematic evaluation when applicable. The indirect estimates are based on comparison with the locus of members of nearby moving groups and open clusters, as done in \citet{desidera2015}. 

For stars with ambiguous membership to groups, we opted for a conservative approach.
In the case of objects with probable membership, we adopted the group age but extended the
possible range of values (minimum and maximum values in Table \ref{t:ages}) to include
the values resulting from the analysis based on other methods, applied depending on the spectral type
of the star.
Object with a low probability of membership were considered field objects, adapting the
age limits to the group age in the adopted range.

The adopted ages are listed in Table \ref{t:ages}. Further details on individual targets are provided in Appendix \ref{a:notes}.

\subsection{Stellar masses}

Stellar masses were derived using the PARSEC isochrones \citet{bressan2012} models and the PARAM interface (see above). As done in \citet{desidera2015} for the objects with ages derived from moving group membership or indirect methods, we restricted the allowed age range for the determination of the stellar mass, assuming uniform prior in the selected age range. This allowed us to consider, among the overlapping isochrones within the error bars of the input parameters, only those consistent with the age estimate. This effect is small but not negligible (typically a few hundredths of a solar mass). For stars with M spectral type the resulting stellar masses appear rather low. This is likely due to issues in the atmospheric models for cool objects in the PARSEC tracks \citep{chengirardi2014}. Therefore, for these stars we derived the stellar masses from the models by \citet{baraffe2015} for the appropriate age of the star.
The masses are listed in Table \ref{t:massteff}.

\subsection{Presence of disks}
\label{sec:disks}

We also checked in the literature for the presence of dusty disks surrounding the stars in the sample. Of the 150 objects, 73 show infrared (IR) excesses, interpreted as the clear presence of dust in the system. Twenty-nine of these were detected and spatially resolved with instruments such as HST/NICMOS, VLT/SPHERE, Gemini/GPI, Herschel/PACS, and  ALMA (see Table \ref{t:massteff}). 
Disks identified as double-belts from SED fitting \citep{chen2014} have a special flag in Table \ref{t:massteff}. Finally, seven objects were classified as potential disk hosts since the IR excesses were only marginally detected, while in two cases there are indications that the observed emission is likely associated with contaminants (background sources). 
The absolute frequency of stars with disks in our sample, in particular those with spatially resolved disks, appears higher than typically found in the literature. The priority enhancement for some stars with disks described in Sect.~\ref{sec:P0} is certainly one of the reasons
for such a high occurrence.
Our selection of well-isolated objects (single stars or components of wide binaries)  with broad dynamical room for extended disks is another
likely reason \footnote{The frequency of stars with detected IR excess is much lower (below 10\%) in the sample of 78 new binaries detected
in the whole SHINE survey (Bonavita et al., submitted).}.
It should be considered that our flag includes the detection of IR excess
at any wavelength, then including a large variety of disk temperatures and configurations. 
Finally, while our original selection criteria do not concern the presence of disks, it is also possible that some indirect bias is at work. This may happen if stars with IR excess were more carefully scrutinized for youth and membership in moving groups and then more likely to be included in our target list. This may be the case for a few individual targets; 
however, most of the main sources of our original target compilations 
\citep[e.g., ][]{sacy,wright2004} are completely 
unrelated to the presence of IR excesses or resolved disks, so we think this
bias is minor, if present at all.

\section{Sample properties}
\label{s:properties}

\begin{figure}[ht]
    \centering
    \includegraphics[width=0.45\textwidth]{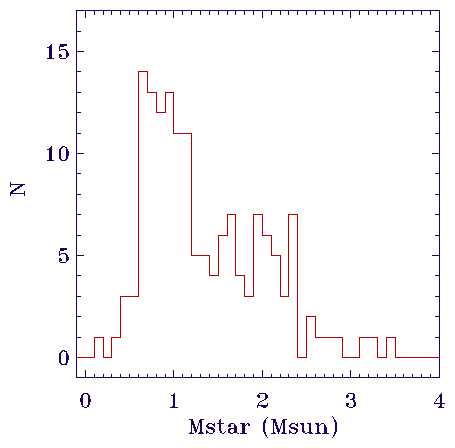}
    \caption{Histogram of stellar masses for the F150 sample}
    \label{fig:histmass}
\end{figure}

\begin{figure}[ht]
    \centering
    \includegraphics[width=0.45\textwidth]{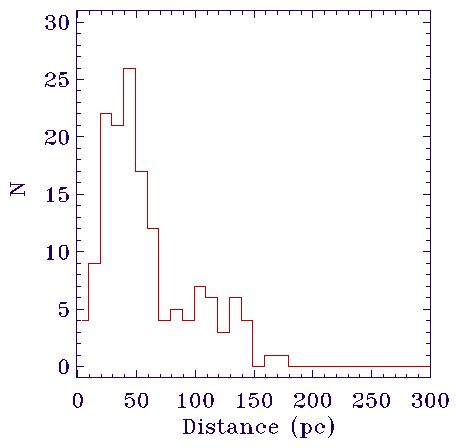}
    \caption{Histogram of distances for the F150 sample}
    \label{fig:histdist}
\end{figure}

\begin{figure}[ht]
    \centering
    \includegraphics[width=0.45\textwidth]{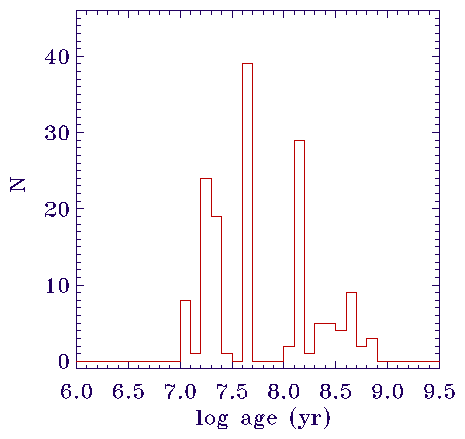}
    \caption{Histogram of stellar ages for the F150 sample}
    \label{fig:histage}
\end{figure}

Figures~\ref{fig:histmass}-\ref{fig:histage} show the distributions for some of the key astrophysical parameters for our F150 sample. Figures~\ref{fig:cumulhmag}, \ref{fig:cumulmass}, and \ref{fig:cumulage} show the cumulative distributions along with the comparison with the GPIES sample (Sect.~\ref{sec:gpiescomparison}).

The bumps in the age distribution are due to the large fraction of members in young MGs. The median age value is 45 Myr, with 
90\% limits of 11 and 450 Myr. 
The adopted ages are typically similar to those originally considered in the sample selection process, although improved thanks to the availability of data on individual targets and better ages of the MGs.
Only in a handful of cases were the revised ages found to be  $>1$ Gyr. These targets
were removed from the present work, being old interlopers in the original sample (tidallylocked binaries, Li-rich giants, or stars with badly measured age indicators;  details will be provided in forthcoming works).

The median mass is 1.15 $M_{\odot}$, with
90\% limits of 0.57 and 2.37 $M_{\odot}$. 
Most of the early-type stars (mass $\ge 1.5 M_{\odot}$) are members of Sco-Cen groups. 
The broad mass range of the sample will allow us to investigate in Paper III the mass dependence of the frequency and properties of substellar companions. 
The analysis will be extended to higher stellar masses by the BEAST survey, targeting B-type stars in Sco-Cen \citep{janson2019}.

The median distance is 48\,pc, with 90\% limits of 11 and 137\,pc. 
The peak in the distance distribution between 100 and 150\,pc is due to the inclusion of Sco-Cen members. 
At 150\,pc, the inner working angle of SPHERE allows us to access separation of $\ge$ 20\,au for the presence of planetary companions.

The stars with resolved disks and detectable IR excess have a different mass distributions with respect to those without these features, being more massive (median values 1.41 versus 0.94 $M_{\odot}$), while
the age distributions of stars with and without disks are similar. 
Considering the inhomogeneity of our census of disks  concerning, for example, the wavelengths of the observations, sensitivity to IR excess with respect to the stellar photosphere, and sensitivity to spatially resolved disks, we do not investigate the origin of these features. 
While our original survey sample has no specific biases linked to the presence of disks or IR excess, which were never considered in the selection process, some of the stars with disks were observed with increased priority because of the presence of the disks themselves (see Sect.~\ref{sec:P0}). This effect will be mitigated by the end of the survey, allowing more robust statistical inferences.

As already noted in \citet{vigan2017}, the majority of young nearby stars have metallicity  values close to solar, making the sample of stars searched for planets via direct imaging somewhat different with respect to those considered by RVs and transits, which span a broader range of age and metallicity.
The available data \citep[e.g.,][]{vianaalmeida2009} suggest a slightly subsolar metallicity for stars in nearby young associations, at odds with expectations from galactic chemical evolution models. Recent results indicate that the standard chemical abundance analysis  might be biased for young stars, because of an overestimation of microturbulent velocities somewhat linked to stellar activity \citep{reddy2017,baratella2020,spina2020}. Since a new analysis of chemical composition of members in various moving groups with these new methods is not yet available, we assumed in the following a solar metallicity for all the targets.

\subsection{Comparison with GPIES and other surveys}
\label{sec:gpiescomparison}

\subsubsection{Properties of individual targets}

In the F150 sample, 67 out of 150 stars were also observed with GPI and were included in the GPIES early statistical analysis \citep{nielsen2019}. In Fig.~\ref{fig:gpi_age} we compare the adopted ages; 49 of the 67 overlapping targets are members of moving groups, for which the adopted individual age is equal (TW Hya, $\eta$ Cha, Tuc-Hor, Columba, Carina, Argus, AB Dor MG) or differ by a very small amount, 1 or 2 Myr ($\beta$ Pic MG, Sco-Cen groups). As a result, the median age difference is equal to zero, and we can infer that the age scales in the two studies are very similar. Nevertheless, there are moderately large discrepancies for some individual field objects, for which ages are more uncertain. In a few cases for which the membership to groups is ambiguous (see Appendix \ref{a:notes} for details on individual objects), we adopted the stellar age derived independently of the group membership constraints (typically much older than the group ages) with lower limits encompassing the group ages. In most of these cases, \citet{nielsen2019} adopted the membership to the groups and the corresponding ages, resulting in fairly large discrepancies. We also note that the ages adopted in \citet{nielsen2019} for the components in the Fomalhaut system (749 Myr for A and 200 Myr for B) bracket our adopted common value for the system (440 Myr).

The comparison of the adopted masses in Fig.~\ref{fig:gpi_mass} also shows fairly good agreement over the whole range of masses considered by the programs. There is a small systematic difference,  median delta of 0.03 $M_{\odot}$ with our masses being smaller. There is perfect agreement for the adopted distances, derived from the same sources (Gaia DR2 and Hipparcos).

\subsubsection{Sample comparison}

\begin{figure}[ht]
    \centering
    \includegraphics[width=0.45\textwidth]{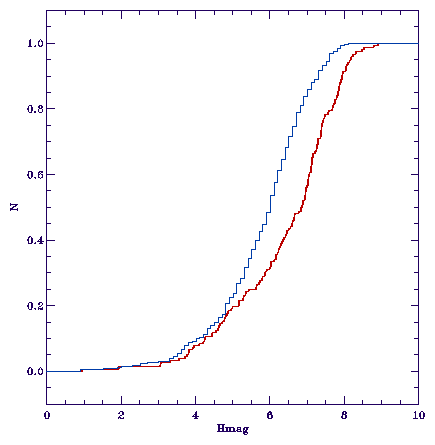}
    \caption{Cumulative distribution of H-band magnitude for the stars in our sample (red line) and that of \citet{nielsen2019} (blue line).}
    \label{fig:cumulhmag}
\end{figure}

\begin{figure}[ht]
    \centering
    \includegraphics[width=0.45\textwidth]{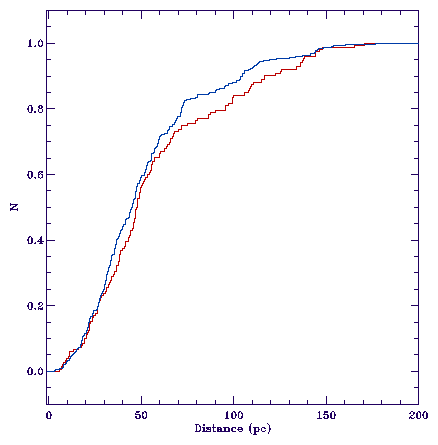}
    \caption{Cumulative distribution of distance for the stars in our sample (red line)
    and that of \citet{nielsen2019} (blue line).}
    \label{fig:cumuldist}
\end{figure}

\begin{figure}[ht]
    \centering
    \includegraphics[width=0.45\textwidth]{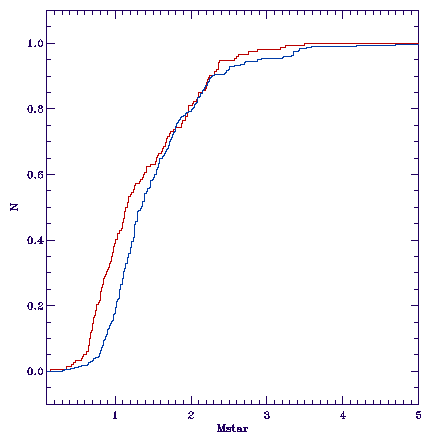}
    \caption{Cumulative distribution of masses for the stars in our sample (red line)
    and that of \citet{nielsen2019} (blue line).}
    \label{fig:cumulmass}
\end{figure}

After the comparison of the individual stellar parameters, we compared the distributions, in order to reveal differences between the two samples. A first highly significant difference concerns the stellar magnitude (Fig. \ref{fig:cumulhmag}). This can be understood due to the differences between the AO systems of GPI and SPHERE, the latter working well to fainter magnitudes. This allowed us to choose a fainter magnitude limit. As a result, our sample includes a larger fraction of low-mass stars and extends to slightly larger distances (Fig. \ref{fig:cumuldist}). The median masses of SHINE-F150 and GPIES being 1.15 and 1.34 $M_{\odot}$, respectively, with stars below 1 $M_{\odot}$ representing 40\% of the sample for SHINE and just 19\% for GPIES (Fig. \ref{fig:cumulmass}). Additional differences concern the high-mass tail, where we stopped at about 3 $M_{\odot}$. There are only three stars with masses higher than this value in our final determination, (i.e., 2\% of the sample, with a maximum value of 3.48 $M_{\odot}$). On the other hand, GPIES extends up to 9 $M_{\odot}$ with 4.5\% of the targets more massive than 3 $M_{\odot}$.

\begin{figure}[ht]
    \centering
    \includegraphics[width=0.45\textwidth]{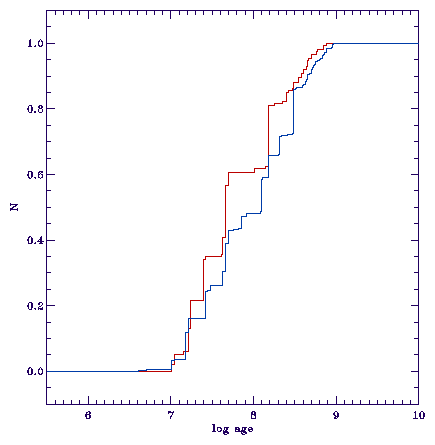}
    \caption{Cumulative distribution of log (age) for the stars in our sample (red line)
    and that of \citet{nielsen2019} (blue line).}
    \label{fig:cumulage}
\end{figure}

There are also some differences in the age distribution (median ages of 45 Myr for SHINE and 125 Myr for GPIES, Fig. \ref{fig:cumulage}) . The age difference is mostly explained by the larger fraction of field early-type stars (typically intermediate age) in the GPIES sample and by the larger fraction of young low-mass MG members in our sample, due to the fainter magnitude limit.

Both teams avoided close visual binaries within the field of view of the high-contrast instruments.
As  the  field of view of GPIES is smaller with respect to SHINE/SPHERE, binaries with 3-6 arcsec projected separation, not present in our sample, are included in GPIES. Finally, GPIES included in their statistics a sample of spatially unresolved binaries. The corresponding planets searched in these systems are circumbinary. Instead, we excluded these systems from the present study, although unrecognized spectroscopic binaries might still be present because of the lack of RV monitoring for a fraction of our targets. As spectroscopic binaries represent about 10\% of the GPIES sample, this difference could have some impact on the statistical results.

\begin{figure}[ht]
    \centering
    \includegraphics[width=0.45\textwidth]{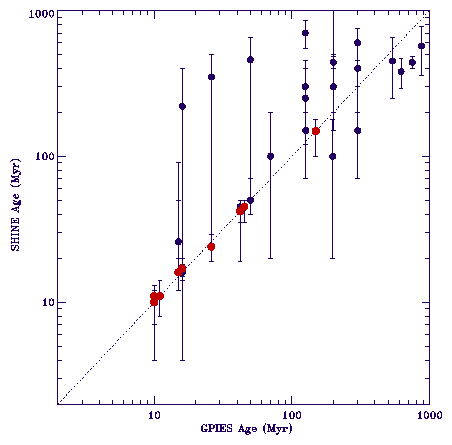}
    \caption{Comparison of ages derived in the present work and in \citet{nielsen2019}. Black circles refer to individual stars and red circles to moving groups (typically several targets for each group). The error bars refer to the minimum and maximum
    age values from Table \ref{t:ages}}.
    \label{fig:gpi_age}
\end{figure}

\begin{figure}[ht]
    \centering
    \includegraphics[width=0.45\textwidth]{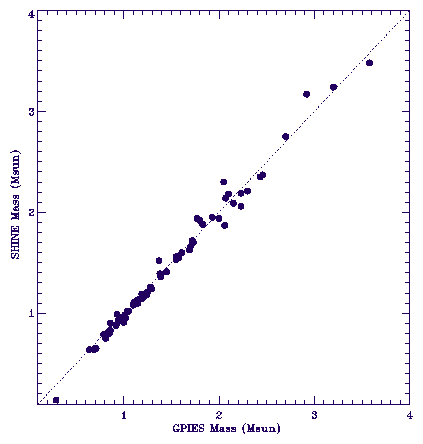}
    \caption{Comparison of stellar masses derived in the present work and in \citet{nielsen2019}.}
    \label{fig:gpi_mass}
\end{figure}

\section{Summary and conclusions}
\label{s:conclusion}

We described SHINE, the largest direct imaging survey for exoplanets at VLT performed as part of SPHERE GTO. We detailed the process of sample selection and the priority ranking scheme. The survey is focused on young nearby stars, with spectral types from A to M. Known binaries within the field of view of the SPHERE-IRDIS camera (6 arcsec) are excluded, as are all known spectroscopic binaries (though not all targets were thoroughly searched).

A subsample of 150 stars with first epoch observations done before February, 2017 was defined for a preliminary statistical assessment of the frequency of planets and brown dwarfs in wide orbits (5-300\,au). This paper presents the characterization of the individual targets and of this subsample as a whole. The companion paper (Langlois et al., in press)  presents the observations, data processing,  identification, and classification of companion candidates, while \citet{vigan2020} presents the statistical analysis of the frequency of substellar companions and its dependence on stellar mass.

We exploited a variety of methods (kinematics and membership to groups, isochrone, lithium, rotation, and activity) to infer the stellar age and other stellar parameters. The median age value is 45 Myr, with 90\% limits of 11 and 450 Myr. The median stellar mass is 1.15 $M_{\odot}$, with 90\% limits of 0.57 and 2.37 $M_{\odot}$. A comparison with GPIES early statistical analysis \citep{nielsen2019} shows no large systematic differences in the age scales between the two studies, but significant differences in the mass distribution and binary properties.

\begin{acknowledgements}

    SPHERE is an instrument designed and built by a consortium consisting of IPAG (Grenoble, France), MPIA (Heidelberg, Germany), LAM (Marseille, France), LESIA (Paris, France), Laboratoire Lagrange (Nice, France), INAF - Osservatorio di Padova (Italy), Observatoire de Gen\`eve (Switzerland), ETH Z\"urich (Switzerland), NOVA (Netherlands), ONERA (France) and ASTRON (Netherlands) in collaboration with ESO. SPHERE was funded by ESO, with additional contributions from CNRS (France), MPIA (Germany), INAF (Italy), FINES (Switzerland) and NOVA (Netherlands). SPHERE also received funding from the European Commission Sixth and Seventh Framework Programmes as part of the Optical Infrared Coordination Network for Astronomy (OPTICON) under grant number RII3-Ct-2004-001566 for FP6 (2004--2008), grant number 226604 for FP7 (2009--2012) and grant number 312430 for FP7 (2013--2016). 

This research has made use of the SIMBAD database and Vizier services, operated at CDS, Strasbourg, France and of the Washington Double Star Catalog maintained at the U.S. Naval Observatory. 
This work has made use of data from the European Space Agency (ESA) mission
{\it Gaia} (\url{https://www.cosmos.esa.int/gaia}), processed by the {\it Gaia}
Data Processing and Analysis Consortium (DPAC,
\url{https://www.cosmos.esa.int/web/gaia/dpac/consortium}). Funding for the DPAC
has been provided by national institutions, in particular the institutions
participating in the {\it Gaia} Multilateral Agreement.
This paper includes data collected with the TESS mission, obtained from the MAST data archive at the Space Telescope Science Institute (STScI). Funding for the TESS mission is provided by the NASA Explorer Program. STScI is operated by the Association of Universities for Research in Astronomy, Inc., under NASA contract NAS 5–26555.
This paper has made use of data products available in ESO archive. Program ID: 
60.A-9036(A); 
072.C-0488(E)  (PI Mayor),
074.C-0364(A)  (PI Robichon),
074.C-0037(A)  (PI Gunther),
075.C-0202(A)  (PI Gunther),
075.C-0689(A)  (PI Galland),
076.C-0010(A)  (PI Gunther),
077.C-0012(A)  (PI Gunther),
077.C-0295(D)  (PI Galland),
078.D-0245(C)  (PI Dall),
079.C-0046(A)  (PI Gunther),
080.D-0151(A)  (PI Hatzes),
080.C-0712(A)  (PI Desort),
180.C-0886(A)  (PI Bonfils),
082.C-0718(B)  (PI Bonfils),
082.C-0427(A)  (PI Doellinger),
082.C-0390(A)  (PI Weise),
183.C-0437(A)  (PI Bonfils),
083.C-0794(A)  (PI Chauvin),
084.C-1039(A)  (PI Chauvin),
184.C-0815(B)  (PI Desort),
089.C-0732(A)  (PI Lo Curto),
191.C-0873(D)  (PI Bonfils),
192.C-0224(A)  (PI Lagrange),
097.C-0864(B)  (PI Lannier),
098.C-0739(A)  (PI Lagrange),
099.C-0205(A)  (PI Lagrange),
099.C-0458(A)  (PI Lo Curto),
1101.C-0557(A)  (PI Lagrange).
We have used data from the WASP public archive in this research. The WASP consortium comprises of the University of Cambridge, Keele University, University of Leicester, The Open University, The Queen's University Belfast, St. Andrews University and the Isaac Newton Group. Funding for WASP comes from the consortium universities and from the UK's Science and Technology Facilities Council.
Based on data retrieved from the SOPHIE archive at Observatoire de Haute-Provence (OHP), available at \url{http://atlas.obs-hp.fr/sophie/}

S.D., V.D., D.M. and R.G. acknowledge the support by INAF/Frontiera through the "Progetti Premiali" funding scheme of the Italian Ministry of Education, University, and Research.   
AV acknowledges funding from the European Research Council (ERC) under the European Union's Horizon 2020 research and innovation programme (grant agreement No. 757561). AML
acknowledges funding from Agence Nationale de la Recherche (France) under
contract number ANR-14-CE33-0018. JC was supported by SC Space Grant and Fulbright Colombia.
M.B. acknowledges funding by the UK Science and Technology Facilities Council (STFC) grant no. ST/M001229/1.

\end{acknowledgements}

\bibliography{f100}
\bibliographystyle{aa}

\begin{appendix}
\section{Notes on individual objects}

\label{a:notes}
Together with relevant notes on individual targets, in  Figs\,\ref{fig:HIP490_fig}--\,\ref{fig:HIP118008}  we summarize the results of our periodogram analysis.  In the top left panel we plot magnitudes versus TESS Julian Date, unless differently specified.
In the top middle panel we plot the Lomb$-$Scargle
periodogram with the spectral window function  (red dotted line) and power level corresponding to FAP = 0.01\% and 0.1\% (horizontal dashed line),
and we indicate the peak corresponding to the rotation period. In the top right panel we plot the CLEAN periodogram. In the bottom panel we plot
the light curve phased with the rotation period. The solid line represents the sinusoidal fit.

\begin{description}

\item {\bf HIP 490 = HD 105}
We measured for the first time the rotation period from the TESS photometric time series  
(Fig.\,\ref{fig:HIP490_fig}). The light curve exhibits a significant evolution of amplitude in subsequent rotation cycles.

\begin{figure}[htbp]
    \centering
    \includegraphics[width=5.7cm,angle=90]{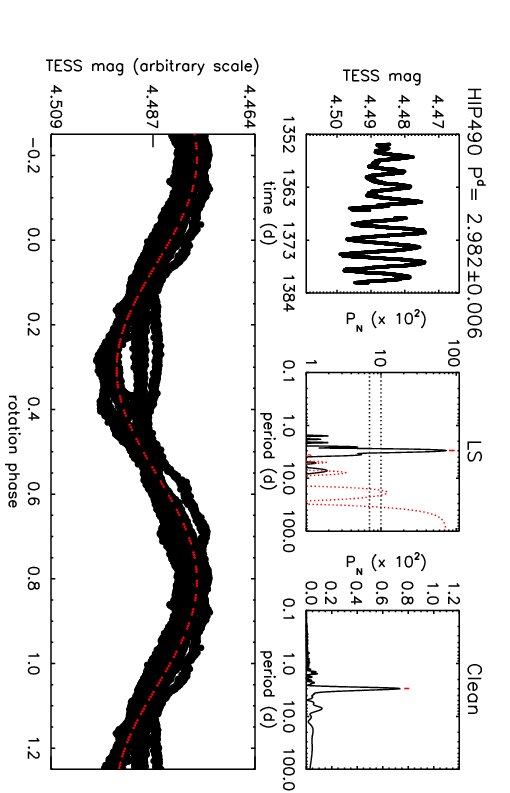}
    \caption{Photometric time sequence and periodogram for HIP490 = HD 105}
    \label{fig:HIP490_fig}
\end{figure}

\item {\bf HIP 682 = HD 377} 
Star with resolved debris disk \citep{choquet2016}. We measured for the first time the rotation period from the All Sky Automated Survey (ASAS)  photometric time series (Fig.\,\ref{fig:HIP682}). This allowed us to refine the age estimate of the target, which results very close to that of the Pleiades. The star does not result as a member of any known young moving group. The inclination estimated from rotation period, $v\sin{i}$, and stellar radius is compatible within the error with that of the disk (85$\pm$5$^\circ$). It was set as P0 target for disk characterization purposes.

\begin{figure}[htbp]
    \centering
    \includegraphics[width=9cm,angle=0, trim = 100 50 0 0]{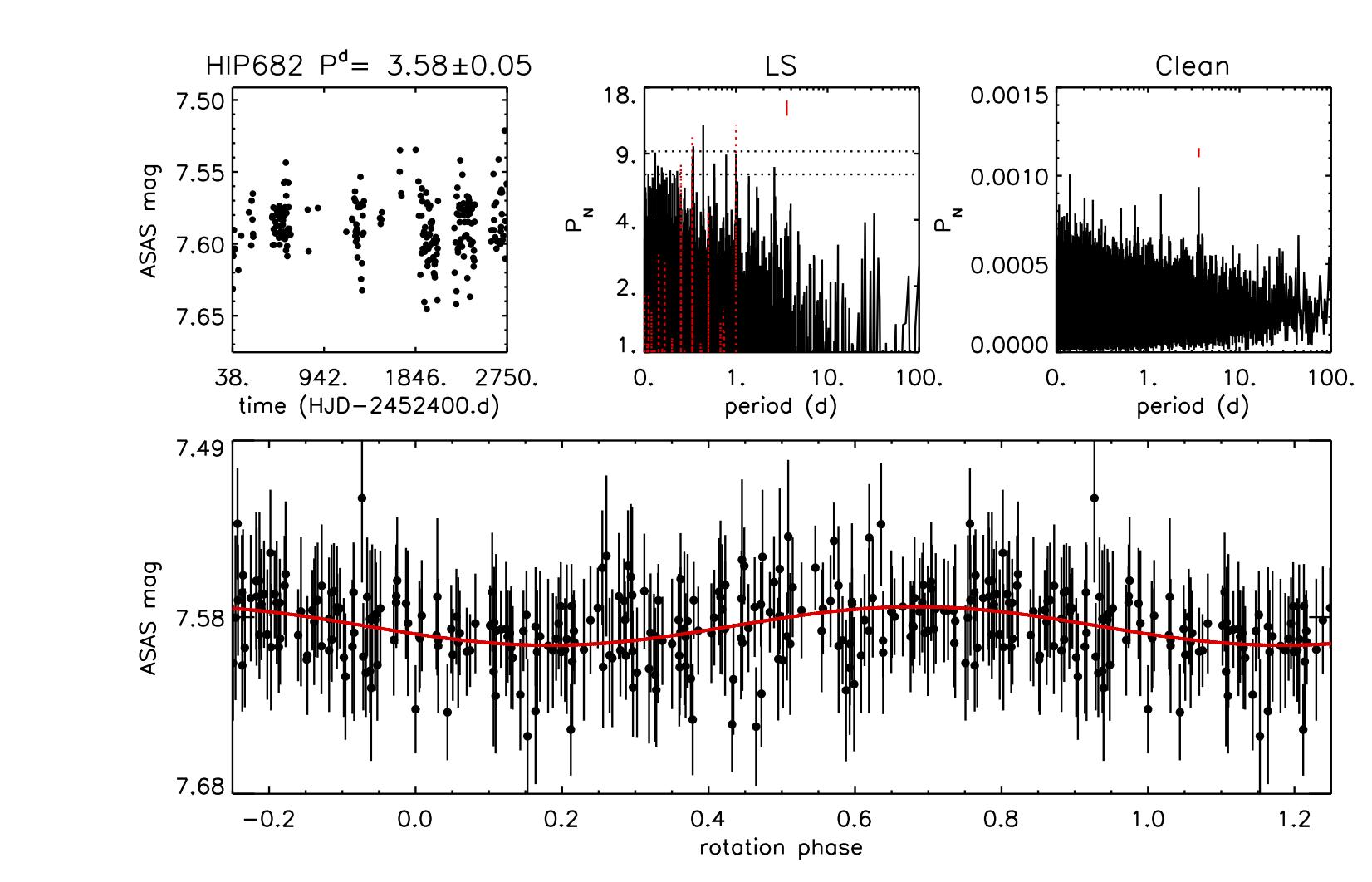}
    \caption{Photometric time sequence and periodogram for HIP682 = HD 377}
    \label{fig:HIP682}
\end{figure}

\item {\bf HIP 1113 = HD 987}
The  photometric  rotation period first measured by \citet{Messina10} is confirmed by our analysis of the TESS data
(Fig.\,\ref{fig:HIP1113}). The TESS data revealed one flare event superimposed on a quite stable light curve. 

\begin{figure}[htbp]
    \centering
    \includegraphics[width=5.7cm,angle=90]{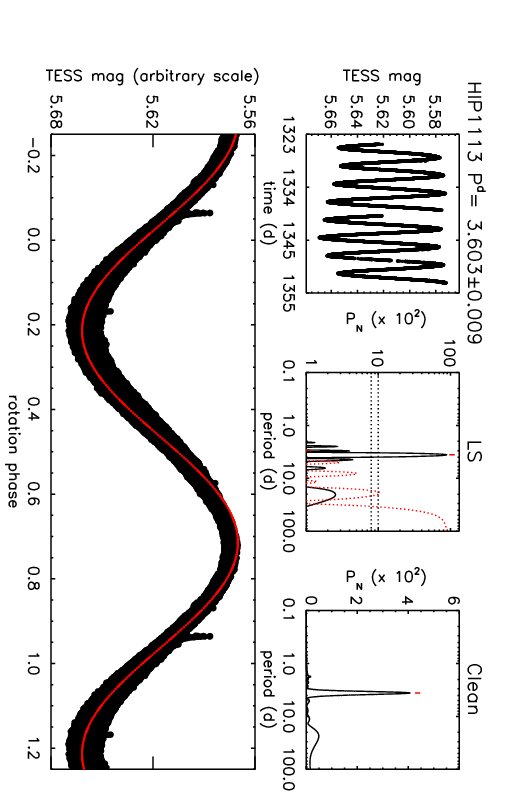}
    \caption{Photometric time sequence and periodogram for HIP1113}
    \label{fig:HIP1113}
\end{figure}

\item {\bf 2MASS J00172353-6645124}
The  photometric  rotation period first measured by \citet{Messina17} is confirmed by our analysis of the TESS data
(Fig.\,\ref{fig:2MASSJ0017-6645}). The TESS data revealed, superimposed on a quite stable light curve, the presence of multiple flare events, that support the young age estimated for this M-type star. 

\begin{figure}[htbp]
    \centering
    \includegraphics[width=5.7cm,angle=90]{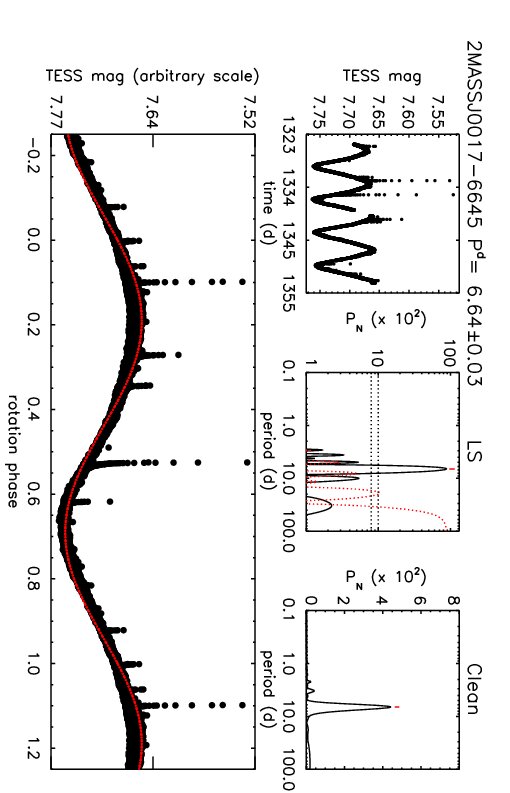}
    \caption{Photometric time sequence and periodogram for 2MASSJ0017-6645}
    \label{fig:2MASSJ0017-6645}
\end{figure}

\item {\bf HIP 1481 = HD 1466}
We measured for the first time the rotation period from the TESS photometric time series
(Fig.\,\ref{fig:HIP1481}). 

\begin{figure}[htbp]
    \centering
    \includegraphics[width=5.7cm,angle=90]{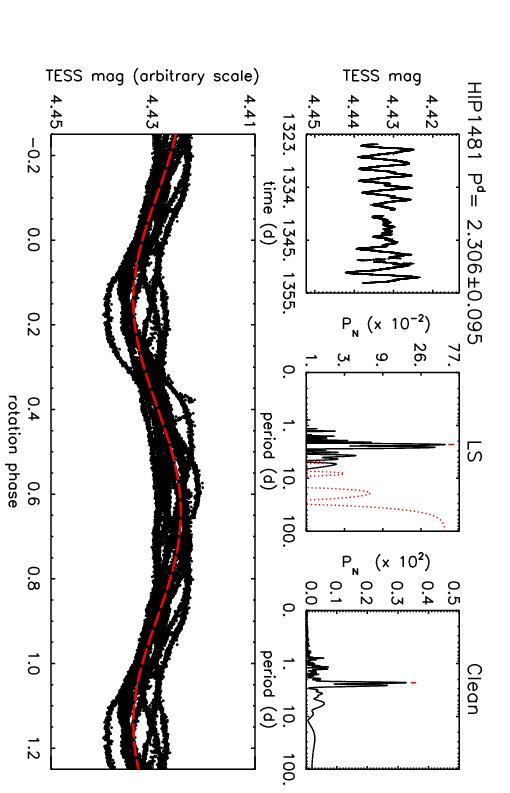}
    \caption{Photometric time sequence and periodogram for HIP1481}
    \label{fig:HIP1481}
\end{figure}

\item {\bf HIP 1993 = CT Tuc}
The  photometric  rotation period first measured by \citet{Messina10} is confirmed by our analysis of the TESS data
(Fig.\,\ref{fig:HIP1993}). The TESS data revealed, superimposed on a quite stable light curve, the presence of multiple flare events, that support the young age estimated for this M-type star. 

\begin{figure}[htbp]
    \centering
    \includegraphics[width=5.7cm,angle=90]{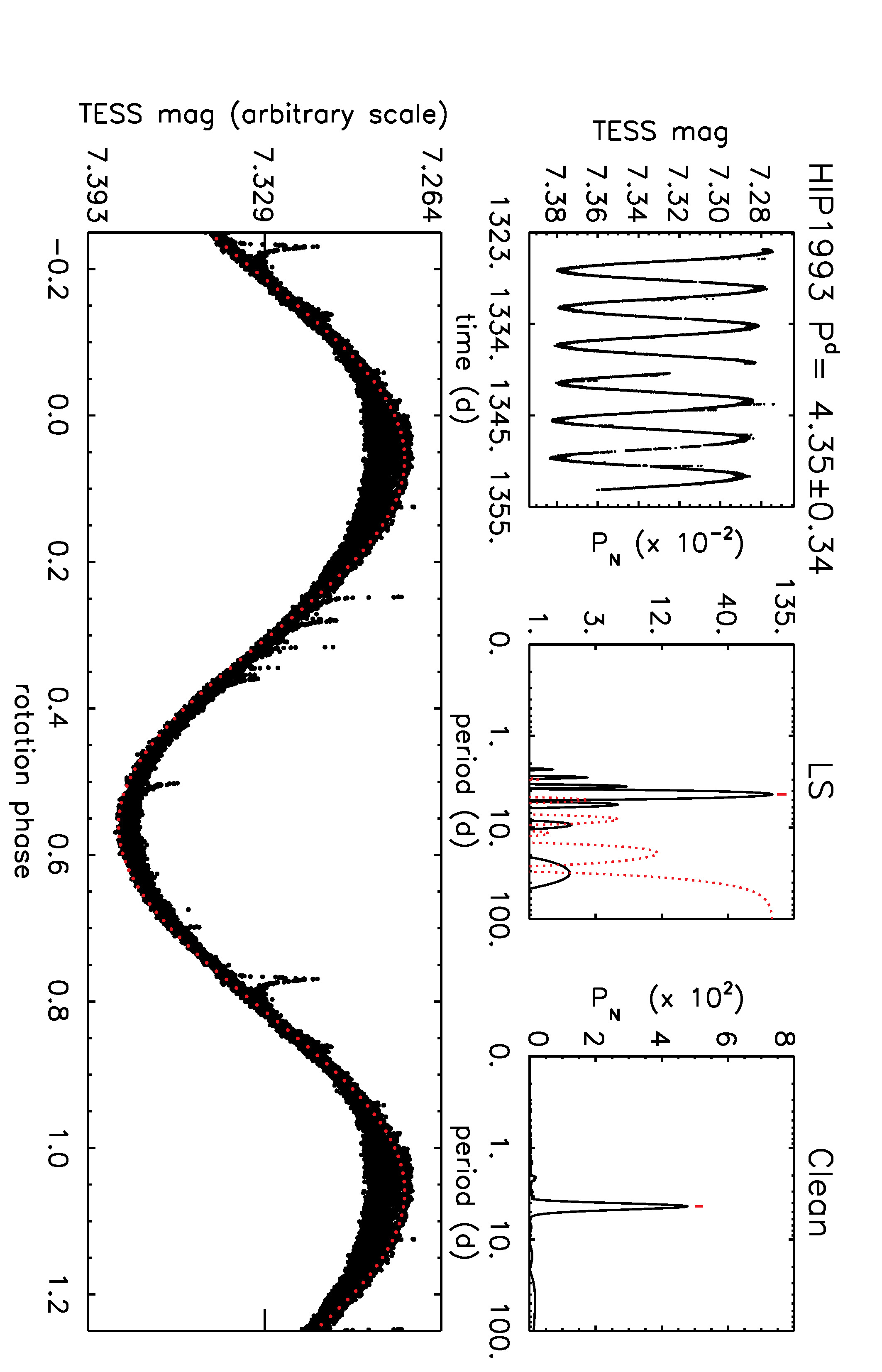}
    \caption{Photometric time sequence and periodogram for HIP1993}
    \label{fig:HIP1993}
\end{figure}

\item {\bf HIP 2578 = HD 3003}
Member of Tuc-Hor association. It is comoving with the quadruple system HIP 2484+HIP 2487 (both of which are also visual binaries) at 25350 au projected separation. Masses of the components in Table \ref{t:wide} taken from \citet{tokovinin2008}.

\item {\bf HIP 6276 = BD-12 243}
We measured for the first time the rotation period from the TESS photometric time series
(Fig.\,\ref{fig:HIP6276}), which is in rough agreement with the earlier measurement (P = 6.40\,d) by \citet{Wright11}.  TESS data revealed, superimposed on a quite stable light curve, the presence of multiple flare events. 

\begin{figure}[htbp]
    \centering
    \includegraphics[width=5.7cm,angle=90]{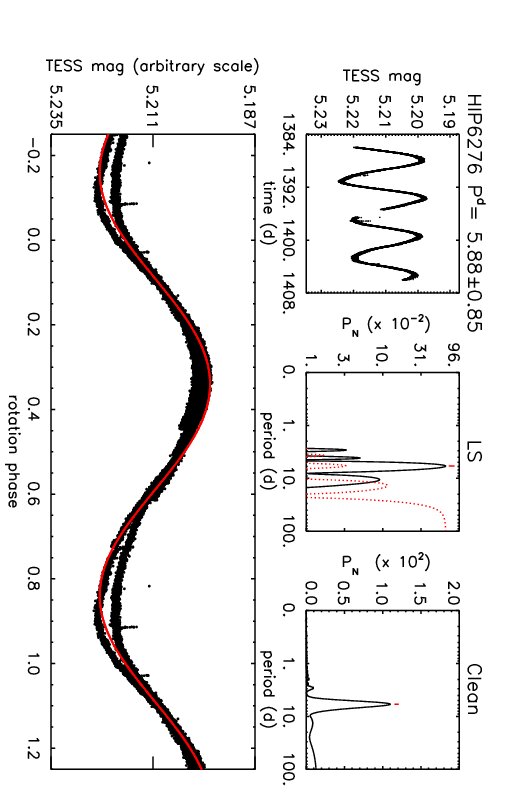}
    \caption{Photometric time sequence and periodogram for HIP6276}
    \label{fig:HIP6276}
\end{figure}

\item {\bf HIP 6485 = HD 8558}
The photometric rotation period first measured by \citet{Messina10} is confirmed by the TESS data 
(Fig.\,\ref{fig:HIP6485}) and is in agreement with the measurements by \citet{Oelkers18}.  TESS data reveal signficant evolution of light curve amplitude and numerous flare events. 

\begin{figure}[htbp]
    \centering
    \includegraphics[width=5.7cm,angle=90]{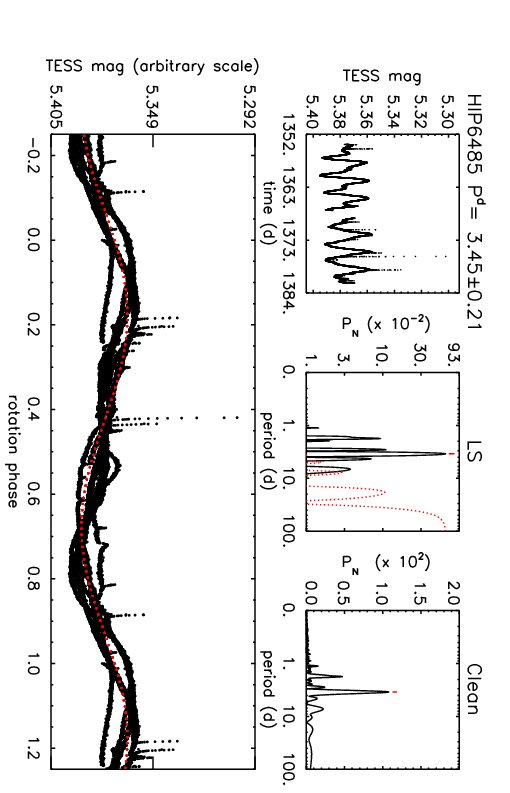}
    \caption{Photometric time sequence and periodogram for HIP6485}
    \label{fig:HIP6485}
\end{figure}

\item {\bf HIP 6856 = HD 9054}
We measured for the first time the rotation period from the TESS photometric time series
(Fig.\,\ref{fig:HIP6856}), which revealed numerous flare events. 

\begin{figure}[htbp]
  \centering
    \includegraphics[width=5.7cm,angle=90]{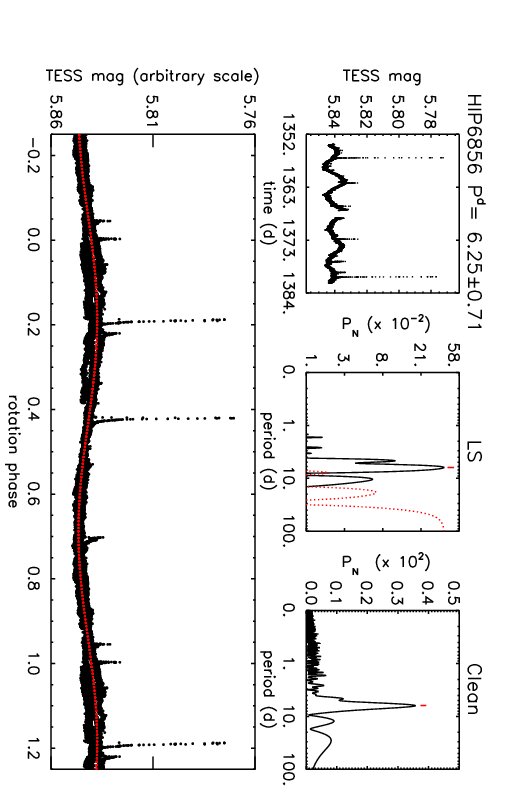}
    \caption{Photometric time sequence and periodogram for HIP6856}
    \label{fig:HIP6856}
\end{figure}

\item {\bf TYC 8047-0232-1}
Star with brown dwarf companion discovered by \citet{chauvin2005gsc}.
The  photometric  rotation period first measured by \citet{Messina10} is confirmed by our analysis of the TESS data
(Fig.\,\ref{fig:TYC804702321}).

\begin{figure}[htbp]
    \centering
    \includegraphics[width=5.7cm,angle=90]{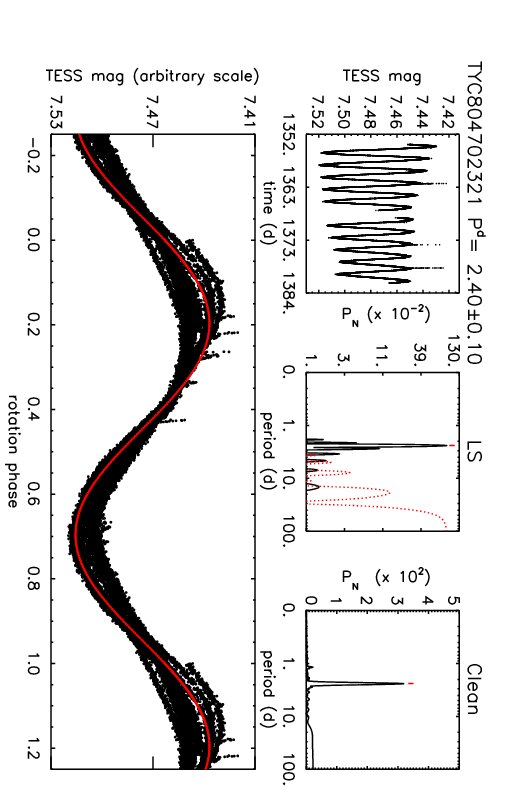}
    \caption{Photometric time sequence and periodogram for TYC 8047-0232-1}
    \label{fig:TYC804702321}
\end{figure}


\item {\bf HIP 11360 = HD 15115}
The revised RV from \cite{desidera2015} supports the membership to the Tuc-Hor association (93.7\% using BANYAN). The star has a spatially resolved edge-on debris disk \citep{kalas2007,engler2019}. We measured for the first time the rotation period from the TESS photometric time series (Fig.\,\ref{fig:HIP11360}).

\begin{figure}[htbp]
    \centering
    \includegraphics[width=5.7cm,angle=90]{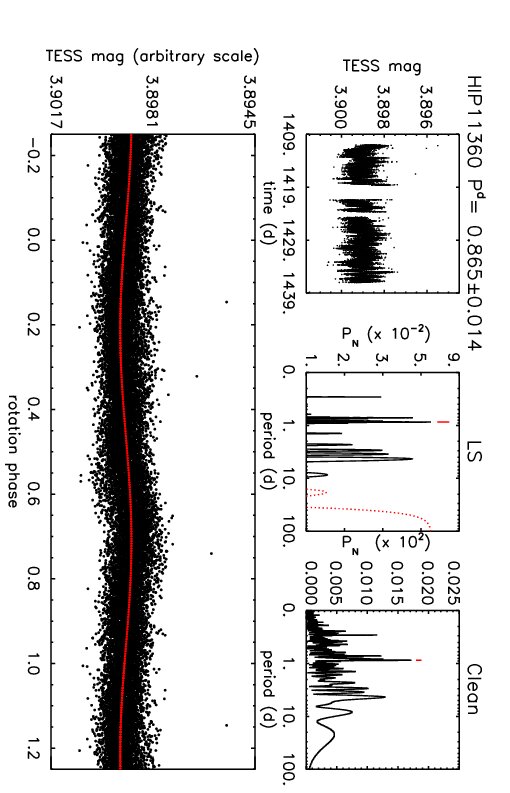}
    \caption{Photometric time sequence and periodogram for HIP11360}
    \label{fig:HIP11360}
\end{figure}


\item {\bf HIP 13402 = HD 17925 = EP Eri} 
Flagged as RS CVn variable, with spectral types K1+K2 and period 6.85d in \citet{rodriguez2015}.
However, HARPS observations available in ESO archive allow us to rule out the presence of close stellar companions (rms = 28 m s$^{-1}$ from 42 RVs over 1200 days). We then kept the star in the sample.
The photometric rotation period first measured by \citet{messina2001} is confirmed by our analysis of the TESS data
(Fig.\,\ref{fig:HIP13402}). 

\begin{figure}[htbp]
    \centering
    \includegraphics[width=5.7cm,angle=90]{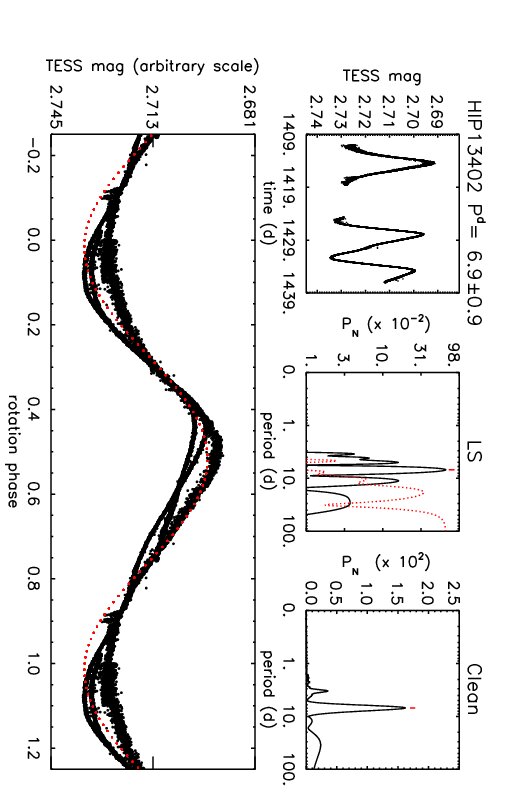}
    \caption{Photometric time sequence and periodogram for HIP13402}
    \label{fig:HIP13402}
\end{figure}

\item {\bf HIP 14551 = HD 19545} 
Originally classified as a member of Tuc-Hor \citep{zuckerman2011}, the updated analysis indicates membership in the Columba association. The wide companion, UCAC4 311-003056, has been also classified as a member of the Columba association \citep{gagne2018}.  
The small but formally significant differences in parallax and proper motion make it possible that the two stars do not form a true binary system and  are only projected very close on the sky (59 arcsec). 

\item {\bf TYC 7026-0325-1}
The  photometric  rotation period first measured by \citet{Messina10} is confirmed by our analysis of the TESS data
(Fig.\,\ref{fig:TYC702603251}). 

\begin{figure}[htbp]
    \centering
    \includegraphics[width=5.7cm,angle=90]{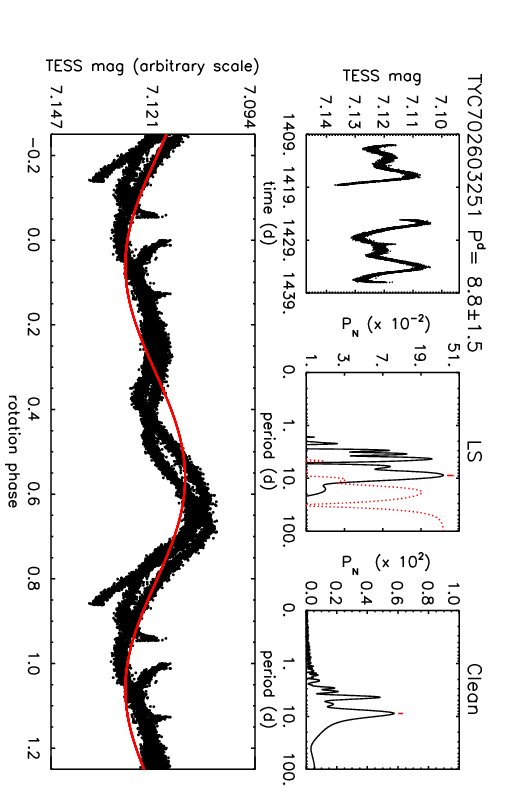}
    \caption{Photometric time sequence and periodogram for TYC 7026-0325-1}
    \label{fig:TYC702603251}
\end{figure}

\item {\bf HIP 15457 =  HD 20630 = $\kappa$ Cet}  
The  photometric  rotation period first measured by \citet{messina2001} is confirmed by our analysis of the TESS data
(Fig.\,\ref{fig:HIP15457}). The presence of numerous flare events are detected.

\begin{figure}[htbp]
    \centering
    \includegraphics[width=5.7cm,angle=90]{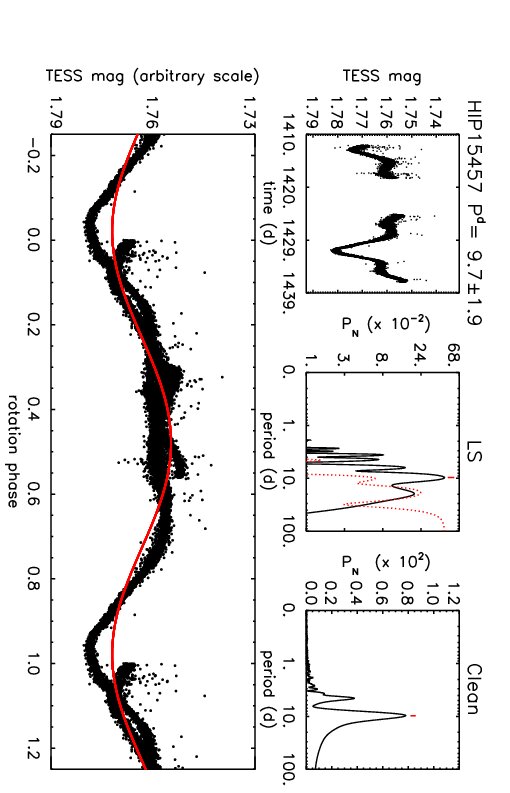}
    \caption{Photometric time sequence and periodogram for HIP15457 ($\kappa$ Cet)}
    \label{fig:HIP15457}
\end{figure}

\item {\bf TYC 8060-1673-1}
The  photometric  rotation period first measured by \citet{Messina10} is confirmed by our analysis of the TESS data
(Fig.\,\ref{fig:TYC806016731}). 

\begin{figure}[htbp]
    \centering
    \includegraphics[width=5.7cm,angle=90]{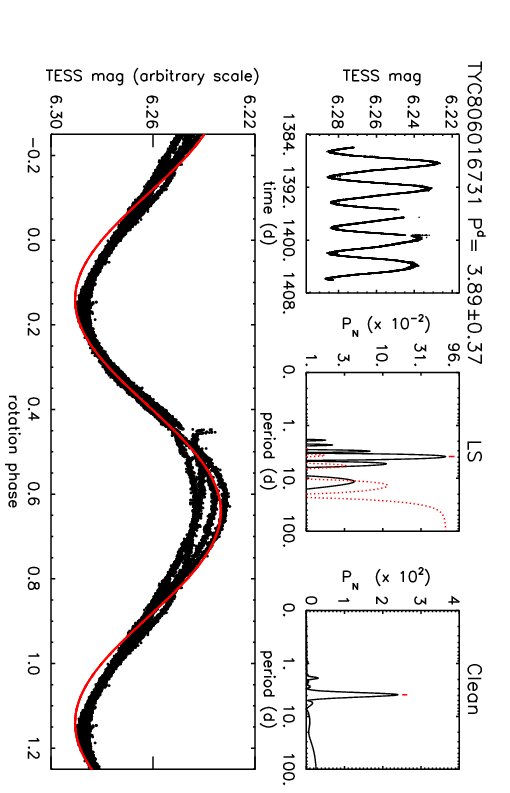}
    \caption{Photometric time sequence and periodogram for TYC 8060-1673-1}
    \label{fig:TYC806016731}
\end{figure}

\item {\bf HIP 17764 = HD 24636}
We measured for the first time the rotation period from the TESS photometric time series
(Fig.\,\ref{fig:HIP17764}).

\begin{figure}[htbp]
    \centering
    \includegraphics[width=5.7cm,angle=90]{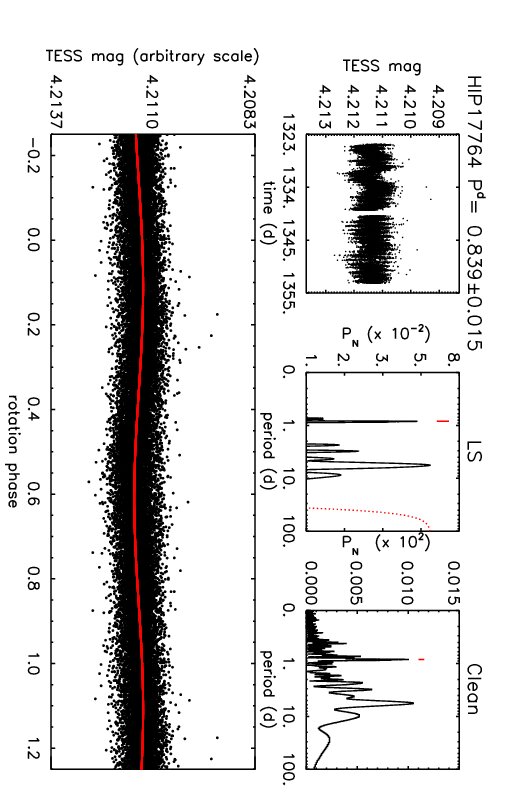}
    \caption{Photometric time sequence and periodogram for HIP17764}
    \label{fig:HIP17764}
\end{figure}

\item {\bf HD 25284B}
We measured for the first time the rotation period from the TESS photometric time series
(Fig.\,\ref{fig:HD25284B}). We note that according to SIMBAD coordinates, HD\,25284A and B are unresolved in the TESS photometry. As shown in the figure, both Lomb$-$Scargle and Clean detected three significant periods that are in order of decreasing power P = 4.54\,d, P = 2.26\,d, and P = 0.31\,d. Considering that HD\,25284B has $v\sin{i}$ = 6.9\,km\,s$^{-1}$, its rotation period should be P = 4.54\,d to reconcile with stellar radius and projected rotational velocity. Following similar reasoning,  considering that HD\,25284A has $v\sin{i}$ = 69.8\,km\,s$^{-1}$, its rotation period should be P = 0.31\,d. The remaining period P = 2.26\,d,  half of the primary period, may  arise from the double-dip shape of the light curve. 

\begin{figure}[htbp]
    \centering
    \includegraphics[width=5.7cm,angle=90]{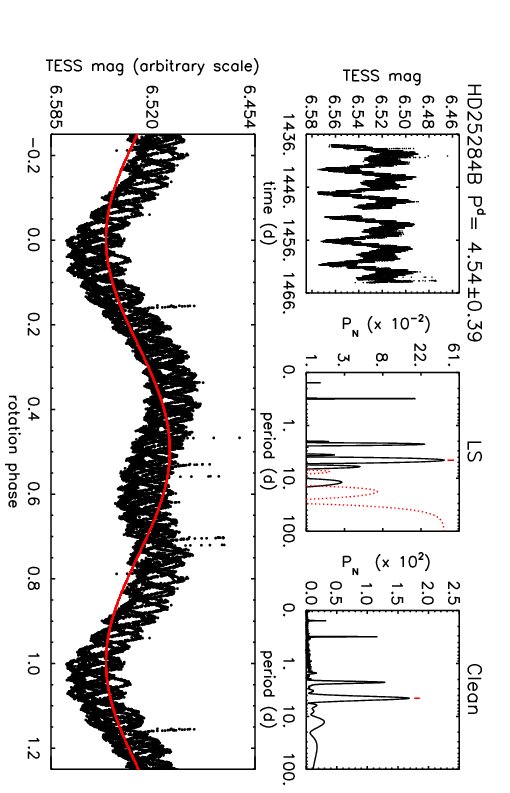}
    \caption{Photometric time sequence and periodogram for HD25284B}
    \label{fig:HD25284B}
\end{figure}

\item {\bf TYC 5882-1169-1 = BD-15 705}  
Originally classified as a member of the Columba association, the updated analysis indicates membership in the Tuc-Hor association.
The  photometric  rotation period first measured by \citet{Messina10} is confirmed by our analysis of the TESS data
(Fig.\,\ref{fig:TYC588211691}). More flare events are detected in the TESS time series. 

\begin{figure}[htbp]
    \centering
    \includegraphics[width=5.7cm,angle=90]{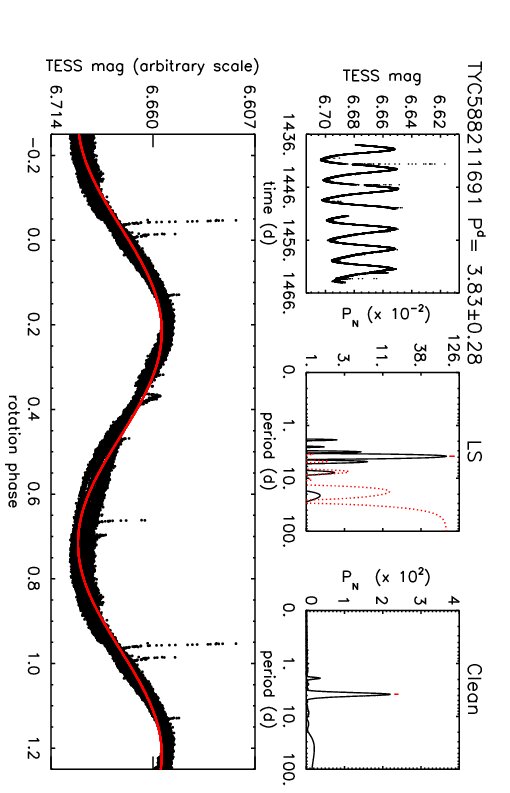}
    \caption{Photometric time sequence and periodogram for TYC 5882-1169-1}
    \label{fig:TYC588211691}
\end{figure}

\item {\bf 51 Eri = HIP 21547}
The star has a planetary companion discovered by \cite{macintosh2015}.
The periodogram analysis of the TESS data shows evidence of multi-periodicity with the most powerful peak at P = 1.84$\pm$0.06\,d and the secondary power peak in agreement with the early rotation period measurement by \citet{Koen02} (Fig.\,\ref{fig:HIP21547}-\ref{fig:51_Eri_HIP}-\ref{fig:51_Eri_mascara}).  Considering the F0IV spectral type, it is likely that most periodicities arise from pulsations rather than variability induced by undiscovered close companions. 

\begin{figure}[htbp]
    \centering
    \includegraphics[width=5.7cm,angle=90]{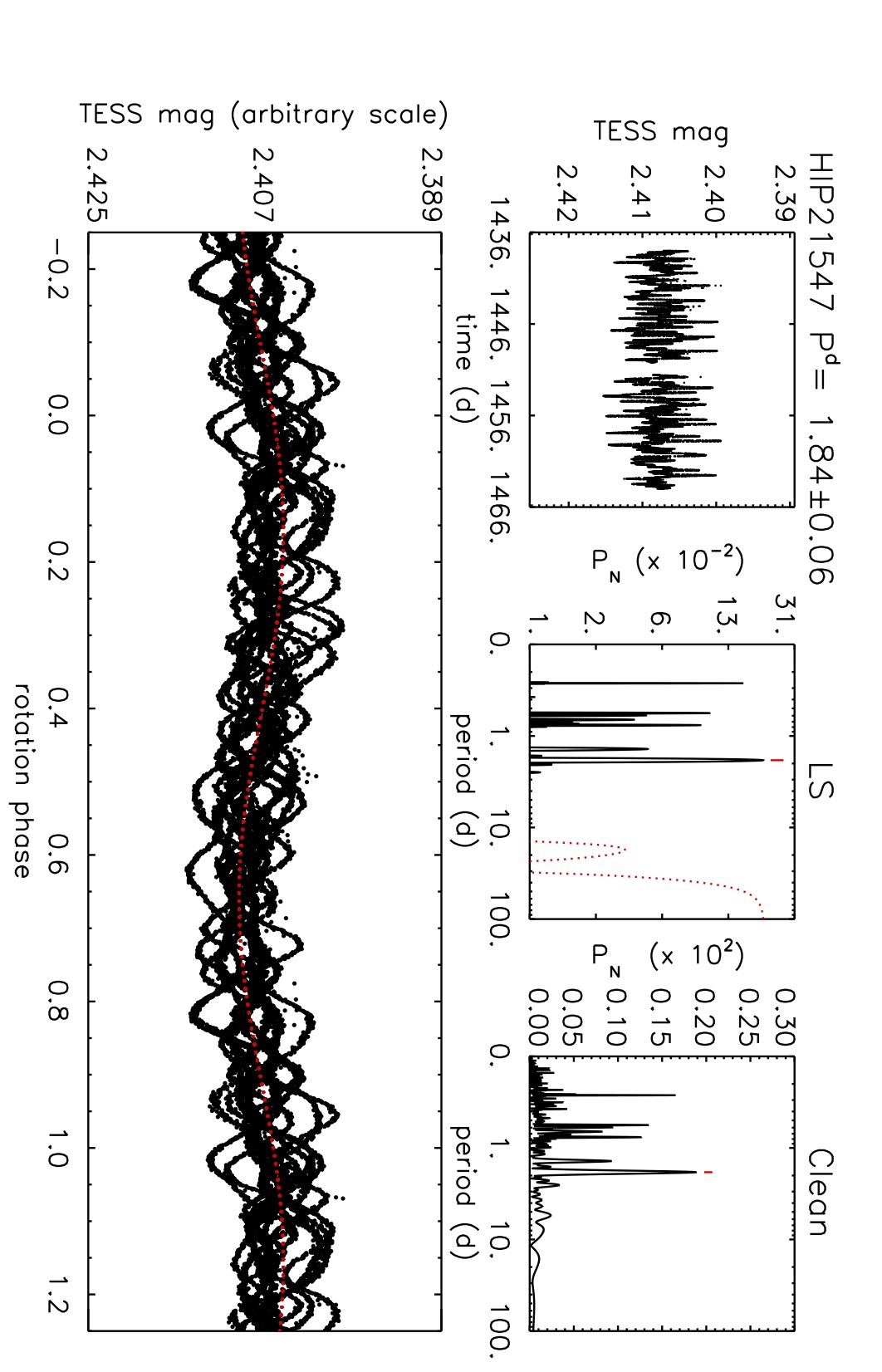}
    \caption{Photometric time sequence and periodogram for 51 Eri (HIP21547)}
    \label{fig:HIP21547}
\end{figure}

\begin{figure}[htbp]
    \centering
    \includegraphics[width=5.7cm,angle=90]{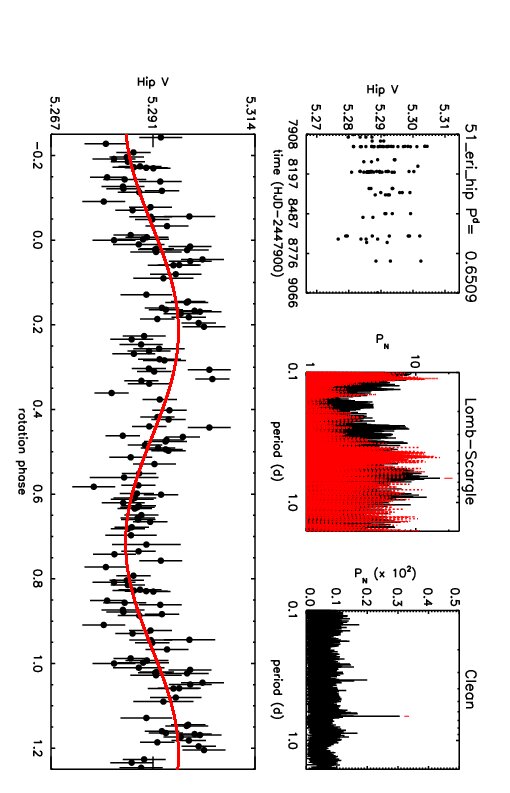}
    \caption{Photometric time sequence and periodogram for 51 Eri (HIP21547) (from Hipparcos)}
    \label{fig:51_Eri_HIP}
\end{figure}

\begin{figure}[htbp]
    \centering
    \includegraphics[width=5.7cm,angle=90]{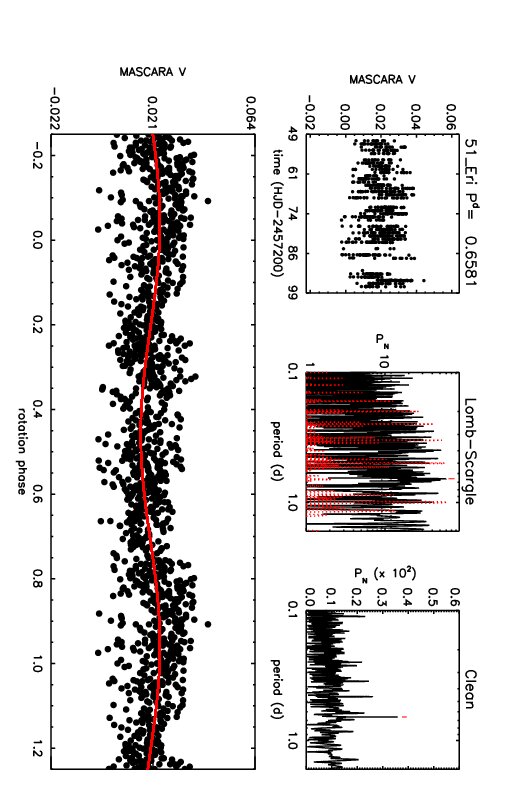}
    \caption{Photometric time sequence and periodogram for 51 Eri (HIP 21547) (from the Mascara survey)}
    \label{fig:51_Eri_mascara}
\end{figure}

\item {\bf HIP 22226 = HD 30447}
Star with debris disk spatially resolved by \citet{soummer2014}. The star has a close pair of faint comoving companions with very similar astrometric parameters at 622$^{\prime\prime}$ = 50100 au projected separation (Gaia DR2 4881308710762664576 and Gaia DR2 4881308710764495744, unique entry in 2MASS, 2MASS J04463413-2627559, $\Delta G$=0.15 mag). The very wide separation is larger than the typical limit for binaries. These objects were flagged as members of Columba in \citet{gagne2018}. The masses of the two components are close to 0.2~M$_\odot$. We measured for the first time the rotation period from the TESS photometric time series (Fig.\,\ref{fig:HIP22226}).

\begin{figure}[htbp]
    \centering
    \includegraphics[width=5.7cm,angle=90]{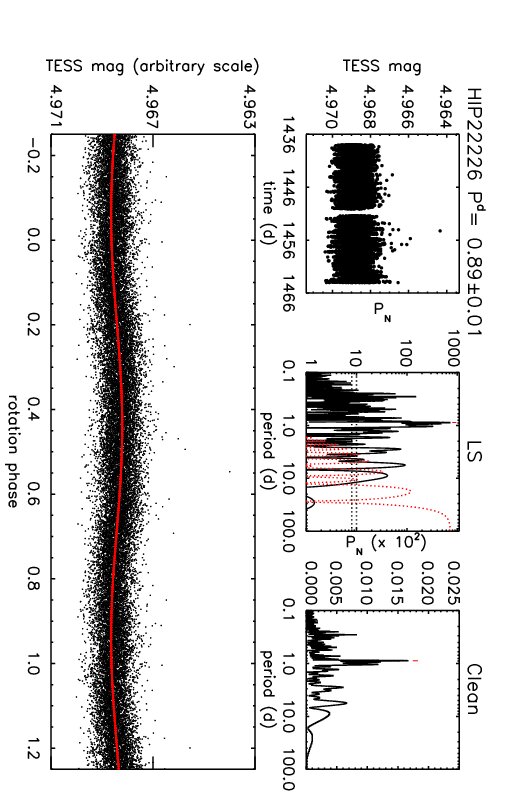}
    \caption{Photometric time sequence and periodogram for HIP 22226}
    \label{fig:HIP22226}
\end{figure}

\item {\bf HIP 22295}
The  photometric  rotation period first measured by \citet{Kiraga12} is confirmed by our analysis of the TESS data
(Fig.\,\ref{fig:HIP22295}). The TESS data show a rapidly evolving double-dip light curve. 

\begin{figure}[htbp]
    \centering
    \includegraphics[width=5.7cm,angle=90]{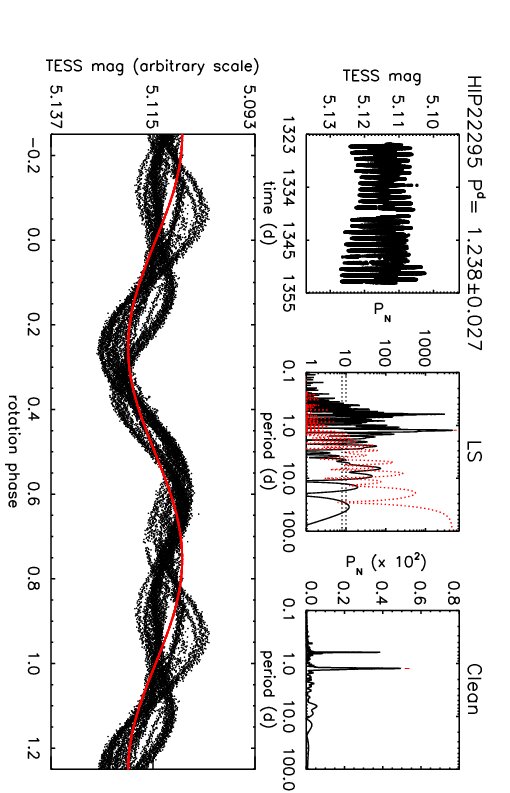}
    \caption{Photometric time sequence and periodogram for HIP 22295}
    \label{fig:HIP22295}
\end{figure}

\item {\bf TYC 5899-0026-1}
We measured for the first time the rotation period from the TESS photometric time series
(Fig.\,\ref{fig:TYC589900261}). The TESS data reveal TYC\,5899-0026-1 to be an M3 star with very intense flare activity. 

\begin{figure}[htbp]
    \centering
    \includegraphics[width=5.7cm,angle=90]{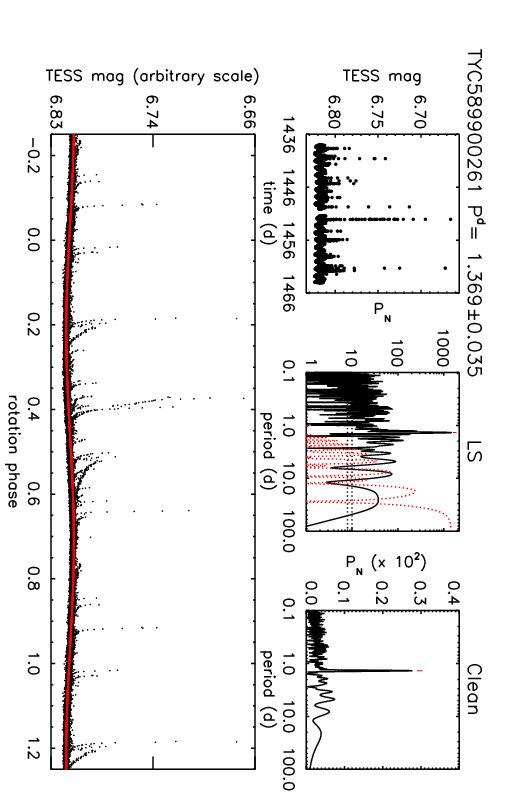}
    \caption{Photometric time sequence and periodogram for TYC 5899-0026-1}
    \label{fig:TYC589900261}
\end{figure}

\item {\bf HIP 23200}
The  photometric  rotation period first measured by \citet{Messina10} is confirmed by our analysis of the TESS data
(Fig.\,\ref{fig:HIP23200}). The TESS data reveal more flare events superimposed on a very stable light curve.

\begin{figure}[htbp]
    \centering
    \includegraphics[width=5.7cm,angle=90]{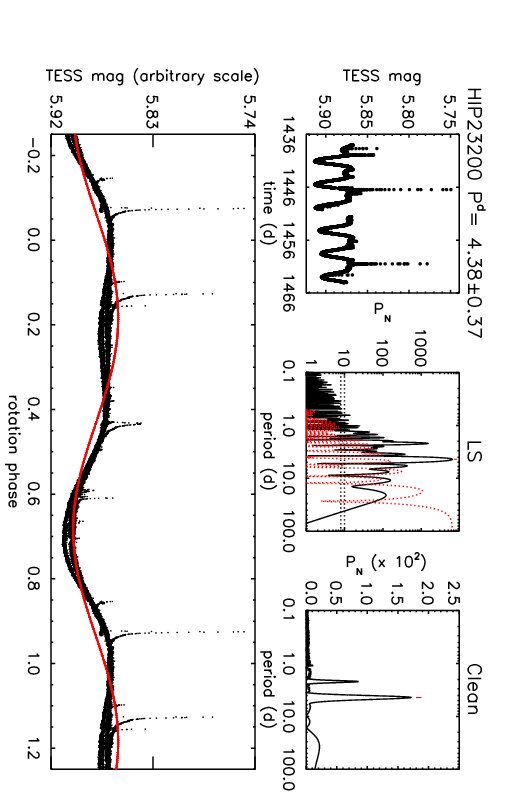}
    \caption{Photometric time sequence and periodogram for HIP 23200}
    \label{fig:HIP23200}
\end{figure}

\item {\bf HIP 23309}
The  photometric  rotation period first measured by \citet{Messina10} is confirmed by our analysis of the TESS data.
(Fig.\,\ref{fig:HIP23309}). The TESS data reveal more flare events. 

\begin{figure}[htbp]
    \centering
    \includegraphics[width=5.7cm,angle=90]{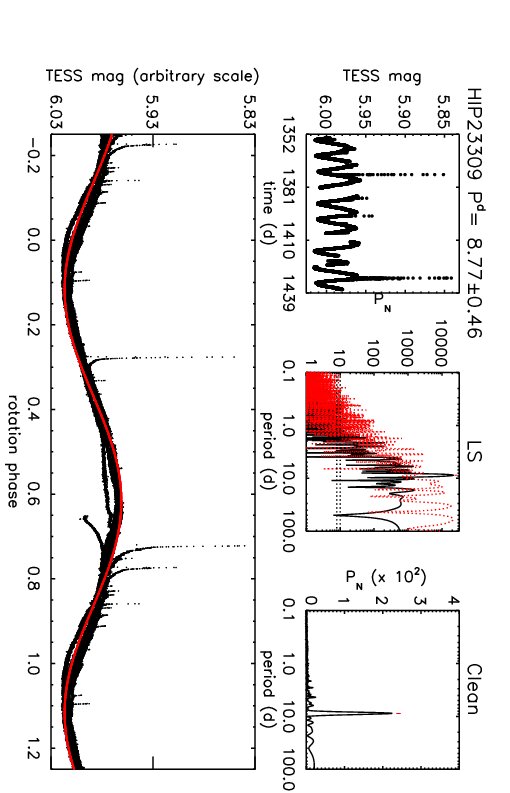}
    \caption{Photometric time sequence and periodogram for HIP 23309}
    \label{fig:HIP23309}
\end{figure}

\item {\bf HIP 24947}
We measured for the first time the rotation period from the TESS photometric time series.
(Fig.\,\ref{fig:HIP24947}).

\begin{figure}[htbp]
    \centering
    \includegraphics[width=5.7cm,angle=90]{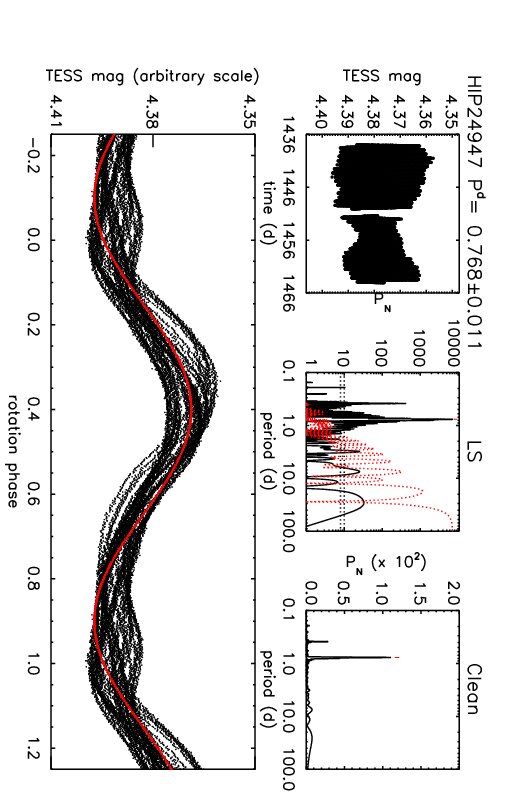}
    \caption{Photometric time sequence and periodogram for HIP 24947}
    \label{fig:HIP24947}
\end{figure}

\item {\bf HIP 25283}
The  photometric  rotation period first measured by \citet{Messina10} is confirmed by our analysis of the TESS data. 
(Fig.\,\ref{fig:HIP25283}).

\begin{figure}[htbp]
    \centering
    \includegraphics[width=5.7cm,angle=90]{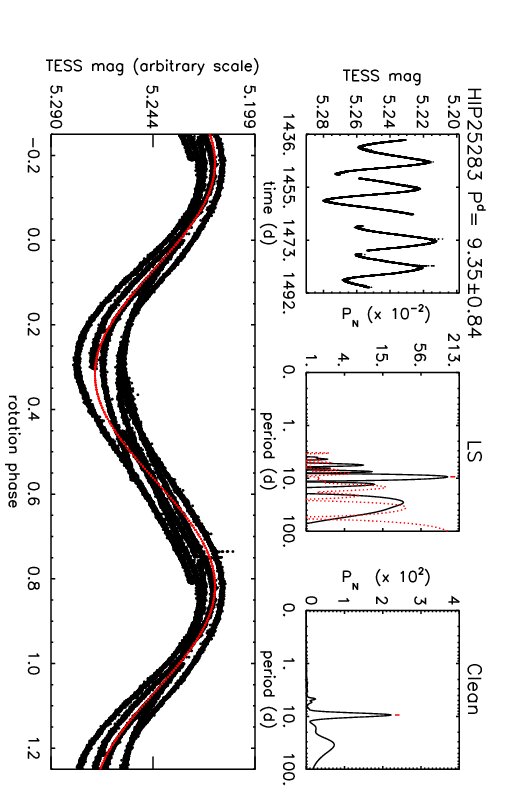}
    \caption{Photometric time sequence and periodogram for HIP 25283}
    \label{fig:HIP25283}
\end{figure}

\item {\bf HIP 25544 = HD 36435} 
Age indicators (Li, rotation period, R$_X$, R$^{'}_{HK}$) converge on an age similar to or possibly slightly older than the Hyades. We adopted 700$\pm$150~Myr.  The  photometric  rotation period first measured by \citet{Koen02} is confirmed by our analysis of the TESS data (Fig.\,\ref{fig:HIP25544}).
The light curve exhibits a significant evolution from single-dip to double-dip shape.

\begin{figure}[htbp]
    \centering
    \includegraphics[width=5.7cm,angle=90]{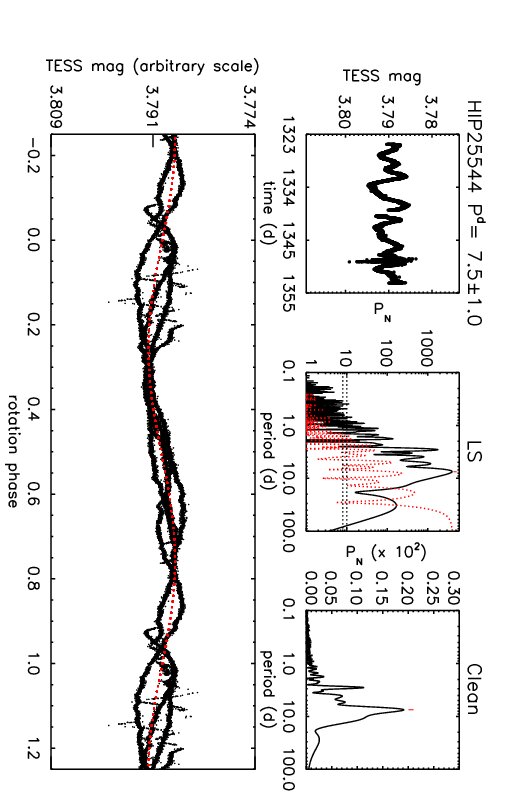}
    \caption{Photometric time sequence and periodogram for HIP 25544}
    \label{fig:HIP25544}
\end{figure}

\item {\bf $\zeta$ Lep = HIP 27288 = HD 38678}
Early-type star with spatially resolved debris disk \citep{moerchen2007}.
It was proposed as a member of Castor MG by \citet{barrado1998}. 
An age of few hundred Myr was derived from isochrone fitting by 
several authors \citep[e.g., ][]{su2006,david2015}.
However, the star was also proposed as a member of $\beta$ Pic MG by \citet{nakajima2012}.
The $\beta$ Pic membership and age was also adopted by \citet{nielsen2019}.
BANYAN $\Sigma$ returns a membership probability of 26.9\% with the
\citet{vl07} astrometric parameters (adopted because of the lower errors with respect to Gaia due to
very bright magnitude; Gaia values yield a similar value, 24.6\%). 
We then considered the $\beta$ Pic membership uncertain, and we adopted the 
age from isochrones, extending the minimum age to include the $\beta$ Pic MG age.


\item {\bf TYC 7084-0794-1 = CD-35 2722}
Star with BD companion  \citep{wahhaj2011}. The star was not moved to special targets (P0).
The  photometric  rotation period first measured by \citet{Messina10} is confirmed by our analysis of the TESS data
(Fig.\,\ref{fig:TYC708407941}). Numerous flare events are detected in the TESS time series. 

\begin{figure}[htbp]
    \centering
    \includegraphics[width=5.7cm,angle=90]{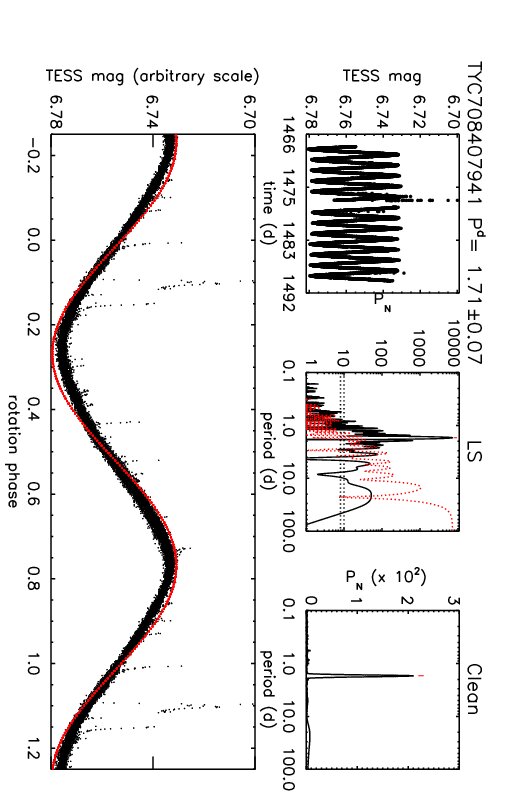}
    \caption{Photometric time sequence and periodogram for TYC 7084-0794-1}
    \label{fig:TYC708407941}
\end{figure}

\item {\bf HIP 30030 = HD 43989}  
Originally classified as a member of Tuc-Hor; the updated analysis indicates membership in the Columba association. Using TESS photometric time series, we measured a rotation period P = 1.361$\pm$0.042\,d with very high confidence (Fig.\,\ref{fig:HIP30030}), which supersedes the earlier determination of P = 1.16\,d measured by \citet{Cutispoto99}. 

\begin{figure}[htbp]
    \centering
    \includegraphics[width=5.8cm,angle=90]{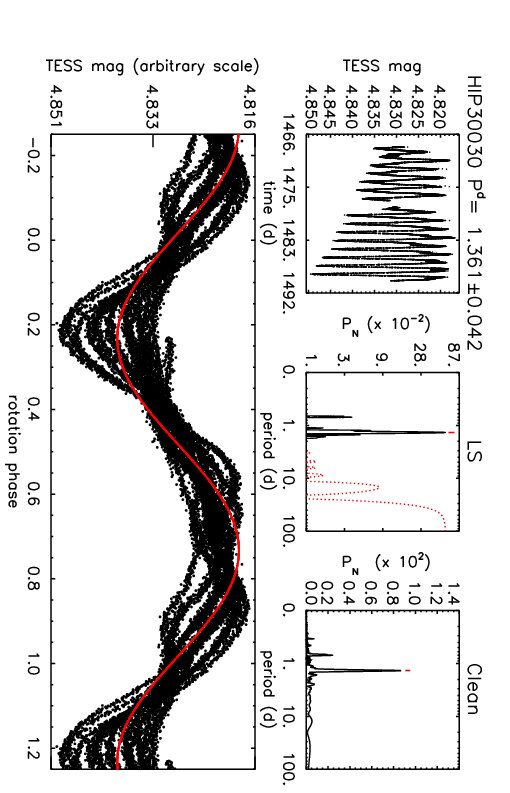}
    \caption{Photometric time sequence and periodogram for HIP 30030}
    \label{fig:HIP30030}
\end{figure}

\item {\bf HIP 30034 = HD 44627 = AB Pic}
Star with substellar companion close to the edge of the IRDIS field of view discovered by \citet{chauvin2005}.
The  photometric  rotation period first measured by \citet{Messina10} is confirmed by our analysis of the TESS data
(Fig.\,\ref{fig:HIP30034}). 

\begin{figure}[htbp]
    \centering
    \includegraphics[width=5.7cm,angle=90]{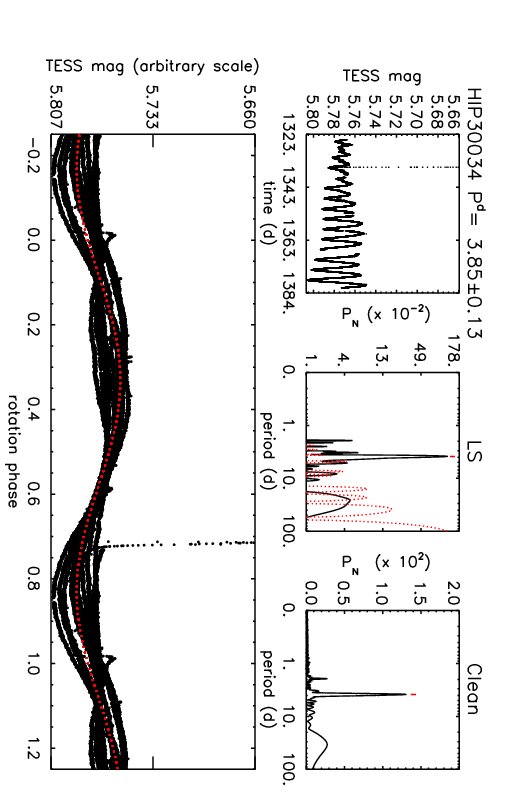}
    \caption{Photometric time sequence and periodogram for HIP 30034}
    \label{fig:HIP30034}
\end{figure}

\item {\bf HIP 30314}
We measured for the first time the rotation period from the TESS photometric time series
(Fig.\,\ref{fig:HIP30314}).

\begin{figure}[htbp]
    \centering
    \includegraphics[width=5.7cm,angle=90]{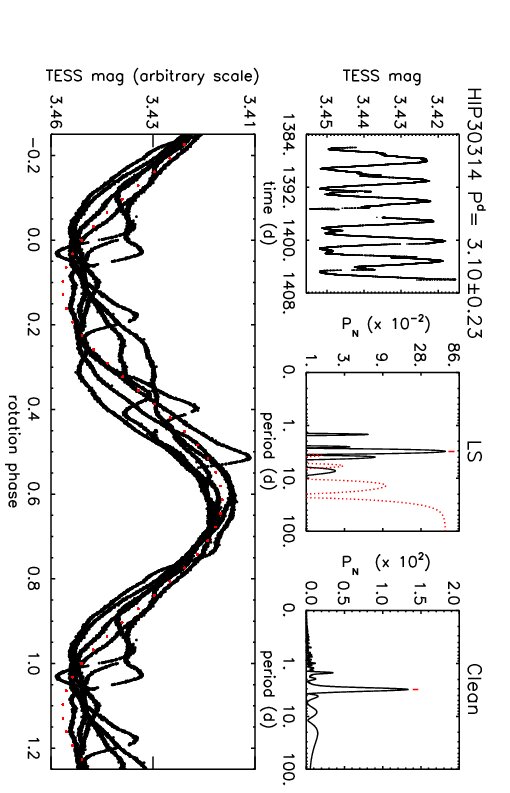}
    \caption{Photometric time sequence and periodogram for HIP 30314}
    \label{fig:HIP30314}
\end{figure}

\item {\bf GSC 8894-0426}
The  photometric  rotation period first measured by \citet{kiraga2012} is confirmed by our analysis of the TESS data
(Fig.\,\ref{fig:GSC88940426}). Numerous flare events are detected in the TESS time series. 

\begin{figure}[htbp]
    \centering
    \includegraphics[width=5.7cm,angle=90]{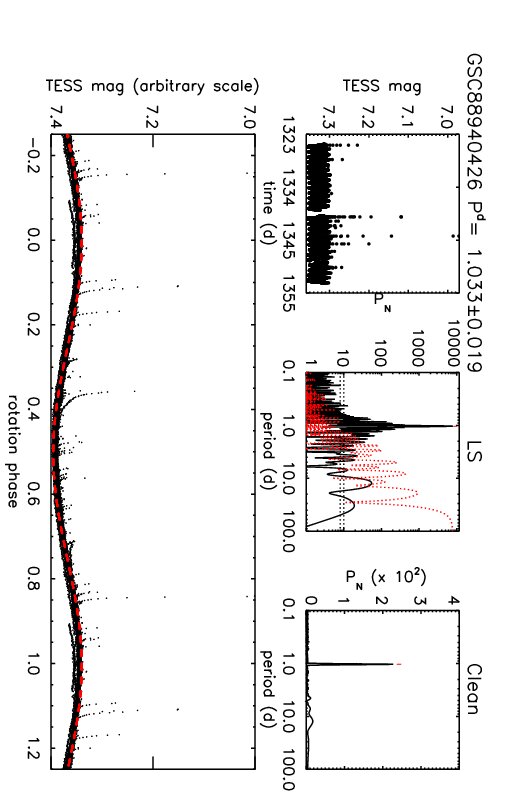}
    \caption{Photometric time sequence and periodogram for GSC 8894-0426}
    \label{fig:GSC88940426}
\end{figure}

\item {\bf TYC 7617-0549-1}
The  photometric  rotation period first measured by \citet{Messina10} is confirmed by our analysis of the TESS data
(Fig.\,\ref{fig:TYC761705491}).

\begin{figure}[htbp]
    \centering
    \includegraphics[width=5.7cm,angle=90]{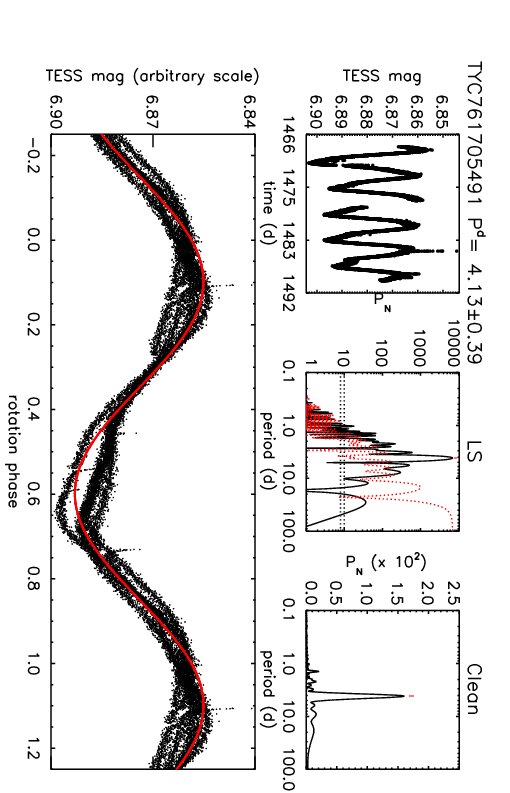}
    \caption{Photometric time sequence and periodogram for TYC 7617-0549-1}
    \label{fig:TYC761705491}
\end{figure}

\item {\bf HIP 31878}
HIP 31711 at 18000 au is a probable wide companion. The discrepancy in the astrometric parameters in Gaia DR2 is likely linked to the binarity of HIP 31711. The  photometric  rotation period first measured by \citet{Messina10} is confirmed by our analysis of the TESS data (Fig.\,\ref{fig:HIP31878}).

\begin{figure}[htbp]
    \centering
    \includegraphics[width=5.7cm,angle=90]{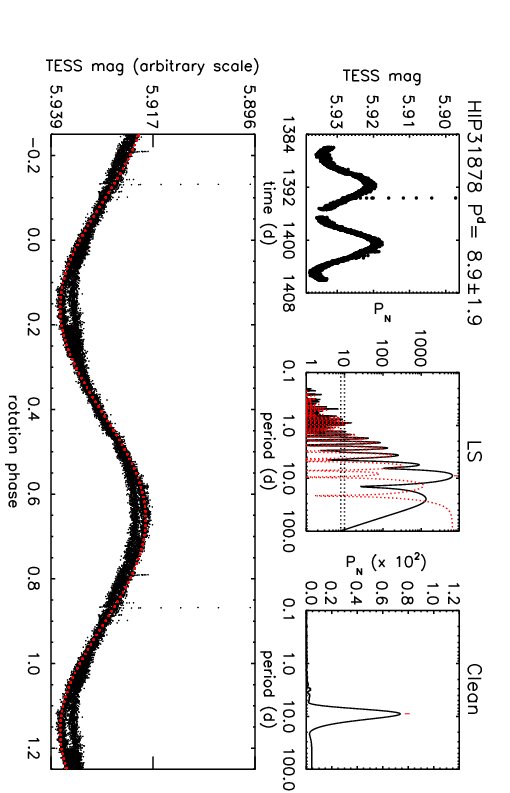}
    \caption{Photometric time sequence and periodogram for HIP 31878}
    \label{fig:HIP31878}
\end{figure}

\item {\bf HIP 32235}
The  photometric  rotation period first measured by \citet{Messina10} is confirmed by our analysis of the TESS data
(Fig.\,\ref{fig:HIP32235}). 

\begin{figure}[htbp]
    \centering
    \includegraphics[width=5.7cm,angle=90]{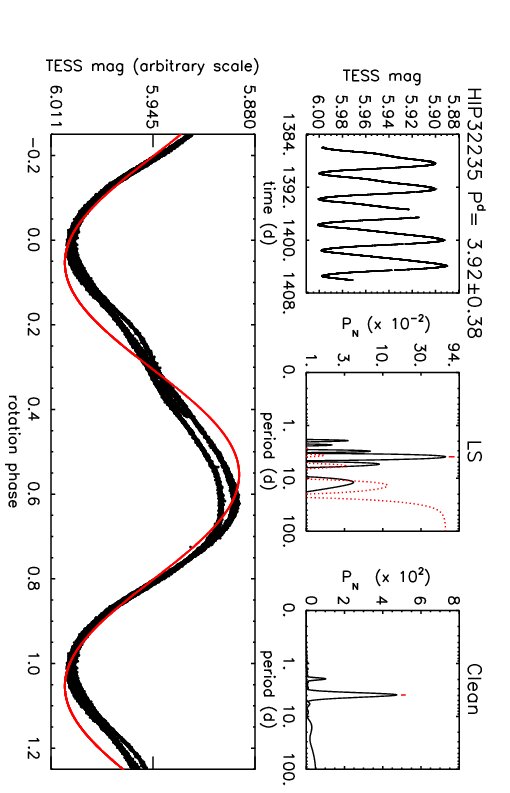}
    \caption{Photometric time sequence and periodogram for HIP 32235}
    \label{fig:HIP32235}
\end{figure}

\item {\bf HD 51797 = TYC 8118-0871-1}
The  photometric  rotation period first measured by \citet{Messina10} is confirmed by our analysis of the TESS data
(Fig.\,\ref{fig:TYC811808711}). 

\begin{figure}[htbp]
    \centering
    \includegraphics[width=5.7cm,angle=90]{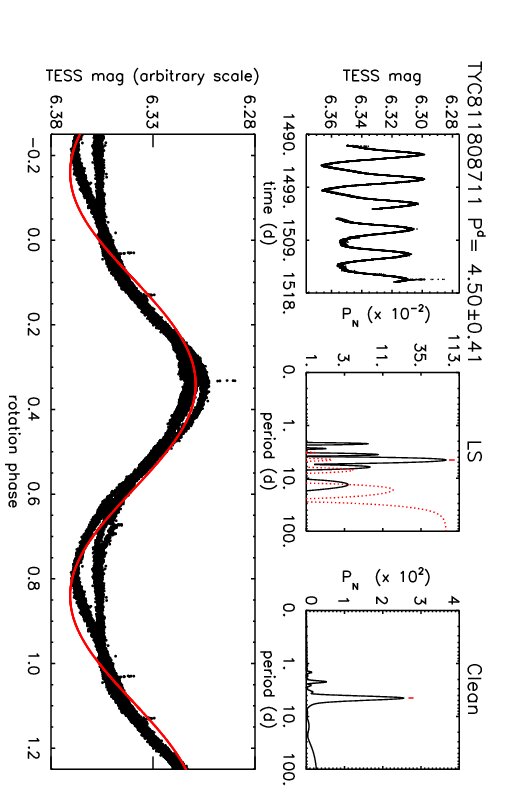}
    \caption{Photometric time sequence and periodogram for TYC 8118-0871-1}
    \label{fig:TYC811808711}
\end{figure}

\item {\bf HIP 33737 = HD 55279} 
Originally classified as a member of Tuc-Hor, the updated analysis indicates membership in the Carina association. The  photometric  rotation period first measured by \citet{Messina10} is confirmed by our analysis of the TESS data
(Fig.\,\ref{fig:HIP33737}). 

\begin{figure}[htbp]
    \centering
    \includegraphics[width=5.7cm,angle=90]{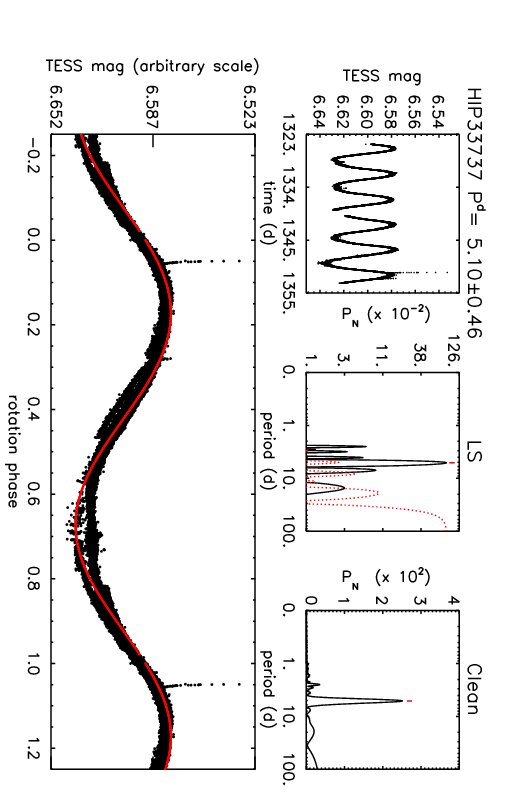}
    \caption{Photometric time sequence and periodogram for HIP 33737}
    \label{fig:HIP33737}
\end{figure}

\item {\bf 2MASS 07065772-5353463}
We measured for the first time the rotation period from the TESS photometric time series
(Fig.\,\ref{fig:2MASSJ0706-5353}). Numerous flare events are detected in the TESS time series. 

\begin{figure}[htbp]
    \centering
    \includegraphics[width=5.7cm,angle=90]{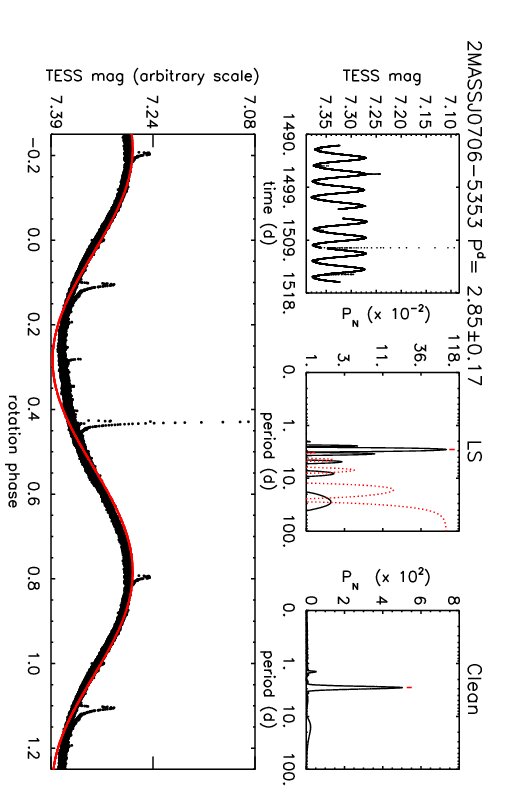}
    \caption{Photometric time sequence and periodogram for 2MASSJ0706-5353}
    \label{fig:2MASSJ0706-5353}
\end{figure}


\item {\bf BD+20 1790 = TYC 1355-214-1 = V429 Gem}
Member of AB Dor MG, with a very high activity level. The presence of a previously claimed hot Jupiter has been refuted by \citet{Carleo18}.

\item {\bf TYC 8128-1946-1 = CD-48 2972} 
Star with a wide companion HIP 36312 = HD 59659 (which is also in the SHINE sample but not observed within the date defining the targets of the present paper). It was originally identified as an Argus member by \cite{torres2008}, which has been confirmed in several works. Independently of the controversy over the existence of the Argus association, the very strong Li line confirms a young age. It is flagged as a possible SB in \cite{desidera2015} based on the marginal RV difference between the two measurements. The recent RV determination by \citet{zuckerman2018}, intermediate between the two previous measurements, does not support the presence of large RV variability. There is a marginal  proper motion difference ($\sim$ 2 $\sigma$) between Gaia DR2 and Tycho2, while the Gaia DR1 and DR2 proper motions do not differ significantly. We kept the star in the sample, and we adopted the Argus membership and age.

\item {\bf HIP 36948 = HD 61005}
Star with spatially resolved debris disk \citep[e.g.,][]{hines2007,olofsson2016}. 
The  photometric  rotation period first measured by \citet{Desidera11} is confirmed by our analysis of the TESS data
(Fig.\,\ref{fig:HD61005}).

\begin{figure}[htbp]
    \centering
    \includegraphics[width=5.7cm,angle=90]{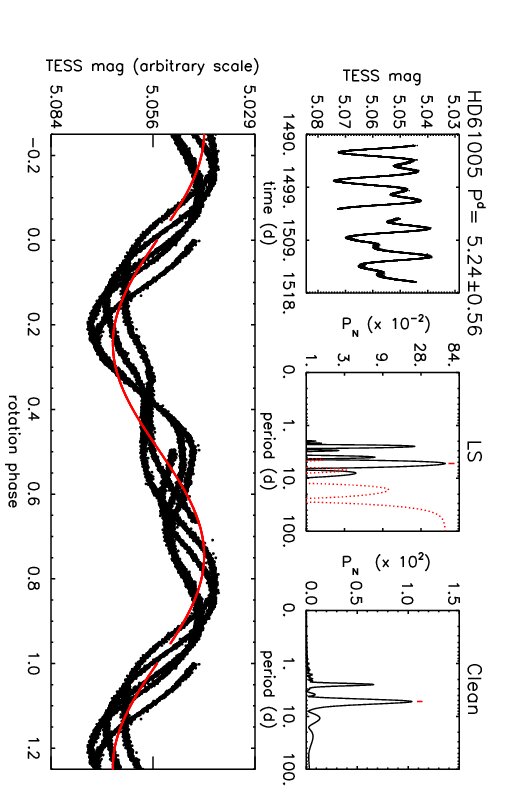}
    \caption{Photometric time sequence and periodogram for HIP36948 (HD 61005)}
    \label{fig:HD61005}
\end{figure}


\item {\bf YZ CMi = HIP 37766 = GJ 285}
The star is identified as a possible $\beta$ Pic MG member in \citet{montes2001}, although it is not considered for membership
in most of the studies on the group.
The updated kinematic analysis using BANYAN $\Sigma$ yields 0\% membership probability.
Considering it as a field object and taking the available results of age indicators into account, we
adopted an age of 100 Myr with lower limit at 20 Myr  and upper limit at 200 Myr.
The  photometric  rotation period first measured by \citet{Chugainov74} is confirmed by our analysis of the TESS data
(Fig.\,\ref{fig:HIP37766}). The TESS time series shows an uninterrupted flare activity. 

\begin{figure}[htbp]
    \centering
    \includegraphics[width=5.7cm,angle=90]{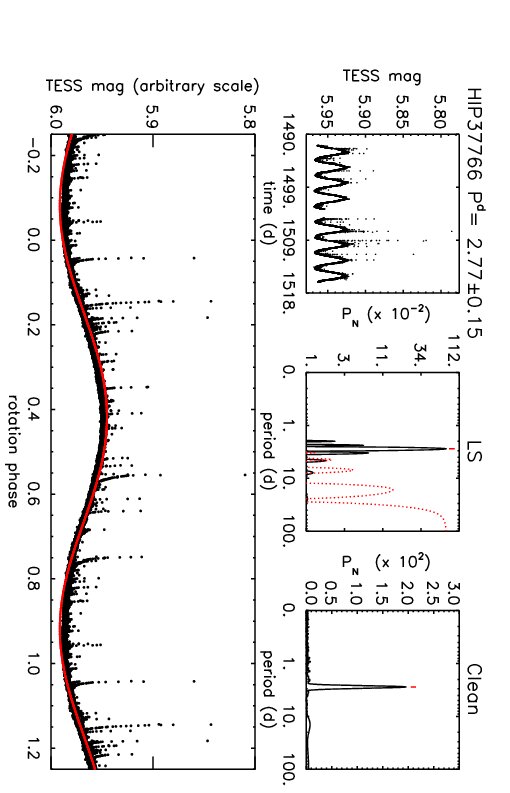}
    \caption{Photometric time sequence and periodogram for YZ CMi = HIP 37766}
    \label{fig:HIP37766}
\end{figure}


\item {\bf HIP 42808}
We measured for the first time the rotation period from the TESS photometric time series
(Fig.\,\ref{fig:HIP42808}). We note some residual instrumental effects around rotation phases $\phi$ = 0.3--0.4. 

\begin{figure}[htbp]
    \centering
    \includegraphics[width=5.7cm,angle=90]{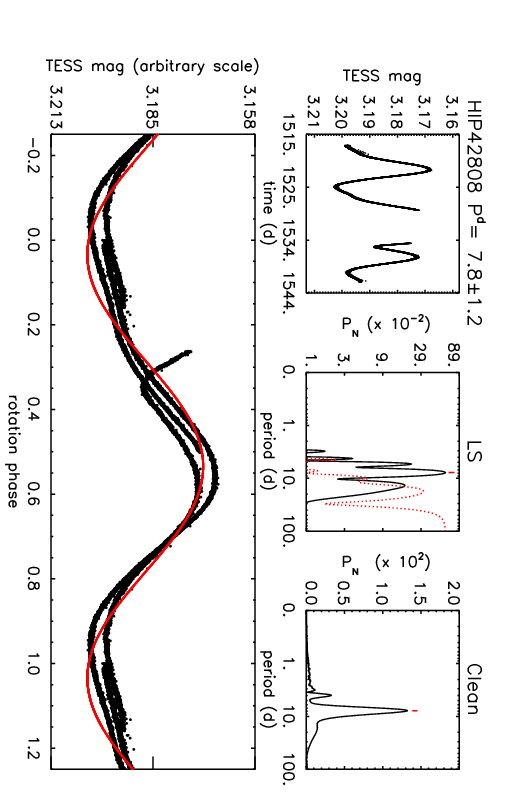}
    \caption{Photometric time sequence and periodogram for HIP 42808}
    \label{fig:HIP42808}
\end{figure}

\item {\bf $\eta$ Cha}
Member of the $\eta$ Cha open cluster \citep{mamajek1999}. Nine comoving objects are found within 10 arcmin in Gaia DR2. We listed in Table \ref{t:wide} the closest one, EK Cha, at a projected separation of 19300 au. 

\item {\bf HD 75505 = RECX13}  
Classified as a probable member of the $\eta$ Cha open cluster \citep{mamajek1999}. This is confirmed by kinematic analysis based on Gaia DR2 \citep{cantat-gaudin2020}. It is also listed among the bona fide members by \citep{gagne2018}. On the other hand, the BANYAN $\Sigma$ online tool yields a low membership probability (9\%). Specific signatures of youth are elusive considering the spectral type of the star (A1V). We consider the star to be a member on the basis of the sky position, parallax, and proper motion, which are very similar to those of the other cluster members.
Nine comoving objects are found within 10 arcmin in Gaia DR2.
We listed in Table \ref{t:wide} the two with a projected separation of less than 20000 au, EH Cha and EI Cha, which are both confirmed members of the cluster.


\item {\bf  V405 Hya = HIP 44526 =   HD 77825}
All age indicators are compatible with an age intermediate between Hyades and Pleiades. We adopted 300 Myr. The star was considered a member of Castor MG, whose existence is uncertain. Our adopted age is in any case close to the typical  value for proposed Castor members. The M2.5 star UCAC4 371-053521, clearly comoving at a projected separation of 6000 au from Gaia DR2, was not previously recognized as a wide companion to V405 Hya. It was known as an active and X-ray emitter source. 
The  photometric  rotation period first measured by \citet{Kiraga12} is confirmed by our analysis of the TESS data
(Fig.\,\ref{fig:HIP44526}). The light curve is affected by residual instrumental effects.

\begin{figure}[htbp]
    \centering
    \includegraphics[width=5.7cm,angle=90]{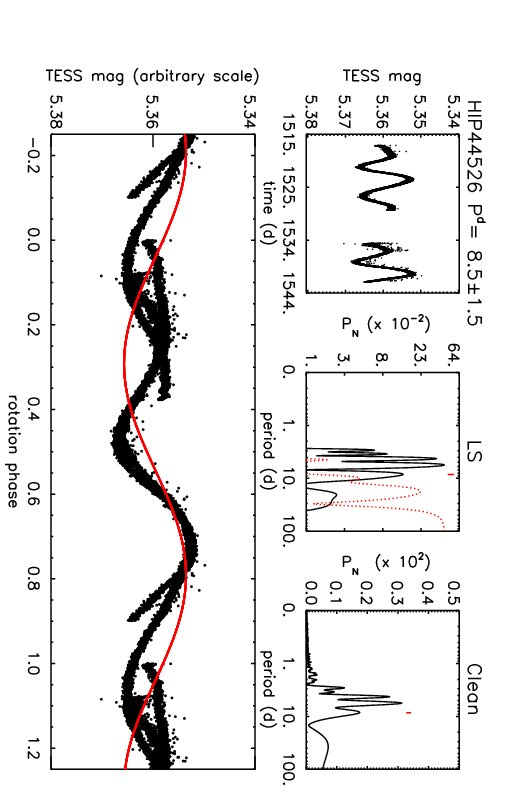}
    \caption{Photometric time sequence and periodogram for HIP 44526}
    \label{fig:HIP44526}
\end{figure}

\item {\bf HIP 47135 = HD 84075}  
G2 star classified as an Argus member in \cite{torres2008}, \cite{zuckerman2011}, \cite{malo2013}, \cite{bell2015}, and \cite{zuckerman2018}. The age indicators are fully compatible with the proposed age for Argus. 
The star has an IR excess with two components \citep{zuckerman2011}. 
We measured the photometric rotation period for the first time using photometric time series collected at the ROAD observatory. Our measurement was subsequently confirmed by our analysis of the TESS data (Fig.\,\ref{fig:HIP47135}-\ref{fig:HIP47135road}).

\begin{figure}[htbp]
    \centering
    \includegraphics[width=5.7cm,angle=90]{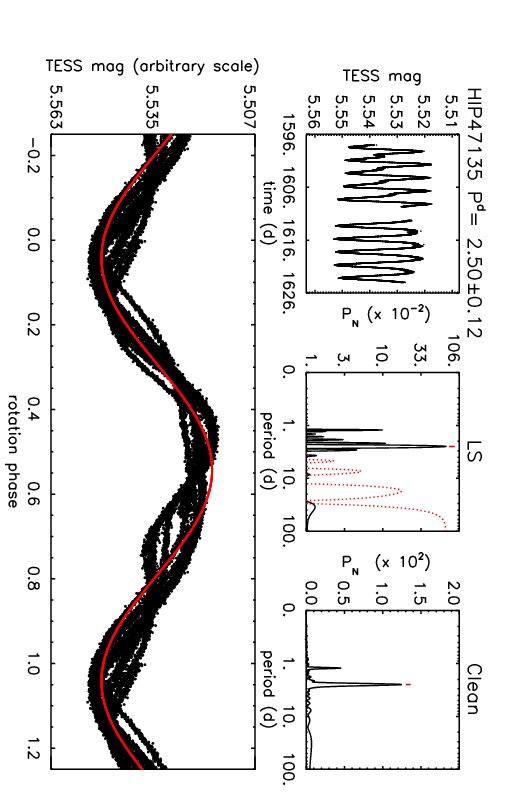}
    \caption{Photometric time sequence and periodogram for HIP 47135}
    \label{fig:HIP47135}
\end{figure}

\begin{figure}[htbp]
    \centering
    \includegraphics[width=5.7cm,angle=90]{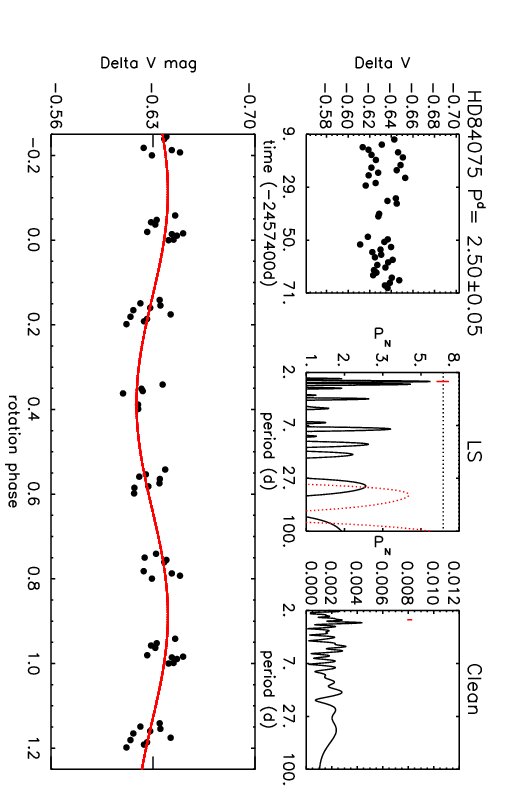}
    \caption{Photometric time sequence and periodogram for HIP 47135 (ROAD data)}
    \label{fig:HIP47135road}
\end{figure}

\item {\bf HIP 50191 = HD 88955 = q Vel} 
A2V star classified as an Argus member in \cite{zuckerman2011}, \cite{bell2015}, and \cite{zuckerman2018}, and with Argus membership supported by BANYAN (probability 98\%). Argus membership is also adopted in \cite{nielsen2019}. Actually, \cite{zuckerman2018} noted the slightly off-sequence position in B$-$V versus $M_{V}$ CMD; instead the star is in a position similar to that of other Argus A-type stars in Gaia CMD. It should be noted that the Hipparcos parallaxes and proper motion have smaller errors than the Gaia values for such a bright star (V=3.85\,mag). Furthermore, the star is classified as a primary standard for A2V spectral type by \cite{pecaut2013} \footnote{\url{http://www.pas.rochester.edu/~emamajek/spt/A2V.txt}}. Adopting their $T_{\rm eff}$ for this spectral type, (8840 K), the isochrone age results 458$\pm$182 Myr, similar to the age reported in the literature using this technique \citep{vican2012,david2015}. The pre-MS age (6$\pm$1 Myr) is instead too young for Argus membership. We consider the post-ZAMS isochrone age reliable as the adopted data appears of high quality and there is no indication of binarity of the object, both at short separation from RV monitoring \citep{lagrange2009} and at larger separation from imaging. We adopted the isochronal age with an error bar extending down to 50 Myr to include the possibility of Argus membership. The star has a significant IR excess  \citep{zuckerman2011}.

\item {\bf TWA 6 = GSC7183-1477 = BX Ant} 
Not member of TWA when using BANYAN Gaia DR2 paramaters. On the other hand, the TWA membership is supported by \cite{lee2019}. In any case, independently of any kinematic evaluation, the very strong lithium unambiguously shows the very young nature of the star. Isochrone fitting yields an age of 10$\pm$3\,Myr, the same as the TWA group.  The  photometric  rotation period first measured by \citet{Lawson05} is confirmed by our analysis of the TESS data
(Fig.\,\ref{fig:TWA6}).

\begin{figure}[htbp]
    \centering
    \includegraphics[width=5.7cm,angle=90]{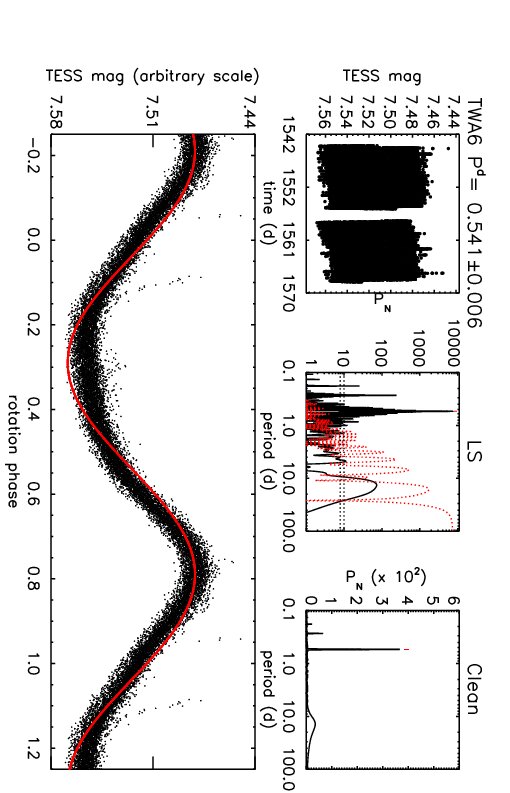}
    \caption{Photometric time sequence and periodogram for TWA6}
    \label{fig:TWA6}
\end{figure}

\item {\bf HIP 51228}
We measured for the first time the rotation period from the TESS photometric time series
(Fig.\,\ref{fig:HIP51228}). 

\begin{figure}[htbp]
    \centering
    \includegraphics[width=5.7cm,angle=90]{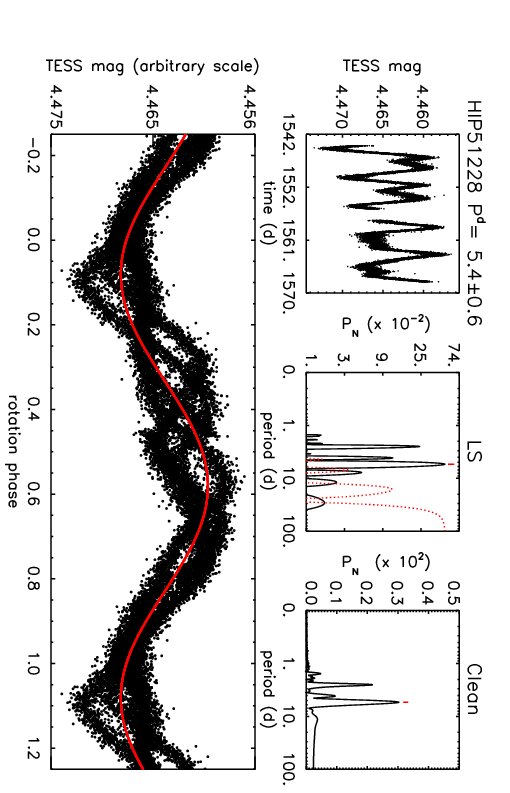}
    \caption{Photometric time sequence and periodogram for HIP 51228}
    \label{fig:HIP51228}
\end{figure}

\item {\bf HIP 51317}
We measured for the first time the rotation period from the K2 Kepler photometric time series.
(Fig.\,\ref{fig:HIP51317})

\begin{figure}[htbp]
    \centering
    \includegraphics[width=5.7cm,angle=90]{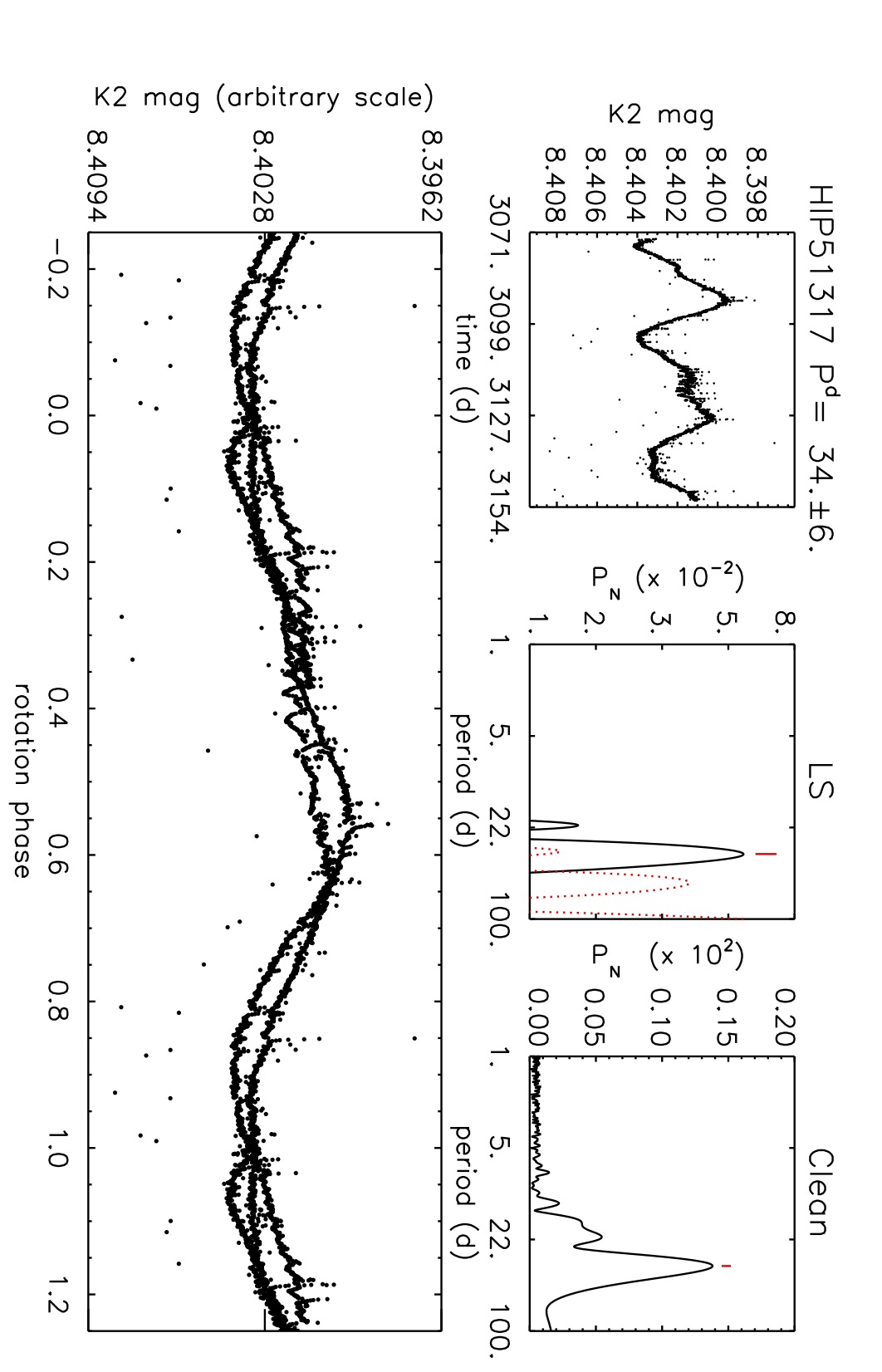}
    \caption{Photometric time sequence and periodogram for HIP 51317}
    \label{fig:HIP51317}
\end{figure}


\item {\bf HD 95086}
Star with planetary companion first discovered by \cite{rameau2013}.
It is a Sco-Cen member according to \cite{dezeeuw1999}. BANYAN returns a 48.5\% probablity of membership in the Carina MG and 33.8\% for LCC \citep[without RV, the value listed in SIMBAD and several catalogs is the astrometric one by][]{madsen2002}. The spatial position is in the outskirts of the Sco-Cen group. The age map by \cite{pecaut2016} yields an age of 26 Myr at the location of HD 95086, clearly older than the bulk of LCC. On the other hand, \citet{schneider2019} recently proposed an age of 22 Myr for the Carina association. The isochrone age gives a lower limit of about 20 Myr for HD95086. Looking in Gaia DR2 for stars with similar position and kinematic parameters, we noticed the F2 star HIP 55334 (HD 98660) with a lower age limit of about 19 Myr (our analysis and \citet{pecaut2012}), consistent with the \cite{pecaut2016} age map. Our tentative conclusion is that HD 95086 is part of a young population that is slightly older than the bulk of LCC and possibly connected to the Carina association, or part of it. We adopted the age from the \cite{pecaut2016} map, with lower and upper limits corresponding to the LCC and Carina.

\item {\bf HIP 54155}
The wide companion HD 96064B has no astrometric solution in Gaia likely due to its close binarity (P=23 yr). The physical association of this triple system is nevertheless confirmed. The  photometric  rotation period first measured by \citet{Cutispoto99} is confirmed by our analysis of the TESS data (Fig.\,\ref{fig:HIP54155}).

\begin{figure}[htbp]
    \centering
    \includegraphics[width=5.7cm,angle=90]{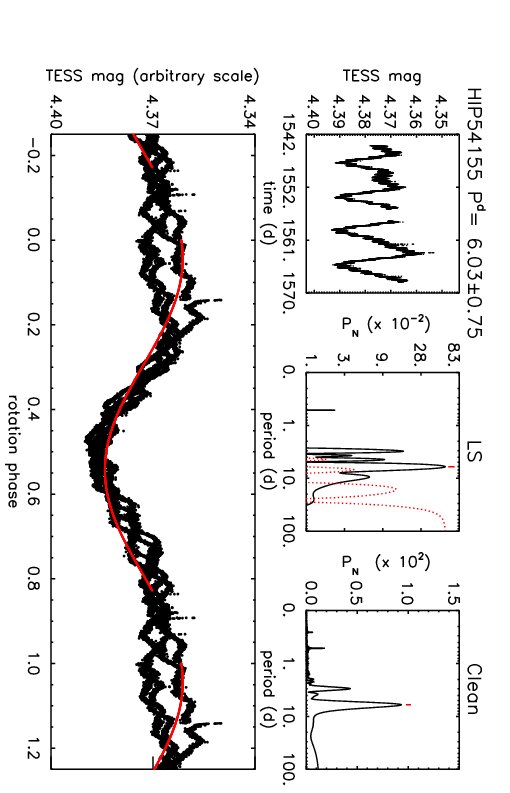}
    \caption{Photometric time sequence and periodogram for HIP 54155}
    \label{fig:HIP54155}
\end{figure}

\item {\bf HIP 54231}
Sco-Cen member according to \cite{dezeeuw1999} and \cite{rizzuto2011}, but with low membership probability from BANYAN. We adopted the LCC age but with the upper limit derived by isochrone fitting (380 Myr). The literature spectral type is A0V, but the colors are more compatible with A1. 
The RV in SIMBAD is the expected value for membership \citep{madsen2002}, not an observational measurement.

\item {\bf HIP 57632 = HD 102647 = $\beta$ Leo}
BANYAN returns an 87.1\% membership probability to Argus, 4.6\% to Carina-Near, and 8.3\% for field. Argus membership is also supported by \citet{zuckerman2011}. We adopted Argus membership. There are no Gaia astrometric data due to the very bright magnitude. The star has a two-belt debris disk.

\item {\bf HIP 58167}
F-type star in Sco Cen. The star has a comoving object (2MASS J11551267-5406215) with very similar astrometric parameters (proper motion difference of 1.6 and 1.2 mas\,yr$^{-1}$, parallax difference of 0.33 mas) in spite of the large projected separation (382$^{\prime\prime}$ corresponding to 41250 au). From the 2MASS magnitudes, a mass as low as 20 Mjup is derived for this object. In Gaia there is another possible comoving object, Gaia DR2 5344340167066548608 at 355" (with a different position angle with respect to 2MASS J11551267-5406215). Its magnitude is extremely faint (G=20.99) and it is not detected in 2MASS. The astrometric parameters are characterized by large errors (3.6 mas on parallax and more than 4 mas yr$^{-1}$
 on the components of proper motion). The object results comoving to HIP 58167 at about the 2 $\sigma$ level. If confirmed, the magnitude fainter than 2MASS J11551267-5406215 (by 2.5 mag) would imply an extremely low-mass object.
We measured for the first time the rotation period from the TESS photometric time series
(Fig.\,\ref{fig:HIP58167}). 

\begin{figure}[htbp]
    \centering
    \includegraphics[width=5.7cm,angle=90]{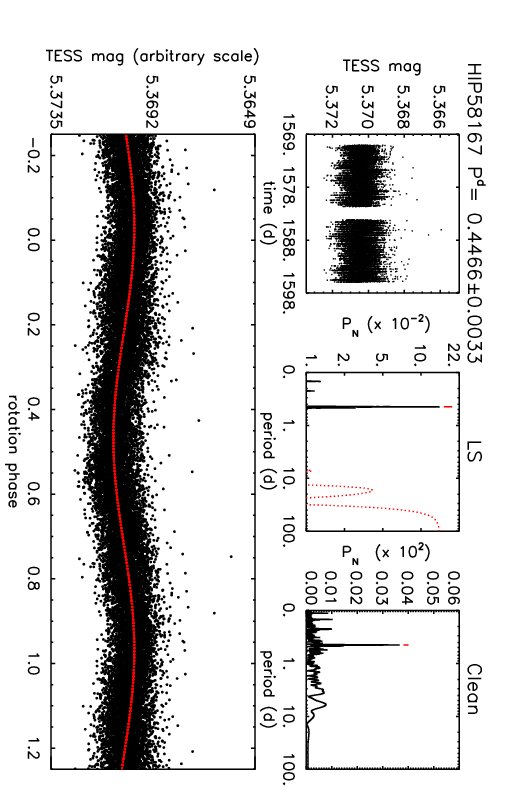}
    \caption{Photometric time sequence and periodogram for HIP 58167}
    \label{fig:HIP58167}
\end{figure}


\item {\bf HIP 60183}
The star is flagged as a member of LCC by \citet{dezeeuw1999} and as a member of one of the LCC sub-groups by \citet{goldman2018}; however it has
however a low membership probability with Banyan.
We adopted the LCC age, but with the upper limit derived from isochrone fitting.

\item {\bf HIP 60459}
We measured for the first time the rotation period from the TESS photometric time series
(Fig.\,\ref{fig:HIP60459}). 

\begin{figure}[htbp]
    \centering
    \includegraphics[width=5.7cm,angle=90]{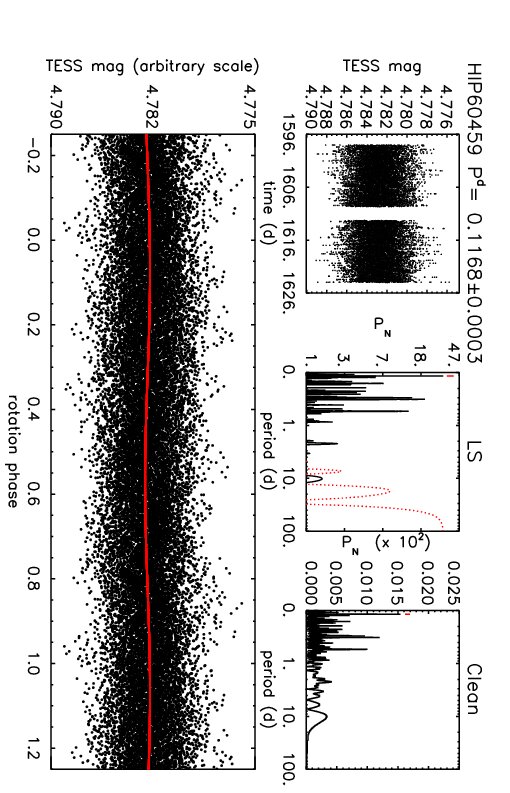}
    \caption{Photometric time sequence and periodogram for HIP 60459}
    \label{fig:HIP60459}
\end{figure}

\item {\bf HD 108767B}
Wide companion to the B-type star $\delta$ Crv = HD108767.
We measured for the first time the rotation period from the TESS photometric time series
(Fig.\,\ref{fig:HD108767B}). 

\begin{figure}[htbp]
    \centering
    \includegraphics[width=5.7cm,angle=90]{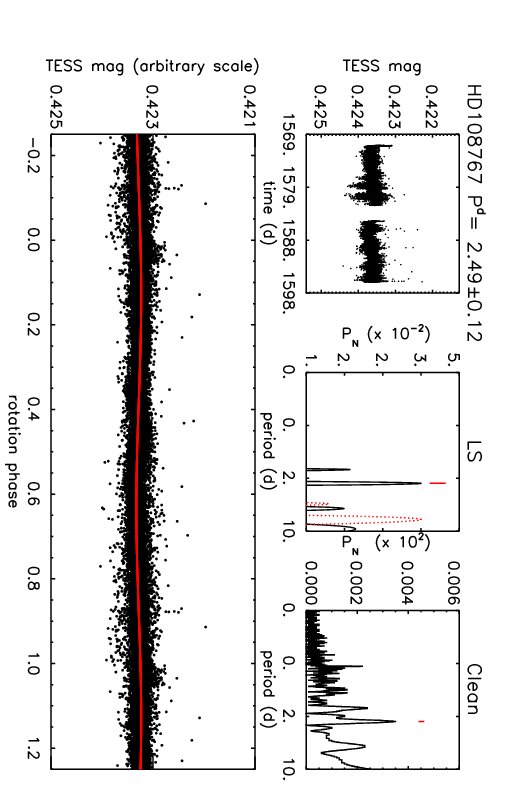}
    \caption{Photometric time sequence and periodogram for HD108767B}
    \label{fig:HD108767B}
\end{figure}

\item {\bf HIP 61468}
The star has a secondary at 15$^{\prime\prime}$, clearly comoving from the Gaia DR2 astrometry. To our knowledge, the binarity has  not been previously reported in the literature. The companion (2MASS J12354637-4101315) is expected to have a spectral type of M3.5 from photometric colors, and is a possible X-ray source from CHANDRA \citep{wang2016}.


\item {\bf HIP 66252}
We measured for the first time the rotation period from K2 photometric time series (Fig.\,\ref{fig:HIP66252}). 

\begin{figure}[htbp]
    \centering
    \includegraphics[width=5.7cm,angle=90]{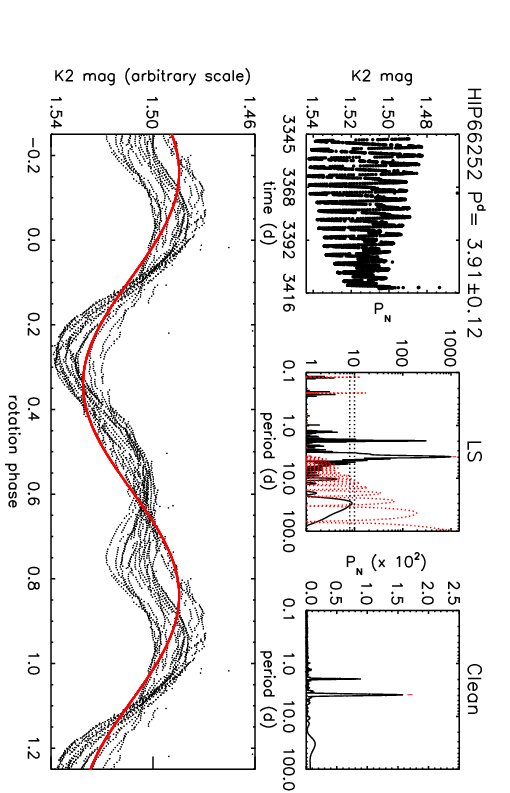}
    \caption{Photometric time sequence and periodogram for HIP 66252}
    \label{fig:HIP66252}
\end{figure}


\item {\bf  TYC 7286-0248-1 = CD-31 11053} 
K star with very strong lithium line identified by \cite{sacy}. It is proposed as a possible member of UCL by \cite{gagne2018} on the basis of Gaia DR1 data, and inclusion of Gaia DR2 makes the case stronger (98\% probability). \citet{damiani2019} also support Sco-Cen membership. The RV difference between SACY and Gaia DR2 (8.8 km\,s$^{-1}$) is not highly significant considering the large error of Gaia RV (4.1\,km\,s$^{-1}$). The discrepancy between photometric colors and spectral type \citep[K3Ve, ][]{sacy} indicates either a significant reddening (E(B$-$V) $\sim$ 0.1) or that the true spectral type is intermediate between K4 and K5. The lithium EW indicates an age at least as young as the $\beta$ Pic MG. A very young age is also supported by the position on the CMD and the isochrone age, in spite of the ambiguity between effective temperature and reddening. These results are fully consistent with the UCL membership, which we then adopted.

\item {\bf HIP 69989 = HD 125451 = 18 Boo}
Mid-F star classified as a possible UMa member by \citet{montes2001} and \cite{king2003}. Instead, BANYAN analysis rejected the membership. We then considered the membership as unconfirmed. The age indicators are  inconclusive considering the spectral type of the star. However, the chromospheric activity and X-ray emission are consistent with Hyades and UMa stars of similar temperature. The position on the CMD is close to ZAMS, with an upper limit of 1.9 Gyr. We adopted the UMa age with the upper limit from isochrone, as membership is uncertain. The star has RV monitoring by \citet{borgniet2019} with SOPHIE. 
We used these data to provide a new value of absolute RV. The star also has an IR excess indicating a debris disk.

\item {\bf HIP 71724}
A stellar companion (mass 146-217 Mjup) at 101 mas was claimed by \citet{Hinkley2015}.
It was not detected in our observations, although it is 
expected to lie beyond the coronagraphic mask.

\item {\bf HIP 71743 = HD 128987 = KU Lib}
G6V star whose age indicators (Li, Prot, X-rays, and RHK) nicely agree on an age close to that of the Hyades or slightly older. We adopted $700\pm 100$~Myr. The star is classified as an extremely wide companion (separation of 1 pc)  to the quadruple system $\alpha$ Lib \citep{caballero2010}. The membership to Castor MG is also proposed in that study.


\item {\bf HIP 73990 = HD 133803}
The two brown dwarf companions at very small separations claimed by \citet{Hinkley2015} are not confirmed by SPHERE observations. See Paper II and Cantalloube et al. (in prep.) for further details. The star has a previously unrecognized wide companion (2MASS J15071795-2929501) at 5230\, au projected separation.

\item {\bf HIP 74824 = HD 135379 = $\beta$ Cir}: 
A3V star with very wide brown dwarf companion \citep{2015MNRAS.454.4476S} and IR excess. Analysis of kinematic parameters yields an 83\% probability of being a member of $\beta$ Pic MG when adopting the Gaia DR2 parameters. However, when adopting VL07  (which has smaller errors because of the very bright magnitude of the star), the membership probability drops to 19\%. The star is slightly brighter than the ZAMS. Isochrone fitting 
yields ages of $450\pm200$~Myr  and 8$\pm$3 Myr assuming post- and pre-MS phases, respectively. The pre-MS age is not compatible with $\beta$ Pic MG. Furthermore, the BD companion $\beta$ Cir B does not show signatures of youth or low gravity \citep{2015MNRAS.454.4476S}. We then adopted the post-MS solution, yielding an age of 450 Myr.

\item {\bf HIP 76063 = HD 138204}
A7 star classified as a Sco-Cen member (UCL subgroup) in \cite{dezeeuw1999} and \citet{rizzuto2011} (probability 55\%). BANYAN $\Sigma$ yield a 26.5\% membership probability on UCL. The distance is significantly closer than the vast majority of the Sco-Cen population, as previously noted by \citet{wright2018}.  
This rules out membership in the core of the Sco-Cen association, but a link with a foreground population of young stars of similar age \citep[see, e.g. the bona-fide young star NZ Lup at 60\, pc][]{boccaletti2019} is possible.
We then adopted the isochrone age (220 Myr), extending the lower limit to encompass the UCL age. \citet{nielsen2019} adopted instead the UCL membership and age.

%
\item {\bf HIP 77457}
Member of US group according to \citet{dezeeuw1999}, while membership is rejected by \citet{pecaut2012}.
Our analysis with BANYAN also gives a low membership probability in US (11\%). We adopted the isochrone age, with
lower limit at the US age to take the possible membership into account.

\item {\bf HIP 77464 = HD 141378}
A-type star, possibly chemically peculiar, with a dual-belt debris disk. The low-mass star 2MASS J15490081-0348147 is a previously unrecognized wide companion at 4240 au.

\item {\bf TYC 7846 1538 1}
The  photometric  rotation period first measured by \citet{Marsden11} is confirmed by our analysis of the TESS data
(Fig.\,\ref{fig:TYC784615381}). 

\begin{figure}[htbp]
    \centering
    \includegraphics[width=5.7cm,angle=90]{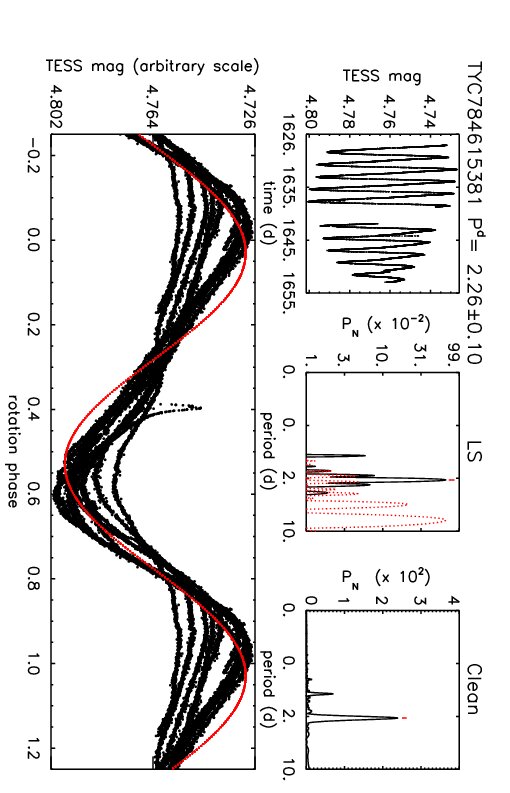}
    \caption{Photometric time sequence and periodogram for TYC 7846-1538-1}
    \label{fig:TYC784615381}
\end{figure}

\item {\bf NZ Lup = HD 141943}
See \citet{boccaletti2019}.

\item {\bf HIP 78099}
The low-mass very wide companion 2MASS J15564019-2309291 was not previously associated with the primary, but was flagged as a bona fide Sco-Cen member \citep{damiani2019} and fast-rotating object \citep{stauffer2018}.

\item {\bf HIP 78196 =  HD 142851}
A stellar companion (mass 98-152 Mjup) at 74 mas was claimed by \citet{Hinkley2015}.
It was not detected in our observations, although it is 
expected to lie beyond the coronagraphic mask.

\item{\bf HIP 78530}
Late B-type object with a brown dwarf companion at very wide separation discovered by \citet{lafreniere2011}.

\item {\bf HIP 78541 = HD 143488 }   
Originally considered to be member of UCL \citet{dezeeuw1999}, it is possibly a field object (85.6\% probability from BANYAN).
We adopted the UCL age with upper limit from our isochrone fitting.


\item {\bf HIP 80591 = HD 148055}
Star member of UCL. 2MASS J16271281-3949144 at 21.9$^{\prime\prime}$ is a low-mass (0.16 $M_{\odot}$) companion, not previously mentioned in the literature as such.

\item {\bf HIP 81084}
We measured for the first time the rotation period from photometric time series we collected at  PEST observatory
(Fig.\,\ref{fig:HIP81084_V}-\ref{fig:HIP81084_I}).

\begin{figure}[htbp]
    \centering
    \includegraphics[width=5.7cm,angle=90]{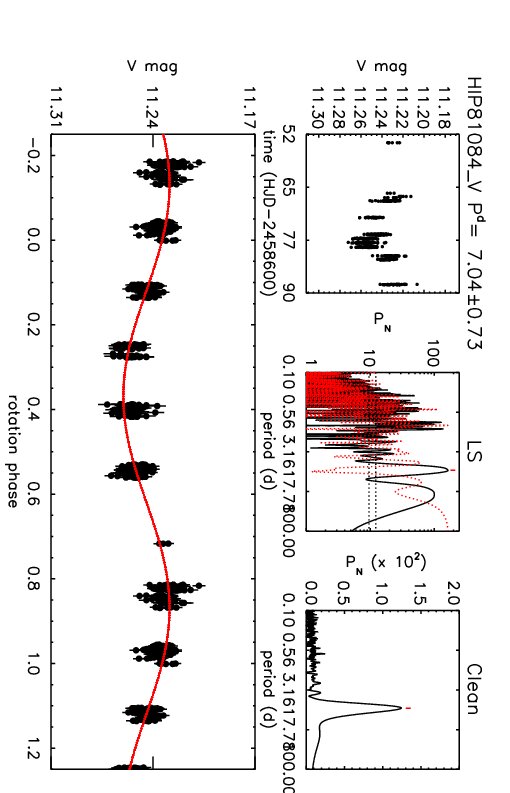}
    \caption{Photometric time sequence and periodogram for HIP81084 (V band)}
    \label{fig:HIP81084_V}
\end{figure}

\begin{figure}[htbp]
    \centering
    \includegraphics[width=5.7cm,angle=90]{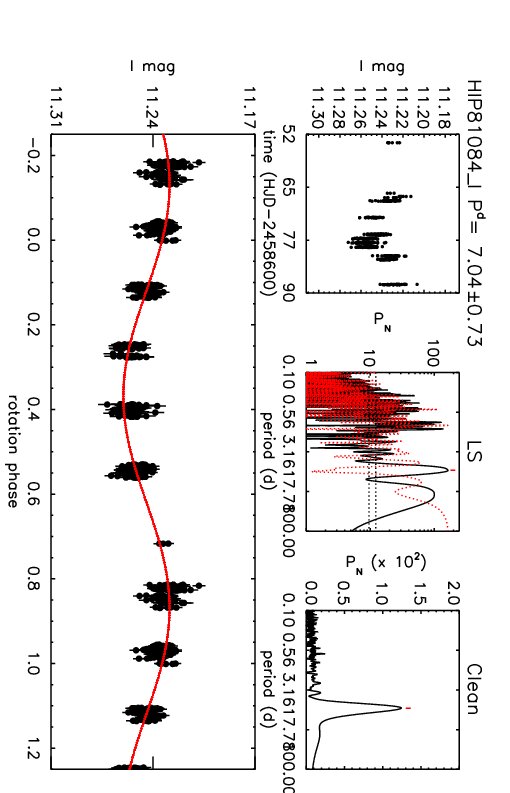}
    \caption{Photometric time sequence and periodogram for HIP 81084 (I band)}
    \label{fig:HIP81084_I}
\end{figure}

\item {\bf  TYC 7879-0980-1 = HD 326277}  
First identified as a young star in \cite{sacy} and more recently classified as a UCL member by \cite{pecaut2016}. Gaia DR2 kinematics coupled to Banyan $\Sigma$ confirms the UCL membership. The age indicators are fully compatible  with this assignment. There is some discrepancy between the \cite{sacy} spectral type (K0IV) and the photometric colors, which would suggest instead a G7 star  \cite[from young stars ][ tables]{pecaut2013}. The isochrone ages for the temperatures corresponding to K0 and G7 are 14-24 Myr respectively, further supporting the young age and bracketing the nominal UCL age. There is a significant (3.1 $\sigma$) proper motion difference between Tycho2 and Gaia DR2, but no other indication of binarity.

\item {\bf HIP 82388}
We measured for the first time the rotation period from the photometric time series we collected at the YCO observatory (Fig.\,\ref{fig:HIP82388}). 

\begin{figure}[htbp]
    \centering
    \includegraphics[width=5.7cm,angle=90]{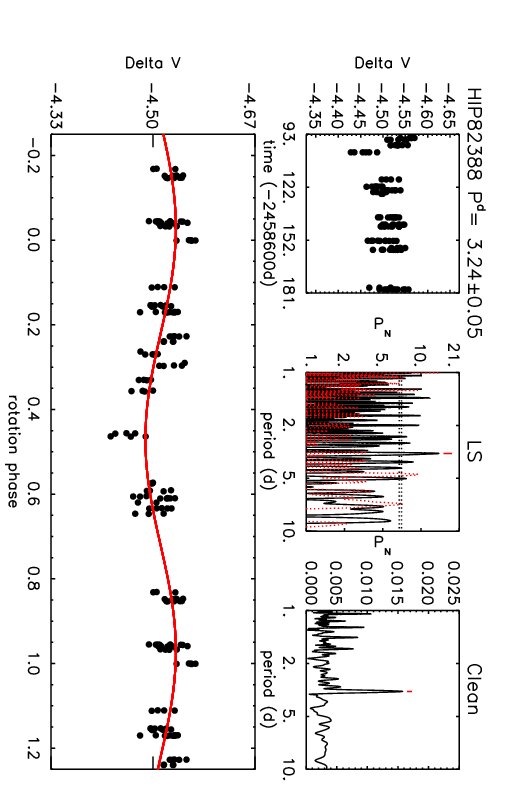}
    \caption{Photometric time sequence and periodogram for HIP 82388}
    \label{fig:HIP82388}
\end{figure}


\item {\bf TYC 7362-0724-1 = HD 156097 } 
Young G5 star with strong lithium identified by \cite{sacy}. The analysis based on the Gaia DR2 parameters yields a 46\% membership probability to UCL and 13\% to $\beta$ Pic MG. Lithium and other indicators are fully compatible with a very young age \citep{desidera2015}. Photometric colors are fully compatible with the G5 spectral classification by \cite{sacy}. Adopting $T_{\rm eff}$\ from the \cite{pecaut2013} tables, we infer an age of 11$\pm$3 Myr, consistent but more accurate than that obtained from indirect methods. We adopted the isochrone age, with UCL age as an upper limit due to the possible membership. We measured for the first time the rotation period from the photometric time series we collected at PEST, and subsequently confirmed by our analysis of the TESS data (Fig.\,\ref{fig:TYC736207241_V}-\ref{fig:TYC736207241_I}-\ref{fig:TYC736207241}).

\begin{figure}[htbp]
    \centering
    \includegraphics[width=5.7cm,angle=90]{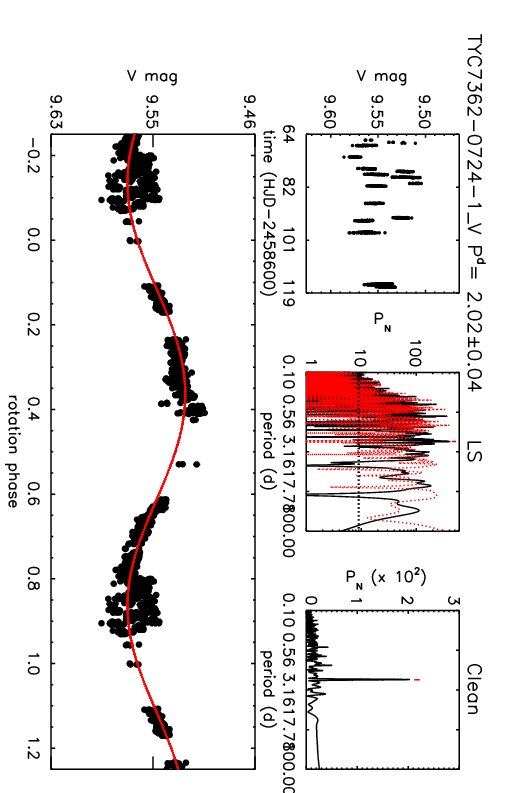}
    \caption{Photometric time sequence and periodogram for TYC 7362-0724-1 (V band; PEST data)}
    \label{fig:TYC736207241_V}
\end{figure}

\begin{figure}[htbp]
    \centering
    \includegraphics[width=5.7cm,angle=90]{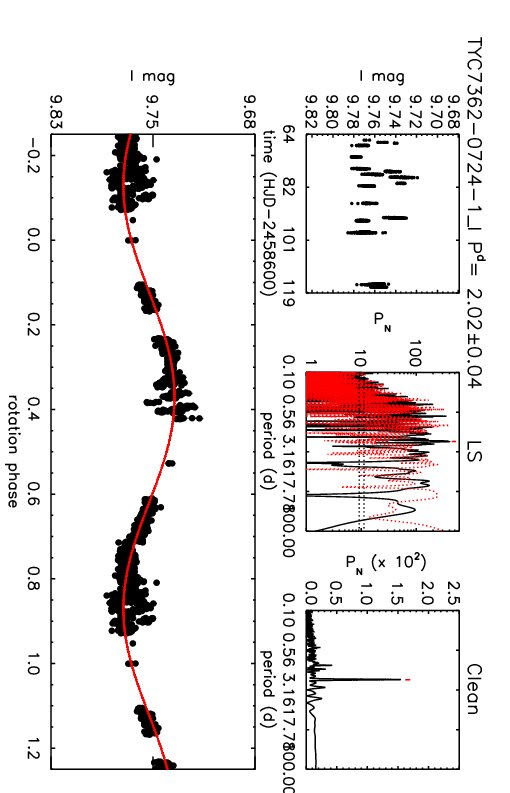}
    \caption{Photometric time sequence and periodogram for TYC 7362-0724-1 (I band; PEST data)}
    \label{fig:TYC736207241_I}
\end{figure}

\begin{figure}[htbp]
    \centering
    \includegraphics[width=5.7cm,angle=90]{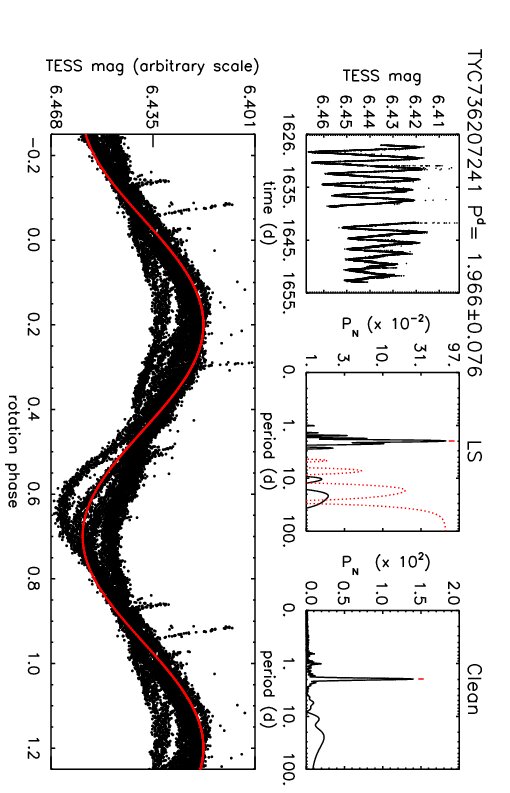}
    \caption{Photometric time sequence and periodogram for TYC 7362-0724-1}
    \label{fig:TYC736207241}
\end{figure}

\item {\bf TYC 8728-2262-1}
The  photometric  rotation period first measured by \citet{Messina17} is confirmed by our analysis of the TESS data
(Fig.\,\ref{fig:TYC872822621}). 

\begin{figure}[htbp]
    \centering
    \includegraphics[width=5.7cm,angle=90]{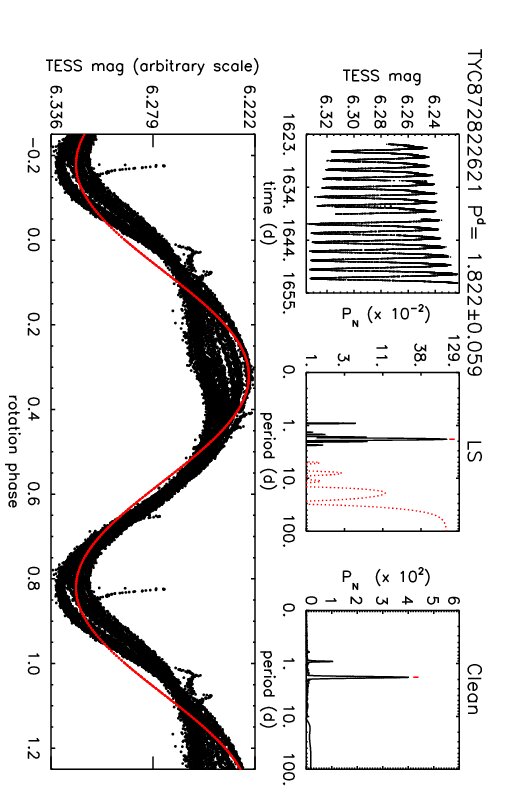}
    \caption{Photometric time sequence and periodogram for TYC 8728-2262-1}
    \label{fig:TYC872822621}
\end{figure}



\item {\bf HIP 86598 = HD 160305}
Star with debris disk spatially resolved from SHINE observations \citep{perrot2019} (see Fig.\,\ref{fig:HIP86598} for the stellar rotation
analysis).

\begin{figure}[htbp]
    \centering
    \includegraphics[width=5.7cm,angle=90]{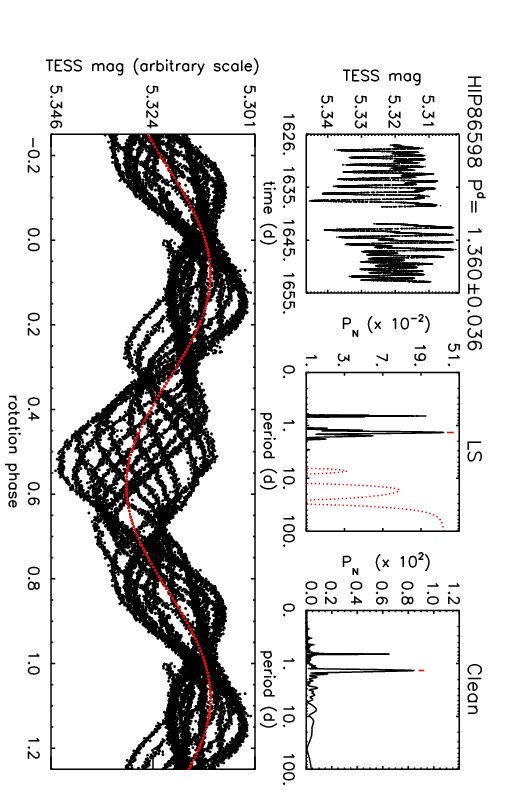}
    \caption{Photometric time sequence and periodogram for HIP 86598}
    \label{fig:HIP86598}
\end{figure}

\item {\bf HIP 88399}
We measured for the first time the rotation period from the TESS photometric time series
(Fig.\,\ref{fig:HIP88399}). 

\begin{figure}[htbp]
    \centering
    \includegraphics[width=5.7cm,angle=90]{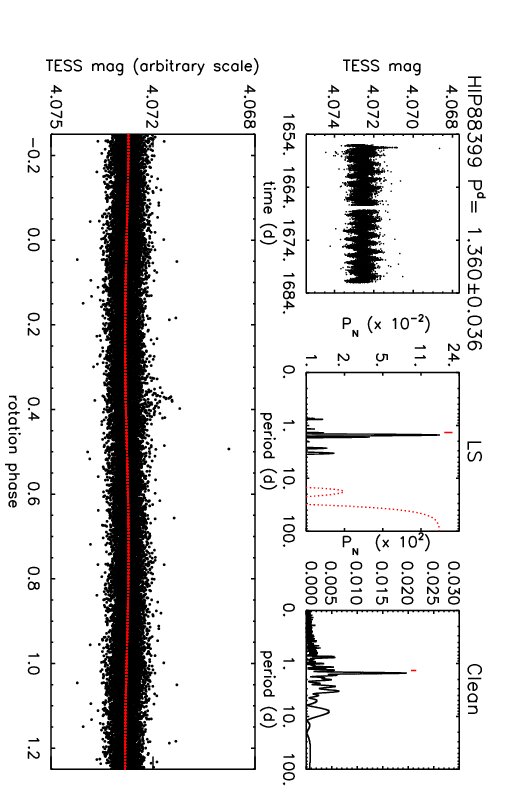}
    \caption{Photometric time sequence and periodogram for HIP 88399}
    \label{fig:HIP88399}
\end{figure}


\item {\bf TYC 9073-0762-1}
The  photometric  rotation period first measured by \citet{Messina10} is confirmed by our analysis of the TESS data.
(Fig.\,\ref{fig:TYC907307621}).

\begin{figure}[htbp]
    \centering
    \includegraphics[width=5.7cm,angle=90]{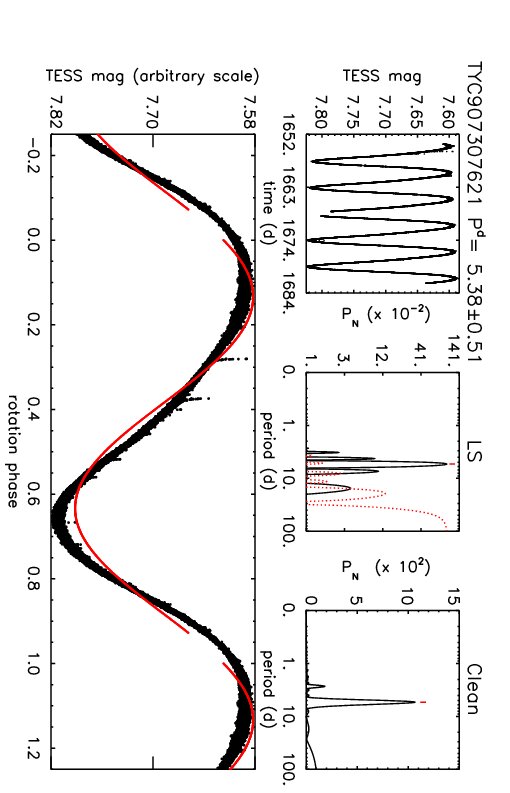}
    \caption{Photometric time sequence and periodogram for TYC 9073-0762-1}
    \label{fig:TYC907307621}
\end{figure}


\item {\bf PZ Tel = HIP 92680 = HD 174429}
Member of $\beta$ Pic MG. It hosts a BD companion (PZ Tel B) 
discovered by \cite{biller2010,mugrauer2010} in a very eccentric orbit \citep{maire2016}. The IR excess detected by \citet{rebull2008} has been shown to be due to a background object \citep{biller2013}.

\item {\bf HIP 92984 = HD 175726}  
The star shows moderate activity and fast rotation. The star has kinematic parameters somewhat similar to UMa although with some differences causing low membership probability in BANYAN $\Sigma$. It also has an IR excess suggesting the presence of a debris disk. The age indicators quite consistently indicate that it is intermediate between the Hyades and Pleiades, independently of UMa membership. We adopted an age of 400$\pm$200 Myr.

\item {\bf HIP 93375}
As suspected in \citet{desidera2015}, Gaia DR2 astrometry shows conclusively that the star UCAC3 123-585870 at 11$^{\prime\prime}$ is not physically associated.
We measured for the first time the rotation period from the ROAD photometric time series in the V and B bands.
(Fig.\,\ref{fig:HIP93375_B}-Fig.\,\ref{fig:HIP93375_V}). 

\begin{figure}[htbp]
    \centering
    \includegraphics[width=5.7cm,angle=90]{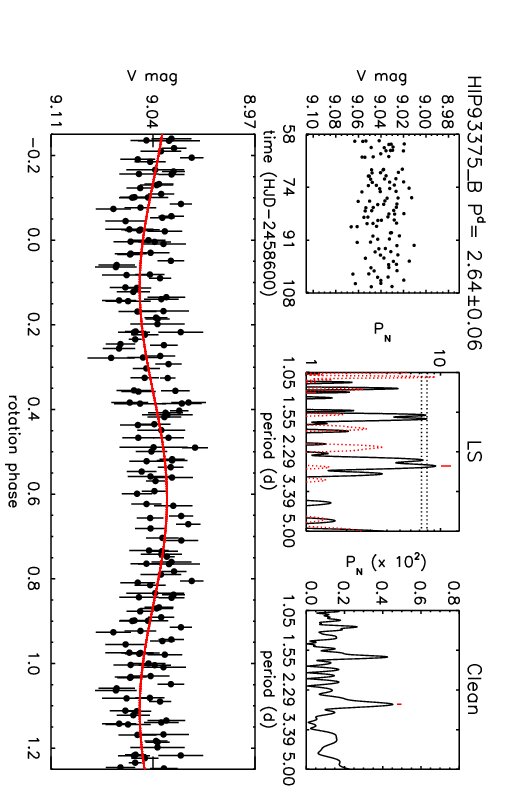}
    \caption{Photometric time sequence and periodogram for HIP 93375 (B band; ROAD data)}
    \label{fig:HIP93375_B}
\end{figure}

\begin{figure}[htbp]
    \centering
    \includegraphics[width=5.7cm,angle=90]{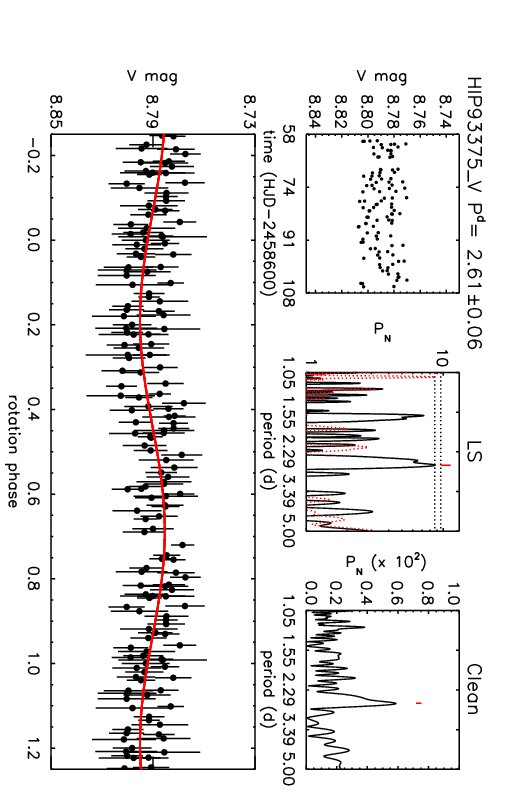}
    \caption{Photometric time sequence and periodogram for HIP 93375 (V band; ROAD data)}
    \label{fig:HIP93375_V}
\end{figure}


\item {\bf TYC 8760-1468-1 = CD-54 8168 } 
Field K2Ve object with very fast rotation and high Li content, similar to the members of the Tuc-Hor association. The RV in RAVE DR5 \citep{rave5} differs by 22 km\,s$^{-1}$ with respect to the SACY value, but the error is very large (6 km\,s$^{-1}$). We thus consider it a suspected SB. The  photometric  rotation period first measured by \citet{Kiraga12} is confirmed by our analysis of the TESS data (Fig.\,\ref{fig:TYC876014681}). 
The kinematic within the \cite{Zuckerman2004}  ``young box'' is compatible with the young age estimated from lithium and rotation.

\begin{figure}[htbp]
    \centering
    \includegraphics[width=5.7cm,angle=90]{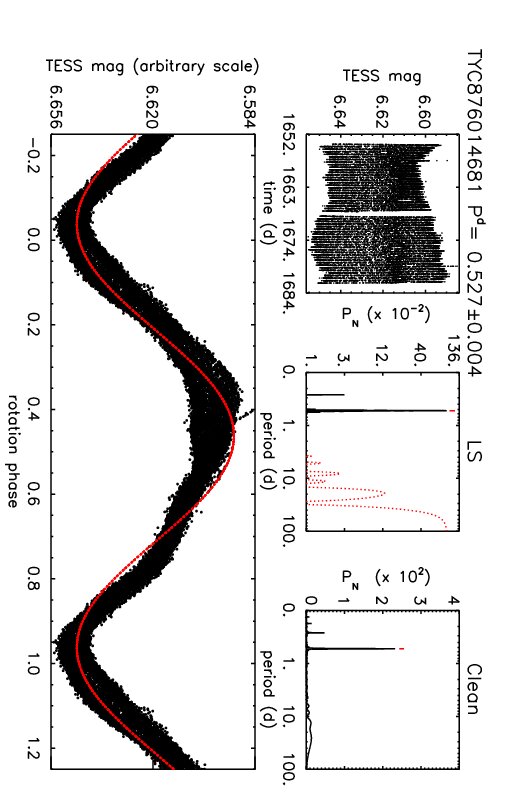}
    \caption{Photometric time sequence and periodogram for TYC 8760-1468-1}
    \label{fig:TYC876014681}
\end{figure}


\item {\bf HIP 95270}
We measured for the first time the rotation period from the TESS photometric time series
(Fig.\,\ref{fig:HIP95270}). 

\begin{figure}[htbp]
    \centering
    \includegraphics[width=5.7cm,angle=90]{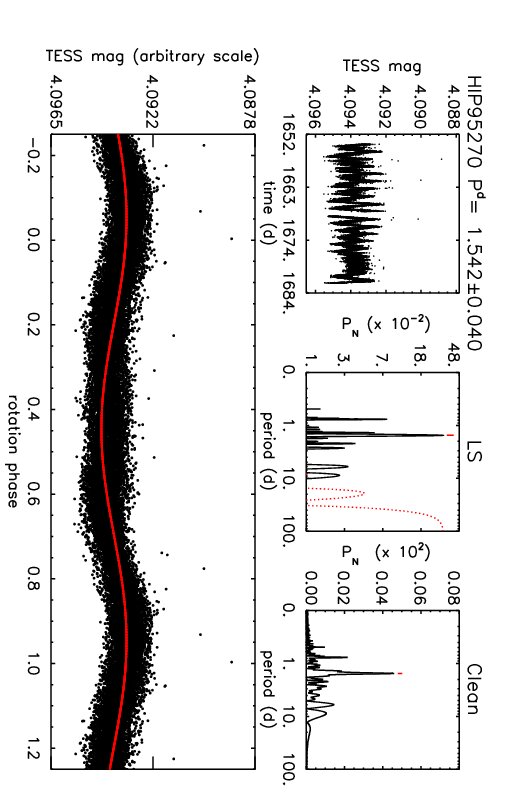}
    \caption{Photometric time sequence and periodogram for HIP 95270}
    \label{fig:HIP95270}
\end{figure}

\item {\bf $\eta$ Tel = HIP 95261 }
The star has a brown dwarf companion discovered by \citet{lowrance2000} and confirmed by \citet{gunther2001}.
It was not promoted as special object (P0) considering the already
available characterization \citep[e.g.,][]{bonnefoy2014}. 


\item {\bf TYC 0486-4943-1} 
Member of AB Dor MG according to \citet{torres2008} and \citet{elliott2014}. It has a low membership probability (19\% for the adopted kinematic parameters) using BANYAN $\Sigma$. The lithium EW is very close to the median locus of AB Dor and Pleiades members, supporting a very similar age. The other age indicators are also compatible with this evaluation. 
\cite{barenfeld2013} found some differences in the chemical composition with respect to AB Dor core members. We adopted an age very close to that of AB Dor MG with slightly increased error bars. From Gaia DR2 a wide companion ({\bf 2MASS J19330197+0345484}) with very similar parallax and proper motion is identified at 28$^{\prime\prime}$ (1968 au). The Gaia DR2 RV is also compatible with that of primary.

\item {\bf TYC 7443-1102-1}
The Herschel IR source is actually identified with two
separate sources at close separation from the star in ALMA  observations \citep{tanner2020}.
This indicates they are likely background objects rather than associated with the star.

\item {\bf HD 189285}  
It is classified as a member of AB Dor in some studies \citep{torres2008}, but BANYAN $\Sigma$ returns a 0.0\% membership probability. The discrepancy was already noticed in \cite{desidera2015} using previous versions of the tool. On the other hand, all the age indicators \citep[see ][for details]{desidera2015}  are fully compatible with membership and \cite{barenfeld2013} found that a chemical composition from several chemical elements is compatible with those of AB Dor core members. The kinematic discrepancy is unlikely to be due to unrecognized binarity as the RV from several sources \citep{desidera2015,gaiadr2,elliott2014,frasca2018} is compatible within the errors, and the SPHERE images do not give any indication of stellar companions. In summary, independently of any membership assignment, we adopted an age close to that of AB Dor MG with slightly increased error bars.
We measured the rotation period from the  photometric time series we collected at the ROAD observatory, which superseded the measurement presented in \citet{desidera2015} and therein flagged as uncertain
(Fig.\,\ref{fig:HD189285_B}-\ref{fig:HD189285_V}). 

\begin{figure}[htbp]
    \centering
    \includegraphics[width=5.7cm,angle=90]{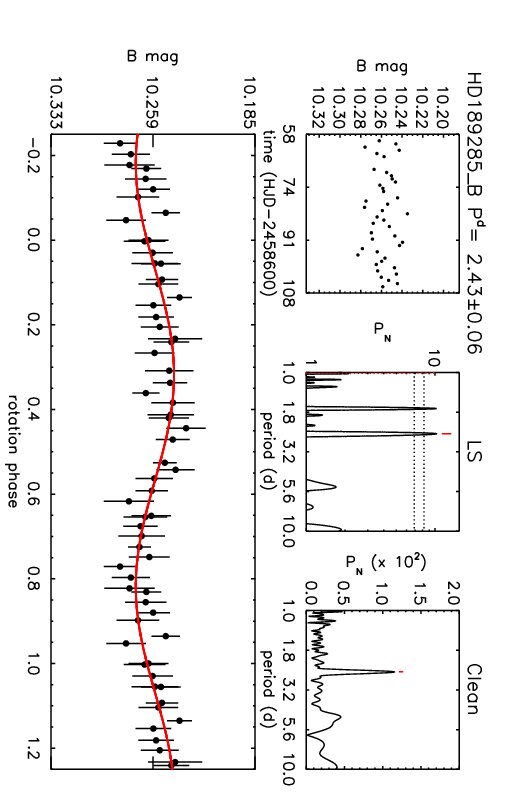}
    \caption{Photometric time sequence and periodogram for HD189285 (B band; ROAD data)}
    \label{fig:HD189285_B}
\end{figure}

\begin{figure}[htbp]
    \centering
    \includegraphics[width=5.7cm,angle=90]{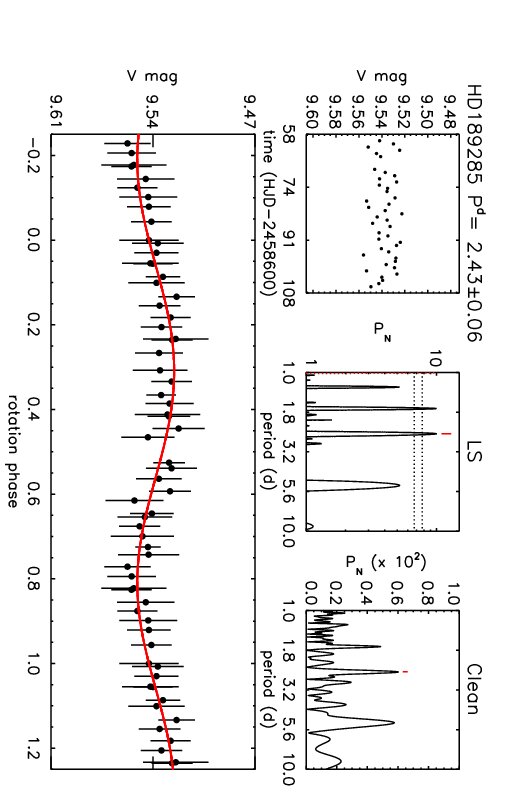}
    \caption{Photometric time sequence and periodogram for HD189285 (V band; ROAD data)}
    \label{fig:HD189285_V}
\end{figure}

\item {\bf HIP 98470 = HD 189245}
A reanalysis of the Hipparcos data allowed us to detect a rotation period P = 0.8662$\pm$0.0003\,d, which, differently from that presented in \citet{desidera2015}, is consistent with the stellar radius and projected rotational velocity (Fig.\,\ref{fig:HIP98470}). 

\begin{figure}[htbp]
    \centering
    \includegraphics[width=5.7cm,angle=90]{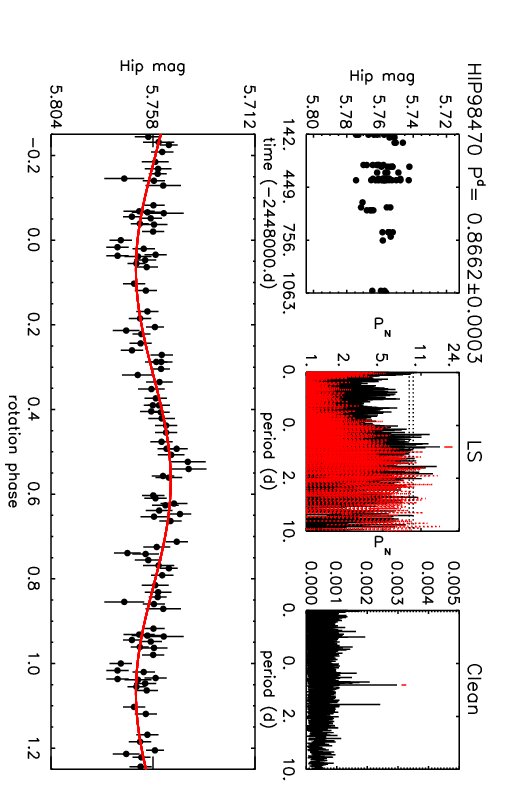}
    \caption{Photometric time sequence and periodogram for HIP 98470}
    \label{fig:HIP98470}
\end{figure}




\item {\bf TYC 8404-0354-1 = CD-52 9381} 
K6Ve star proposed as an Argus member by \cite{torres2008}. Considered as likely older than Argus by \cite{zuckerman2018} on the basis of CMD. Lithium is also lower than expected for a 50 Myr star, and similar to the mean values of Pleaides and AB Dor stars of similar color. The very fast rotation period (0.83 days) is also compatible with this age estimate. Our isochrone analysis confirms the position close to the ZAMS. We adopted 120 (50-200) Myr. The  photometric  rotation period first measured by \citet{Messina11} is confirmed by our analysis of the TESS data (Fig.\,\ref{fig:TYC840403541}). Numerous flare events are detected in the TESS time series. 

\begin{figure}[htbp]
    \centering
    \includegraphics[width=5.7cm,angle=90]{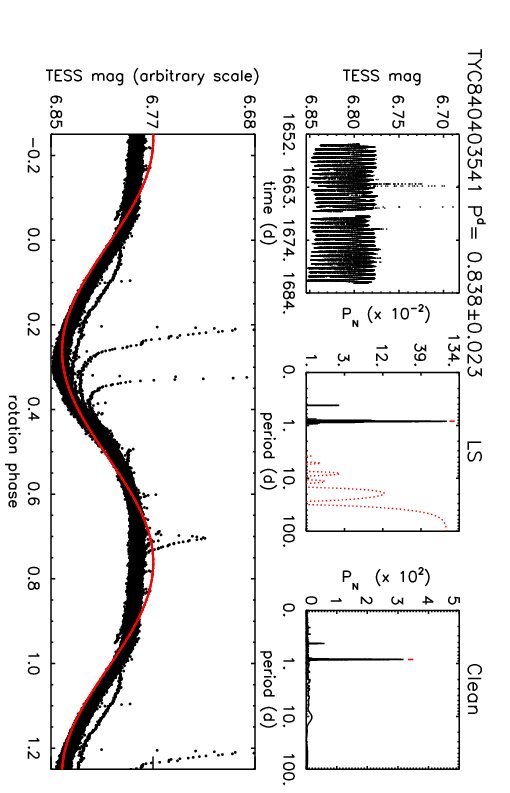}
    \caption{Photometric time sequence and periodogram for TYC 8404-0354-1}
    \label{fig:TYC840403541}
\end{figure}



\item {\bf AU Mic = HIP 102409}
The star has a prominent debris disk seen close to edge-on first spatially resolved by \citet{liu2004}. A transiting planet
at short period has been discovered with TESS \citep{plavchan2020}.
The  photometric  rotation period measured by \citet{Messina10} is confirmed by our analysis of the TESS data
(Fig.\,\ref{fig:HIP102409}). Numerous flare events are detected in the TESS time series.

\begin{figure}[htbp]
    \centering
    \includegraphics[width=5.7cm,angle=90]{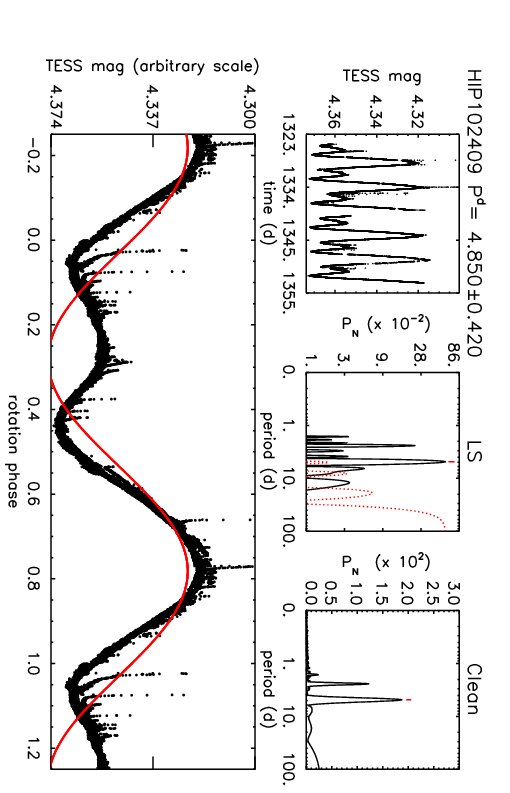}
    \caption{Photometric time sequence and periodogram for HIP 102409}
    \label{fig:HIP102409}
\end{figure}

\item {\bf HIP 102626 = HD 197890 = BO Mic }  
Very fast-rotating star (P=0.380~d; $v\sin{i}$ 128 km\,s$^{-1}$). This characteristic makes the determination of the spectral parameters difficult. \cite{sacy} determined a large Li EW, indicating a very young age. It was proposed by some studies \citep{kraus2014,bell2015} to be a member of the Tuc-Hor association, but a very low probability is returned by the kinematic analysis including Gaia DR2. The controversial membership is linked to the wide dispersion of the astrometric parameters\footnote{When using the Tycho2 long-term proper motion coupled with the Hipparcos parallax, BANYAN returns a membership probability of 82.4\% in Tuc-Hor}. Gaia and Hipparcos parallaxes and even different Hipparcos reductions show large differences (up to 7.5 and 3.4 mas, respectively) while the proper motions derived from Gaia and Hipparcos differ by more than 20 mas yr$^{-1}$. The original Hipparcos reduction includes an astrometric acceleration trend, while Gaia DR2 notes the presence of a large excess of astrometric noise. This suggests the presence of a fairly massive stellar companion that was not revealed in any in the various direct imaging surveys that targeted this object \citep{chauvin2010,galicher2016},  including SHINE-SPHERE (see Paper II). Radial velocities are hardly conclusive because of the extreme $v\sin{i}$ value. \cite{barnes2005} discussed the possible spectroscopic companions compatible with the observational constraints.
The isochrone age results very young when adopting Gaia DR2 parallax (4 Myr for the $T_{\rm eff}$\ corresponding to the K3 spectral type by \cite{sacy}; 8 Myr for the $T_{\rm eff}$\ corresponding to the K2 spectral type indicated by broadband colors), but we consider this highly uncertain as the errors on parallax are possibly underestimated. The minimum radius from the observed rotation period and $v\sin{i}$ is 0.99 $R_{\odot}$. The inclination value of 70 deg proposed by \cite{barnes2005} through the Doppler imaging technique implies R=1.05 $R_{\odot}$, compatible with a pre-main sequence star of early K spectral type. We conclude that membership in Tuc-Hor cannot be ruled out until the spread in the astrometric values and the possibility of binarity are better understood. The very strong lithium line in any case indicates an age younger than 100 Myr. We thus adopted the Tuc-Hor age, with min--max values of 5--100 Myr considering the various indicators.\\
The  photometric  rotation period measured by \citet{Kiraga12} is confirmed by our analysis of the TESS data (Fig.\,\ref{fig:HIP102626}). 

\begin{figure}[htbp]
    \centering
    \includegraphics[width=5.7cm,angle=90]{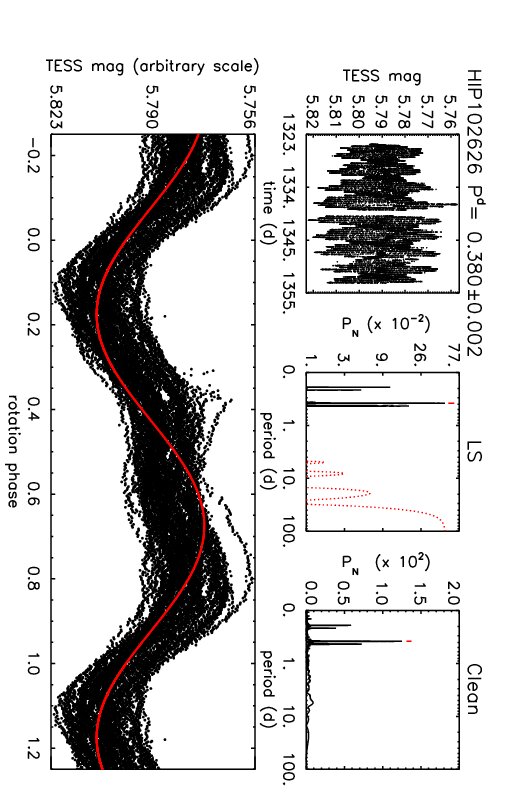}
    \caption{Photometric time sequence and periodogram for HIP 102626}
    \label{fig:HIP102626}
\end{figure}

\item {\bf TYC 1090-0543-1} 
Star with a low membership probability on AB Dor MG (8.8\%) in spite of the previous kinematic assignment \citep{torres2008}. It is a wide companion of {\bf HD199058}, which is itself a tight visual binary \citep{nacolp}, making the system triple. \cite{barenfeld2013} found the HD 199058 chemical pattern to be compatible with that of AB Dor core members. Lithium and the activity and rotation indicators of both components are compatible with those of AB Dor and Pleiades of similar spectral type.

\item {\bf HIP 104365 = HD 201184 = $\chi$ Cap} 
A0V star, flagged as a possible member of Tuc-Hor by \citet{zuckerman2012}. BANYAN $\Sigma$ yields a 20\% membership probability (80\% field). We  adopted the age from isochrone fitting, with the minimum value set at the minimum age of Tuc-Hor. This is in any case close to the lower limit allowed by stellar models. This is a triple system, as there is a close pair of comoving objects \citep{vigan2012} at 9$^{\prime\prime}$ labeled  WDS 21086-2112E and WDS 21086-2112F. There is one corresponding entry in Gaia DR2, Gaia DR2 6832248844207846144, without astrometric parameters, likely because of the multiplicity.


\item {\bf HIP 105388}
The  photometric  rotation period first measured by \citet{Messina10} is confirmed by our analysis of the TESS data
(Fig.\,\ref{fig:HIP105388}). 

\begin{figure}[htbp]
    \centering
    \includegraphics[width=5.7cm,angle=90]{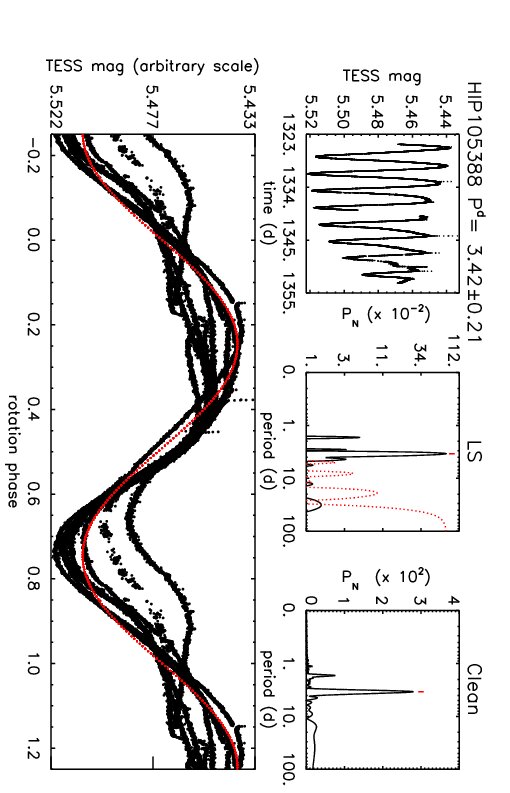}
    \caption{Photometric time sequence and periodogram for HIP 105388}
    \label{fig:HIP105388}
\end{figure}

\item {\bf TYC 9482-121-1}
We measured for the first time the rotation period from the TESS photometric time series
(Fig.\,\ref{fig:TYC94821211}). 

\begin{figure}[htbp]
    \centering
    \includegraphics[width=5.7cm,angle=90]{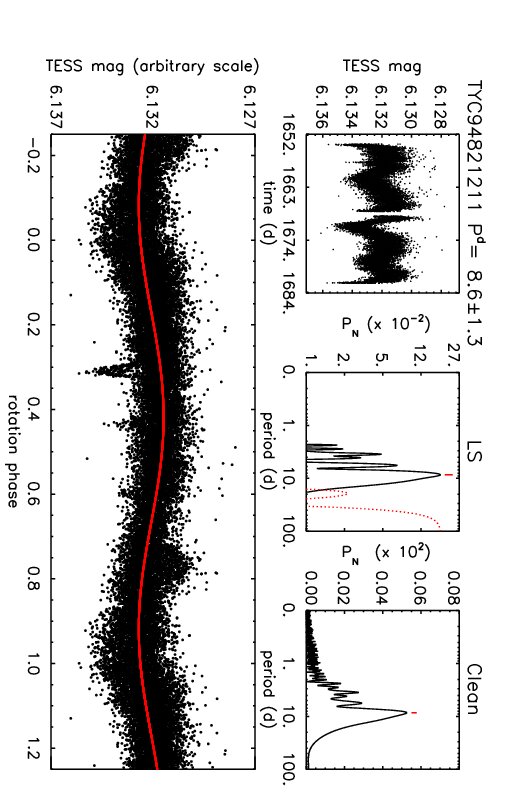}
    \caption{Photometric time sequence and periodogram for TYC 94821211}
    \label{fig:TYC94821211}
\end{figure}

\item {\bf HIP 107345}
The  photometric  rotation period first measured by \citet{Messina10} is confirmed by our analysis of the TESS data
(Fig.\,\ref{fig:HIP107345}). A large amplitude flare is detected in the TESS time series. 

\begin{figure}[htbp]
    \centering
    \includegraphics[width=5.7cm,angle=90]{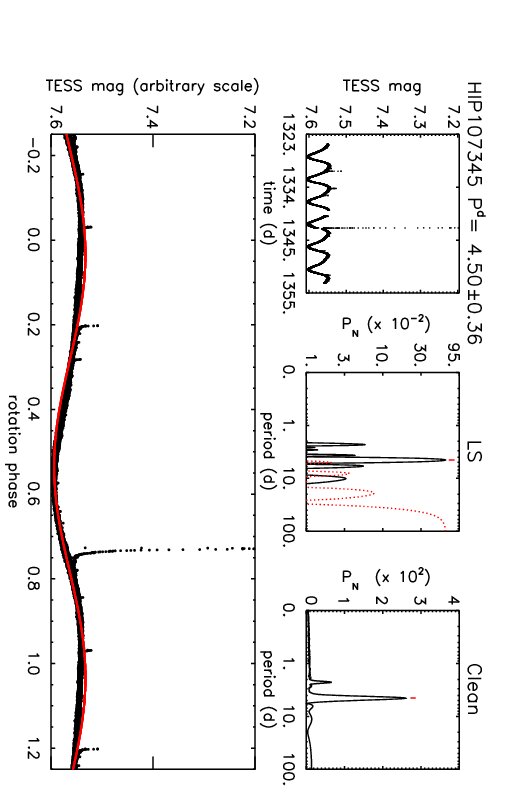}
    \caption{Photometric time sequence and periodogram for HIP 107345}
    \label{fig:HIP107345}
\end{figure}

\item {\bf HIP 107350 = HD 206860 = HN Peg}
The star has a low-mass brown dwarf companion at wide separation.

\item {\bf HIP 107412 = HD 206893} 
Star with a debris disk and with a substellar companion detected
by \citet{milli2017} and characterized by \citet{delorme2017} and
\citet{grandjean2019} (see \cite{delorme2017} for further details on stellar parameters).

\item {\bf 2MASS J22021616-4210329}
The  photometric  rotation period first measured by \citet{kiraga2012} is confirmed by our analysis of the TESS data
(Fig.\,\ref{fig:2MASSJ220-42103}). Numerous flare events are detected in the TESS time series. 

\begin{figure}[htbp]
    \centering
    \includegraphics[width=5.7cm,angle=90]{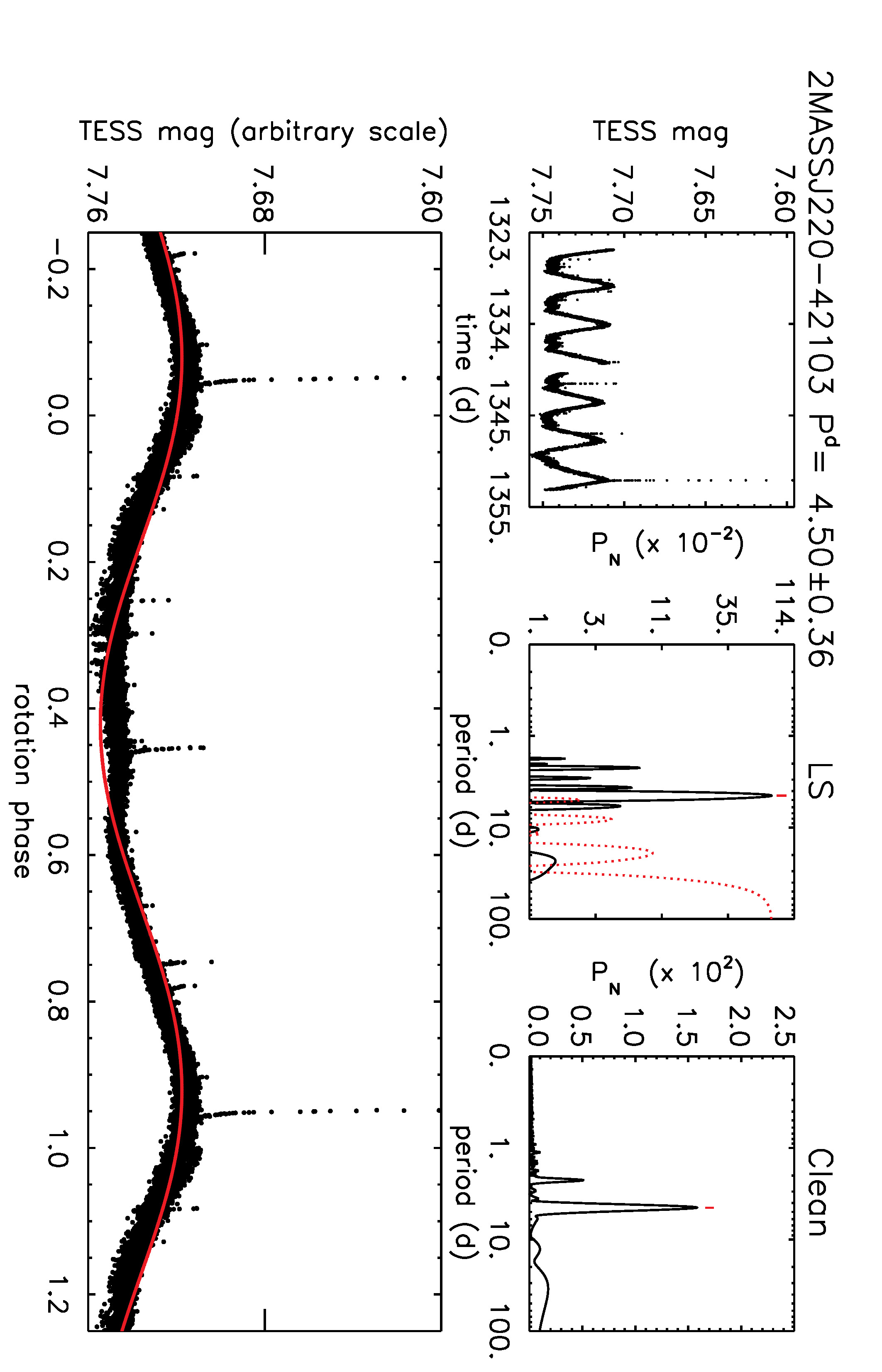}
    \caption{Photometric time sequence and periodogram for 2MASSJ220-42103}
    \label{fig:2MASSJ220-42103}
\end{figure}

\item {\bf TYC 9340-0437-1}
Star with debris disk spatially resolved by Herschel observations \citep{tanner2020}.
The  photometric  rotation period first measured by \citet{Messina10} is confirmed by our analysis of the TESS data
(Fig.\,\ref{fig:TYC934004371}). Numerous flare events are detected in the TESS time series.

\begin{figure}[htbp]
    \centering
    \includegraphics[width=5.7cm,angle=90]{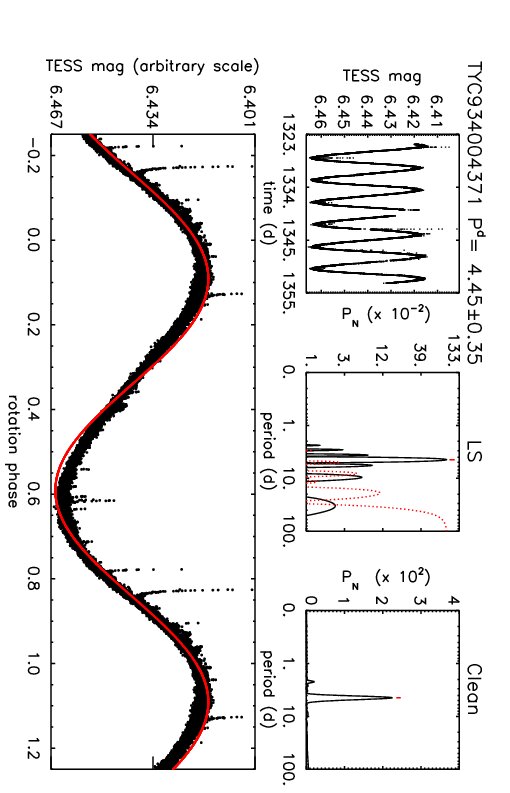}
    \caption{Photometric time sequence and periodogram for TYC 9340-0437-1}
    \label{fig:TYC934004371}
\end{figure}


\item {\bf HIP 113283 = TW PsA = Fomalhaut B}
IR excess at 160 micron detected by \citet{montesinos2016}, but not seen at shorter wavelengths, indicating very cold dust. The star has a significant Gaia--Hipparcos proper motion difference. A dedicated search using imaging and radial velocities did not detect companions responsible for the astrometric signature \cite{derosa2019}. Our even deeper imaging observations confirm this result. The age of the system is from \cite{mamajek2012}.
The  photometric  rotation period first measured by \citet{Busko78} is confirmed by our analysis of the TESS data
(Fig.\,\ref{fig:HIP113283}). 

\begin{figure}[htbp]
    \centering
    \includegraphics[width=5.7cm,angle=90]{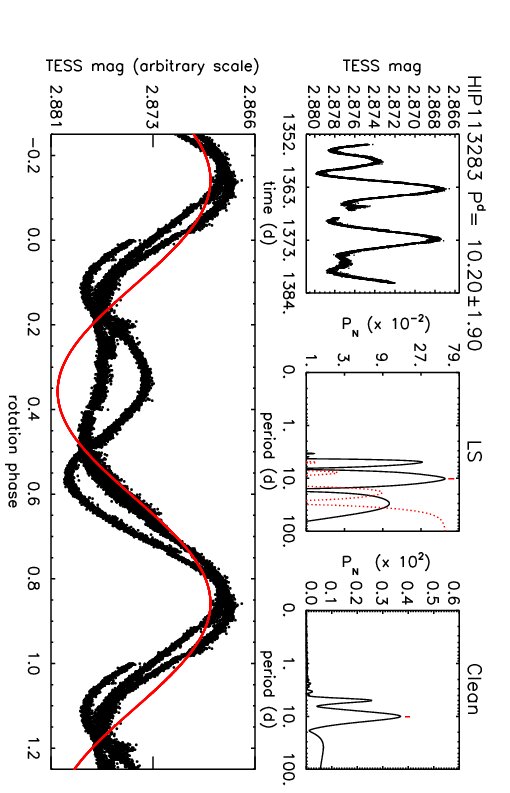}
    \caption{Photometric time sequence and periodogram for HIP 113283}
    \label{fig:HIP113283}
\end{figure}

\item {\bf Fomalhaut = HIP 113368} 
The age of the system is from \cite{mamajek2012}. The controversial planet candidate Fomalhaut b \cite{kalas2008,janson2012,lawler2015} is well outside the field of view of SPHERE.

\item {\bf HIP 114189 = HD 218396 = HR 8799}
Star with the first multi-planetary system detected through
imaging \citep{marois2008,marois2010}.
It was proposed as a member of the Columba association by \cite{marois2010}. BANYAN returns a moderate membership probability (49\%)  to this group. Very recently, \cite{lee2019} proposed it as a probable member of $\beta$ Pic MG. A younger age would imply lower masses for the four planets orbiting the star. This would expand the extremely narrow space of parameters that fit the astrometric data ensuring at the same time dynamical stability \citep{esposito2013,wang2018}. We adopted the Columba age with a lower limit extending to $\beta$ Pic MG age.

\item {\bf HIP 114530}
The  photometric  rotation period first measured by \citet{Messina10} is confirmed by our analysis of the TESS data
(Fig.\,\ref{fig:HIP114530}). It is interesting to note that the light curve clearly shows evidence of a secondary small-amplitude periodicity superimposed on the P = 5.10\,d rotation period. In Fig.\,\ref{fig:HIP114530_2} we show that a period P = 0.3493$\pm$0.0022\,d is detected by both Lomb--Scargle and Clean with a rotational modulation amplitude of about 0.02\,mag. Considering the stability of the light curve phased with this short period,  compared to the short timescale evolution of that phased with the longer period, we suspect that such a short periodicity does not arise from magnetic activity rather than the ellipsoidal effect of a likely close binary star observed within the TESS aperture radius.

\begin{figure}[htbp]
    \centering
    \includegraphics[width=5.7cm,angle=90]{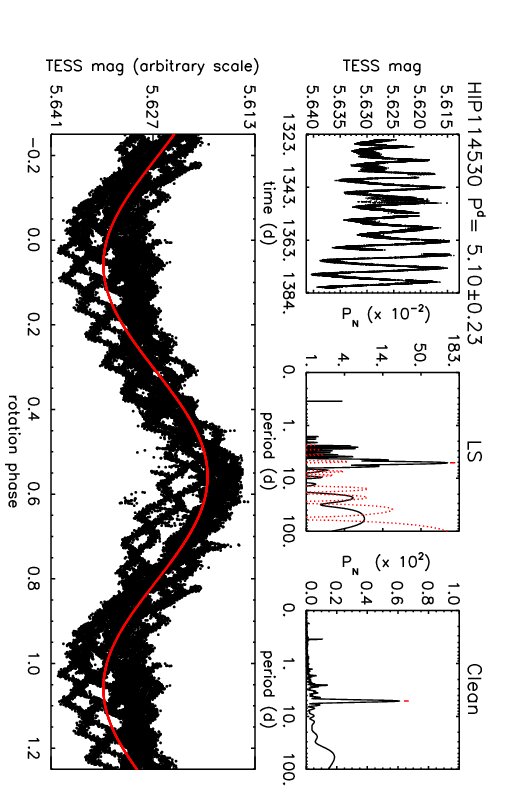}
    \caption{Photometric time sequence and periodogram for HIP 114530}
    \label{fig:HIP114530}
\end{figure}
\begin{figure}[htbp]
    \centering
    \includegraphics[width=8cm,angle=0]{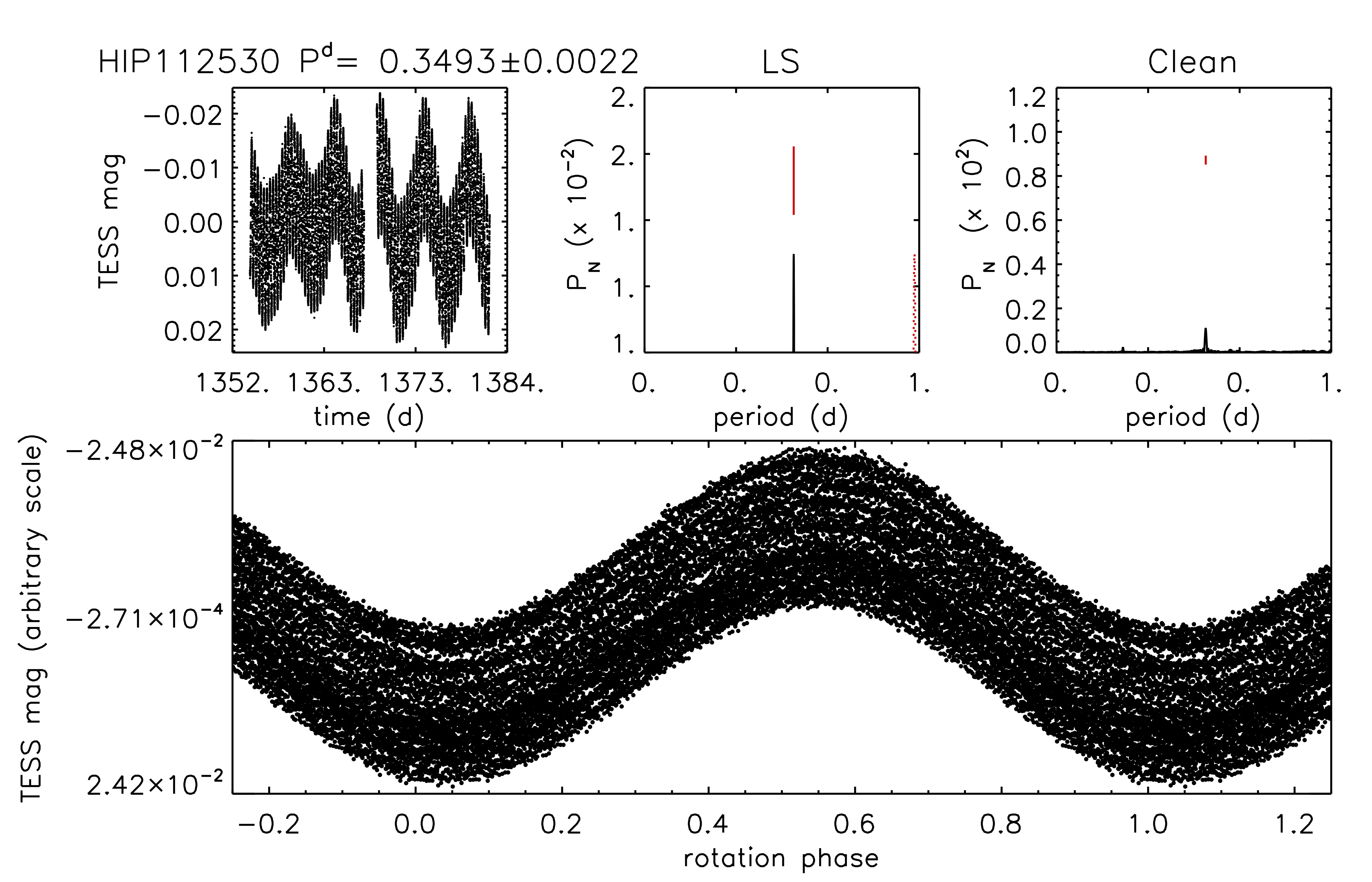}
    \caption{Photometric time sequence and periodogram for HIP 114530 after filtering out the P = 5.10\,d rotational modulation.}
    \label{fig:HIP114530_2}
\end{figure}

\item {\bf HIP 114948}
We measured for the first time the rotation period from the TESS photometric time series
(Fig.\,\ref{fig:HIP114948}).

\begin{figure}[htbp]
    \centering
    \includegraphics[width=5.7cm,angle=90]{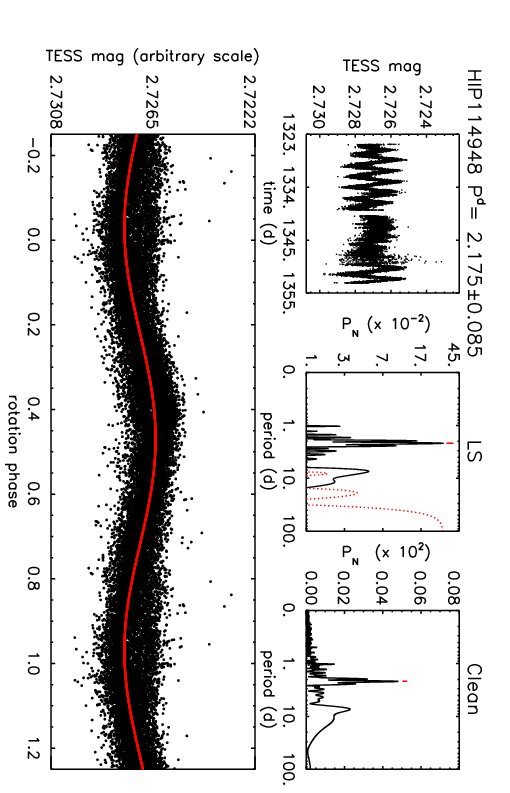}
    \caption{Photometric time sequence and periodogram for HIP 114948}
    \label{fig:HIP114948}
\end{figure}


\item {\bf HIP 118008}
We measured for the first time the rotation period from the TESS photometric time series
(Fig.\,\ref{fig:HIP118008}). 

\begin{figure}[htbp]
    \centering
    \includegraphics[width=5.7cm,angle=90]{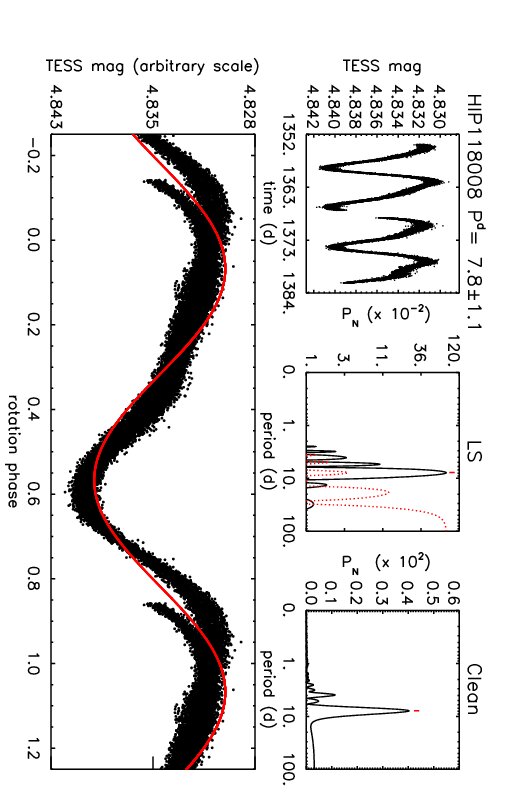}
    \caption{Photometric time sequence and periodogram for HIP 118008}
    \label{fig:HIP118008}
\end{figure}

\end{description}

\clearpage

\clearpage

\end{appendix}

\addtocounter{table}{3}
\clearpage
\onecolumn

\setcounter{table}{4}

 								        
Parallax and proper motion references: Gaia2: Gaia DR2 \citet{gaiadr2}, VL07: \citet{vl07}. 
RV references: Des20: this paper; Car18: \citet{Carleo18}; Cha17: \citet{chauvin2017b}; Che18: \citet{cheetham2018}; Che11: \citet{chen2011}; Des15: \citet{desidera2015}; Duf95: \citet{duflot95}; Ell14: \citet{elliott2014};  Eva67: \citet{evans1967}; Fer08: \citet{fernandez2008}; Gaia2: \citet{gaiadr2};  Gon06: \citet{gontcharov2006}; Kis11: \citet{kiss2011}; Kra14: \citet{kraus2014}; Mal14: \citet{malo2014}; Mam13: \citet{mamajek2013}; Mon01: \citet{montes2001}; Nid02: \citet{nidever2002}; Nor04: \citet{nordstrom2004}; SACY: \citet{sacy}; XHIP: \citet{xhip}; Whi07: \citet{white2007}; Zuk01: \citet{zuckerman2001}; Zuk11: \citet{zuckerman2011}

Source of data for previously unpublished RV determinations. ESO Program ID is provided for HARPS spectra:
HIP 490: 074.C-0037, 075.C-0202, 076.C-0010, 192.C-0224, 099.C-0205;
HIP 1113: 084.C-1039, 192.C-0224;
HIP 1481: 074.C-0037, 192.C-0224;
HIP 1993: 074.C-0037, 075.C-0202, 077.C-0012;
HIP 6276: 192.C-0224, 098.C-0739; 
HIP 6856: 078.D-0245;
HIP 13402: 192.C-0224, 098.C-0739; 
TYC 5882-1169-1: 074.C-0037, 076.C-0010; 
TYC 5899-0026-1: 097.C-0864;
HIP 23200: 192.C-0224, 097.C-0864, 098.C-0739;
HIP 23309: 074.C-0037, 076.C-0010, 192.C-0224;
HIP 25283: 072.C-0488, 192.C-0224, 098.C-0739;
HIP 30034: 074.C-0037, 076.C-0010, 078.D-0245, 192.C-0224, 098.C-0739;
HIP 30314: 083.C-0794, 084.C-1039, 192.C-0224, 098.C-0739;
GSC 8894-0426: 097.C-0864; 
HIP 31878: 192.C-0224, 098.C-0739;
HIP 32235: 074.C-0037, 076.C-0010;
HIP 36948: 184.C-0815, 192.C-0224, 0102.C-0584;
HIP 44526: 192.C-0224;
HIP 47135: 074.C-0037, 076.C-0010;
HIP 51228: 074.C-0364;
HIP 51317: 072.C-0488, 082.C-0718, 183.C-0437, 191.C-0873, 192.C-0224, 099.C-0205;
TWA 7: 074.C-0037, 075.C-0202, 076.C-0010, 077.C-0012, 079.C-0046;
HIP 54155: 074.C-0037, 075.C-0202, 082.C-0427, 082.C-0390;
HIP 66252: 075.C-0202, 192.C-0224;
HIP 69989: \citet{borgniet2019}, reduced spectra available on OHP archive;
HIP 81084: 097.C-0864;
HIP 95270: 083.C-0794, 084.C-1039, 192.C-0224;
HIP 102409: 075.C-0202, 192.C-0224, 098.C-0739, 099.C-0205;
HIP 107350: 192.C-0224, 099.C-0205;
HIP 107412: 098.C-0739, 192.C-0224, 099.C-0205, 1101.C-0557;
HIP 113283: 60.A-9036;
HIP 114530: 192.C-0224;
HIP 114948: 075.C-0689, 077.C-0295, 080.C-0712, 184.C-0815;
HIP 118008: 084.C-1039, 192.C-0224, 099.C-0205
\end{landscape} 



 								        
References for rotation period:
1a: this study (TESS data);
1b: this study (ASAS data);
1c: this study (ROAD data);
1d: this study (K2 data);
1e: this study (PEST data);
1f: this study (Hipparcos data);
1g: this study (YCO data);
2: \cite{Messina10};
3: \cite{Messina17};
4: \cite{messina2001};   
5: \cite{Koen02};
6: \cite{Kiraga12};
7: \cite{Cutispoto03};
8: \cite{Carleo18};    
9: \cite{Messina11};
10: \cite{Desidera11};
11: \cite{Lawson05};
12: \cite{Cutispoto99};
13: \cite{Gaidos00};
14: \cite{Marsden11};
15: \cite{maire2016};
16: \cite{Mosser09};
17: \cite{Folsom16};
18: \cite{Noyes84};
19: \cite{delorme2017};   
20: \cite{Busko78};
21: \cite{Donahue96};
22: \cite{Oelkers18}
23: \cite{Chugainov74} \\

References for $v \sin i$, EW Li and $\log R_{HK}$
Bal95: \citet{baliunas1995};
Bar05: \citet{barnes2005};
Boe88: \citet{boesgaard1988};
Cha17: \citet{chauvin2017};
Che18: \citet{cheetham2018};
Che11: \citet{chen2011};
Cut02: \citet{Cutispoto03};
DaS09: \citet{dasilva2009};
Dav15: \citet{david2015};
Del17: \citet{delorme2017};
Des11: \citet{Desidera11};
Des15: \citet{desidera2015};
Eis13: \citet{eisenbeiss2013};
Ers03: \citet{erspamer2003};
Fav97: \citet{favata1997}
Gai00: \citet{Gaidos00};
Gra03: \citet{gray2003};
Gra06: \citet{gray2006};
Gue07: \citet{gunther2007};
Hen96: \citet{henry1996};
Hoj19: \citet{hojjatpanah2019};
Jef18: \citet{jeffers2018};
Jen06: \citet{jenkins2006};
Jen11: \citet{jenkins2011};
Kin05: \citet{king2005};
Kis11: \citet{kiss2011};
Kra14: \citet{kraus2014};
Isa10: \citet{isaacson2010};
MA10: \citet{martinezarnaiz2010};
Mal13: \citet{malo2013};
Mal14: \citet{malo2014};
Mar14: \citet{marsden2014};
Mes17: \citet{Messina17};
Mor01: \citet{mora2001};
Nor04: \citet{nordstrom2004};
Pal92: \citet{pallavicini1992};
Pri14: \citet{pribulla2014};
Rei03: \citet{reiners2003};
Rei12: \citet{reiners2012};
Roy02: \citet{royer2002};
Roy07: \citet{royer2007};
SACY: \citet{sacy};
Sch07: \citet{scholz2007};
Sch09: \citet{schroeder2009};
Ske08  \citet{skelly2008};
Tak05: \citet{takeda2005};
Ues70: \citet{uesugi1070};
Whi07: \citet{white2007};
Wic03: \citet{wichmann2003}
Wri04: \citet{wright2004};
Zik05: \citet{zickgraf2005};
Zor12: \citet{zorec2012};
Zuk01: \citet{zuckerman2001};
SPH-C: this paper (CORALIE spectra);
SPH-F: this paper (FEROS spectra);
SPH-H: this paper (HARPS spectra)

Source of data for previously unpublished determinations (ESO Program ID for HARPS and FEROS):
HIP 1113: CORALIE RV survey \citep{udry2000};
HIP 6485: 074.C-0037, 075.C-0202, 076.C-0010, 0103.C-0759 (HARPS);
TYC 5882-1169-1: 074.C-0037, 076.C-0010 (HARPS); 
HIP 25283: 074.D-0016 (FEROS);
HIP 30030: 60.A-9036, 192.C-0224 (HARPS);
HIP 30034: CORALIE  RV survey \citep{udry2000};
HIP 31878: 074.D-0016, 083.A-9003 (FEROS);
HIP 32235: 074.C-0037, 076.C-0010 (HARPS);
TYC 1355-214-1:  083.A-9003 (FEROS); 
HIP 44526: CORALIE  RV survey \citep{udry2000};
HIP 66252: 074.D-0016 (FEROS)     
HIP 95270: 083.C-0794, 084.C-1039, 192.C-0224 (HARPS)
HIP 114948: 075.C-0689, 077.C-0295, 080.C-0712, 184.C-0815 (HARPS)
\end{landscape} 


 								        
Disk flags: RES: spatially resolved disks (not only with SPHERE, only at longer wavelength in several cases); IREX: IR excess from SED fitting; IREX2B: IR excess with 
double components from SED fitting, typically from \citet{chen2014}; IREX? possible IR excess; NO: no detectable IR excess; BCK: IR excess is most likely due 
to background sources.
\end{landscape}


 	
Flags or references: MG: age from membership to moving groups, associations or open cluster; Des20: this paper; Boc19: \citet{boccaletti2019}; 
Cha17: \citet{chauvin2017}; Del17: \citet{delorme2017};  Laz18: \citet{lazzoni2018}; Mam12: \citet{mamajek2012}; Vig17: \citet{vigan2017}.
\end{landscape}

 								        
\end{landscape}


\end{document}